\documentclass[usenatbib]{mn2e}
\usepackage{apjfonts,amsfonts,amsmath,amssymb,bm,ctable,verbatim}

\newcommand{\Alf}{{Alfv\'en}}

\newcommand{\bhat}{\hat{\bf b}}

\newcommand{\etal}{et al.}

\newcommand{\acknowledgments}[1]{\begin{small}\section*{Acknowledgments}\end{small}{\noindent #1}\vspace{5pt}}
\newcommand{\datastatement}[1]{\begin{small}\section*{Data Availability Statement}\end{small}{\noindent #1}\vspace{5pt}}
\newcommand{\mun}[1]{\langle \mu^{#1}_{f} \rangle}
\newcommand{\muone}{\mun{1}}
\newcommand{\mutwo}{\mun{2}}
\newcommand{\momentgrad}{\mathcal{G}}
\newcommand{\Seff}{S^{\rm eff}}
\newcommand{\closurefunction}{\mathcal{M}_{2}}

\title[Consistent Closures for CR Dynamics]{A Consistent Reduced-Speed-of-Light Formulation of Cosmic Ray Transport Valid in Weak and Strong-Scattering Regimes}

\author[Hopkins \etal]{
\parbox[t]{\textwidth}{
Philip F.~Hopkins$^1$, Jonathan Squire$^{2}$, Iryna S.\ Butsky$^{3}$
}\vspace*{4pt} \\
$^1$ TAPIR, Mailcode 350-17, California Institute of Technology, Pasadena, CA 91125, USA. E-mail:phopkins@caltech.edu \\
$^2$ Physics Department, University of Otago, 730 Cumberland St., Dunedin 9016, New Zealand \\
$^3$ Astronomy Department, University of Washington, Seattle, WA 98195, USA
}

\date{}
\begin{document}
\maketitle

\begin{abstract}
We derive a consistent set of moments equations for CR-magnetohydrodynamics, assuming a gyrotropic distribution function (DF). Unlike previous efforts we derive a closure, akin to the M1 closure in radiation hydrodynamics (RHD), that is valid in both the nearly-isotropic-DF and/or strong-scattering regimes, {\em and} the arbitrarily-anisotropic DF or free-streaming regimes, as well as allowing for anisotropic scattering and transport/magnetic field structure. We present the appropriate two-moment closure and equations for various choices of evolved variables, including the CR phase space distribution function $f$, number density $n$, total energy $e$, kinetic energy $\epsilon$, and their fluxes or higher moments, and the appropriate coupling terms to the gas. We show that this naturally includes and generalizes a variety of terms including convection/fluid motion, anisotropic CR pressure, streaming, diffusion, gyro-resonant/streaming losses, and re-acceleration. We discuss how this extends previous treatments of CR transport including diffusion and moments methods and popular forms of the Fokker-Planck equation, as well as how this differs from the analogous M1-RHD equations. We also present two different methods for incorporating a reduced speed of light (RSOL) to reduce timestep limitations: in both we carefully address where the RSOL (versus true $c$) must appear for the correct behavior to be recovered in all interesting limits, and show how  current implementations of CRs with a RSOL neglect some additional terms. 
\end{abstract}


\section{Introduction}
\label{sec:intro}

Cosmic rays (CRs) could play a potentially crucial role in the inter-stellar and circum-galactic medium, star and galaxy formation, and our understanding of high-energy astro-particle and plasma physics. In recent years, there has been a surge of interest in attempts to model CR dynamics {\em explicitly} in star, planet, and galaxy simulations -- i.e.\ following the transport and matter interactions of CRs alongside the magnetohydrodynamics (MHD), gravity, and other plasma physics effects in these systems \citep[see e.g.][]{uhlig:2012.cosmic.ray.streaming.winds,Wien13,Sale14,Simp16,Pakm16,2016MNRAS.456..582S,Rusz17,zweibel:cr.feedback.review,Mao18,Giri18,chan:2018.cosmicray.fire.gammaray,Buts18,su:turb.crs.quench,hopkins:cr.mhd.fire2,ji:fire.cr.cgm}. Simultaneously, work has continued on more traditional CR propagation methods that trace CR trajectories as ``tracer particles'' across static analytic galaxy models in order to to understand solar system observables \citep[e.g.][]{2016ApJ...819...54G,2016ApJ...824...16J,2016ApJ...831...18C,2016PhRvD..94l3019K,evoli:dragon2.cr.prop,2018AdSpR..62.2731A}. Ideally, one would simply solve the full Vlasov equation for CRs as a function of position ${\bf x}$ and momentum ${\bf p}$ for each CR species, but the high dimensionality of this equation is prohibitive. Moreover, in planet/star/galaxy formation models the resolution scales are vastly larger than CR gyro-radii for CRs with energies $\lesssim$\,TeV (which contain most of the energy/pressure, and dominate the interactions with the non-relativistic matter). As such, these applications have generally relied on moment-based approaches, where one begins by assuming that the  CR distribution function (DF) $f$ is {gyrotropic} (symmetric around the magnetic-field direction), averages over the micro-scale Lorentz forces and scattering processes, then considers moments of the distribution function in terms of the remaining momentum direction, the pitch angle $\mu$. 

The simplest of these -- ``zeroth moment methods'' -- correspond to pure diffusion models. These involve either assuming nearly-isotropic behavior and solving an isotropic Fokker Plank equation for $f$, or solving a diffusion-like equation, $\partial_{t} q = \nabla \cdot (\boldsymbol{\kappa} \cdot \nabla q) + ...$, for some integrated ``macroscopic'' CR property $q$ (e.g., energy density; the diffusion tensor $\boldsymbol{\kappa}$ should be anisotropic on scales much larger than the gyro radius, $\boldsymbol{\kappa} = \kappa_{\|}\,\bhat\bhat$). But it is well-known that this approximation cannot accurately represent many regimes of interest: the free-streaming or weak-scattering regimes, significantly-anisotropic $f(\mu)$, the trans-\Alf{ic} CR ``streaming'' limit, and others. Moreover, it can produce highly un-physical behavior (e.g.\ super-luminal CR transport), and imposes a severe timestep (and therefore CPU cost) penalty in numerical simulations that explicitly integrate the CRs. Motivated by this, recently \citet{jiang.oh:2018.cr.transport.m1.scheme,chan:2018.cosmicray.fire.gammaray,thomas.pfrommer.18:alfven.reg.cr.transport} proposed two-moment schemes, effectively evolving not just the isotropic part of $f$ but its first moment as well (or equivalently, evolving both CR energy and its flux), which resolve many of these problems. The formulations in \citet{jiang.oh:2018.cr.transport.m1.scheme} and \citet{chan:2018.cosmicray.fire.gammaray} were heuristically motivated by the analogous popular moments methods for radiation hydrodynamics (RHD), but they did not attempt to link these to the actual equations of motion for a gyrotropic CR distribution. \citet{thomas.pfrommer.18:alfven.reg.cr.transport} did make such a link and developed a formalism for further expanding on this; however, their formulation makes some restricting assumptions, e.g.\ that the CRs are ultra-relativistic and that the DF $f(\mu)$ is always nearly-isotropic. Moreover, although all of these works have suggested and adopted the use of a ``reduced speed of light'' (RSOL)  as a method to prevent extremely small numerical timesteps when CRs are free-streaming (again, analogous to the procedure common in  RHD), none have attempted to verify that the RSOL formulation is consistent in all relevant limits of their equations to guarantee accurate steady-state solutions. 

In this manuscript we therefore expand upon this previous work to develop  more general forms of the CR-MHD equations. In application, this work is intended primarily for numerical models of planet, star, and galaxy formation, or the interstellar or circum/inter-galactic medium, where one desires to evolve CR populations explicitly. We make two fundamental assumptions throughout, appropriate for these applications: (1) that the background MHD fluid velocities ${\bf u}$ are non-relativistic (so we can expand to leading-order in e.g. $\mathcal{O}(u/c)$), and (2) that the CRs have a gyrotropic DF with gyro radii/timescales much smaller than the macroscopically resolved scales in the calculation. Importantly, however, we {\em do not} assume that e.g.\ the CR scattering mean-free-paths are short -- akin to e.g.\ kinetic MHD \citep{Kulsrud1983}, we will show that the small-gyro-radius assumption is sufficient for a ``fluid-like'' expansion of the Vlasov equation, provided appropriate closure relations are adopted to truncate the moments expansion.

In \S~\ref{sec:assumptions} we present various assumptions and definitions, and in \S~\ref{sec:derivation} use this to derive the appropriate two-moment equations (\S~\ref{sec:moment.expressions}) and closures governing the CR distribution function (\S~\ref{sec:closures}) or its integrals (CR number or energy density; \S~\ref{sec:number.energy.equations}), as well as the corresponding couplings to the gas equations (\S~\ref{sec:gas.eqns}). In \S~\ref{sec:direct.methods} we alternatively present expressions appropriate for methods that attempt to explicitly evolve the CR pitch-angle distribution directly (\S~\ref{sec:finite.volume}). 
In \S~\ref{sec:examples} we consider a number of test problems to compare various closure assumptions and ``zeroth moment methods'' to exact solutions, summarized in \S~\ref{sec:prop.summary}. In \S~\ref{sec:relations}, we discuss how the formulations here extend previous moments equations in the literature (\S~\ref{sec:prev.moments}) and popular forms of the Fokker-Planck equation (\S~\ref{sec:isotropic.fp}), and relate to analogous RHD expressions (\S~\ref{sec:compare.M1}). We discuss the reduced-speed-of-light (RSOL) approximation in \S~\ref{sec:rsol} and present two possible implementations (\S~\ref{sec:rsol.viable.eqns}), deriving correction terms needed in various limits to ensure reasonable behavior (\S~\ref{sec:rsol.limits}) and reviewing the (dis)advantages of each (\S~\ref{sec:rsol.advantages}). We summarize in \S~\ref{sec:conclusions}. 

For ease of reference, we define variables in Table~1 and collect many of the most important derived equations in Appendix~\ref{sec:equations}.

\begin{footnotesize}
\ctable[caption={{\normalsize Commonly-Used Variables in This Paper}},center 
]{cl}{
}{
\hline\hline
$f$ & CR distribution function (DF) $f \equiv f({\bf x},\,{\bf p},\,t,\,s,\,...)$ \\ 
${\bf p}$, ${\bf v}$ & CR momentum ${\bf p}$, velocity ${\bf v}$ ($p\equiv|{\bf p}|$, $v\equiv|{\bf v}|$) \\
$\mu$ & CR pitch-angle $\mu \equiv \hat{\bf p}\cdot\bhat$ \\
$\bhat$, $v_{A}$ & Magnetic field direction $\bhat \equiv{\bf B}/|{\bf B}|$, \Alf\ speed $v_{A}$ \\
\hline
$c$, $\tilde{c}$ & True ($c$) \&\ ``reduced'' ($\tilde{c}$) speed-of-light (RSOL) \\
$\beta$, $\gamma$ & CR velocity/Lorentz factors $\beta=v/c$, $\gamma=1/\sqrt{1-\beta^{2}}$ \\
${\bf u}$, $\boldsymbol{\beta}_{u}$ & Gas velocity ${\bf u}$, with $\boldsymbol{\beta}_{u} \equiv {\bf u}/c$ \\
$D_{t}$ & Conservative comoving derivative $D_{t} X \equiv \partial_{t} X + \nabla \cdot ({\bf u}\,X)$ \\
\hline
$q$, ${\bf F}_{q}$ & Moments of the DF \&\ associated fluxes (Eqs.~\ref{eqn:moment.q}-\ref{eqn:flux.q}) \\
$n$, $e$, $\epsilon$ & CR number $n$, energy $e$, kinetic energy $\epsilon$ densities \\
$n^{\prime}$, $e^{\prime}$, $\epsilon^{\prime}$ & Differential $n^{\prime}\equiv dn/dp = 4\pi\,p^{2}\,\bar{f}_{0}$, etc. \\ 
$\bar{f}_{n}$ & Pitch-angle moments of the DF: $\bar{f}_{n}\equiv \langle \mu^{n}\,f \rangle_{\mu}$ (Eq.~\ref{eqn:pitch.angle.f}) \\
\hline
$\mun{n}$ & DF-weighted pitch-angle moment $\mun{n} \equiv \bar{f}_{n} / \bar{f}_{0}$ (Eq.~\ref{eqn:pitch.angle.mu}) \\
$\bar{\nu}$ & Pitch-angle averaged scattering rate $\bar{\nu}\equiv \bar{\nu}_{+}+\bar{\nu}_{-}$ (Eq.~\ref{eqn:D.coefficient.defn}) \\
$\bar{D}_{\mu\mu}$, $\bar{D}_{pp}$ & Averaged scattering coefficients $D_{\mu\mu}$, etc. (Eq.~\ref{eqn:D.coefficient.defn}) \\ 
$\bar{v}_{A}$ & Streaming speed $\bar{v}_{A} \equiv v_{A}\,(\bar{\nu}_{+}-\bar{\nu}_{-})/(\bar{\nu}_{+}+\bar{\nu}_{-})$ \\
\hline
$\momentgrad$ & Derivative operator $\momentgrad(X) \equiv \bhat \cdot [\nabla \cdot (\mathbb{D}\,X) ]$ (Eq.~\ref{eqn:gradop.defn}) \\ 
$\mathbb{P}$, $\mathbb{D}$ & CR pressure tensor $\mathbb{P}$ \&\ Eddington-type tensor $\mathbb{D}$ (Eq.~\ref{eqn:eddington.tensor.defn}) \\
$\chi$ & Second-moment function $\chi \equiv (1-\mutwo)/2$ (Eq.~\ref{eqn:chi.defn}) \\ 
$\closurefunction$ & Closure function $\mutwo \approx \closurefunction(\muone)$ (Eq.~\ref{eqn:closure}) \\
\hline\hline
}
\end{footnotesize}

\section{Assumptions \&\ Definitions}
\label{sec:assumptions}

Our starting point is the general focused CR transport equation  \citep[see e.g.][]{skilling:1971.cr.diffusion,1975MNRAS.172..557S,1997JGR...102.4719I,2001GeoRL..28.3831L,2005ApJ...626.1116L,zank:2014.book,2015ApJ...801..112L} as written in polar momentum coordinates:
\begin{align}
\nonumber \frac{1}{c}&\,D_{t} f  + \mu\,\beta\,\bhat \cdot \nabla f - f\,\nabla \cdot \boldsymbol{\beta}_{u} \\ 
\nonumber &+ \left[ 
\frac{1-3\,\mu^{2}}{2}\,\left(\bhat\bhat:\nabla\boldsymbol{\beta}_{u} \right)
- \frac{1-\mu^{2}}{2}\,\nabla\cdot \boldsymbol{\beta}_{u}
-\frac{\mu\,\bhat\cdot {\bf a}}{\beta\,c^{2}}
\right]\,p\,\frac{\partial f}{\partial p} \\ 
\nonumber &+ \left[ 
\beta \nabla \cdot \bhat + \mu\,\nabla \cdot \boldsymbol{\beta}_{u} - 3\mu\,\left(\bhat\bhat:\nabla\boldsymbol{\beta}_{u} \right)
- \frac{2\,\bhat\cdot{\bf a}}{\beta\,c^{2}}
\right]\,\frac{1-\mu^{2}}{2}\,\frac{\partial f}{\partial \mu} \\ 
\label{eqn:focused.cr} &= \frac{1}{c} \frac{\partial f}{\partial t}{\Bigr|}_{\rm coll}.
\end{align}
This describes the evolution of a {\em gyrotropic} CR distribution function (DF) $f$, defined in the co-moving frame (with fluid velocity ${\bf u}$), valid to second order in $\mathcal{O}(u/c)$ (where $c$ denotes the true speed of light throughout). We will consider the CR equations as a continuous function of momentum $p$ or Lorentz factor $\gamma$ for a given CR species $s$ -- i.e.\ it should be understood here that some quantity $\psi$ is actually $\psi_{\gamma,\,s}({\bf x},\,t,\,p,\,s,\,m_{s},\,...)$ for species $s$ with mass $m_{s}$, etc., but we will not write this out for the sake of compact notation.

In Eq.~\ref{eqn:focused.cr}, 
$\mu$ is the CR pitch angle, 
$\bhat \equiv {\bf B}/|{\bf B}|$ is the unit magnetic field vector, 
$\beta \equiv |{\bf v}|/c = v/c$ is the speed of the CRs, 
$\boldsymbol{\beta}_{u} \equiv {\bf u}/c$ the speed of the fluid, 
${\bf a} \equiv d{\bf u}/dt \equiv \partial {\bf u}/\partial t + ({\bf u}\cdot \nabla)\,{\bf u}$ the fluid acceleration, 
${\bf A}:{\bf B} \equiv {\rm Tr}[{\bf A}\cdot{\bf B}]$ denotes the double dot product, 
$D_{t} X \equiv \partial_{t} X + \nabla \cdot ({\bf u}\, X) \equiv \rho\,{\rm d}_{t} (X/\rho)$ is the conservative comoving derivative, 
$\rho$ the fluid density, ${\rm d}_{t} X \equiv \partial_{t} X + ({\bf u}\cdot\nabla) X$, 
$\partial_{t} X \equiv \partial X/\partial t$, 
and
$\partial_{t} f |_{\rm coll}$ denotes the scattering+collisional terms and other loss/injection processes.

We define various integrals of the DF as,
\begin{align}
\label{eqn:moment.q} q = q({\bf x},\,...) &\equiv \int d^{3}{\bf p}\,\psi_{q}\,f = \int\,p^{2}\,dp\,d\mu\,d\phi\,\psi_{q}\,f(\mu,\,p,\,{\bf x},\,...),
\end{align}
where $\phi$ is the phase angle, ${\bf x}$ is the spatial coordinate, and $\psi_{q}$ corresponds to each $q$. So for e.g.\ the volumetric number density $n$, total energy $e$, or kinetic energy $\epsilon$, we have $q=(n,\,e,\,\epsilon)$ with $\psi_{q}=(1,\,E(p),\,T(p))$, respectively, where $E(p)=\gamma\,m\,c^{2}$ and $T(p)=E(p)-m\,c^{2}$ refer to the total and kinetic energy of an {\em individual} CR particle of rest mass $m$. We will consider a single CR species: we can later reconstruct the total DF by summing over different species.
The corresponding fluxes are,
\begin{align}
\label{eqn:flux.q} {\bf F}_{q} = F_{q}\,\bhat &\equiv \int d^{3}{\bf p}\,\psi_{q}\,f\,{\bf v} = \bhat\,\int\, 4\pi\,p^{2}\,dp\left( \frac{1}{2}\,\int d\mu\,\psi_{q}\,f\,\mu\,v \right),
\end{align}
where the alignment with $\pm\bhat$ follows immediately from our assumed gyrotropic DF.
The CR pressure tensor $\mathbb{P}$ is defined as,
\begin{align}
\label{eqn:pressure.tensor} \mathbb{P} &\equiv \int d^{3}{\bf p}\,({\bf p}\,{\bf v})\,f \equiv 3\,P_{0}\,\mathbb{D}, 
\end{align}
where $P_{0} \equiv \int 4\pi\,p^{2}\,dp\,(p\,v/3)\,f$ is a scalar pressure and $\mathbb{D}$ is an Eddington-type tensor of trace unity (specified below). 
We also define the pitch-angle-averaging operations, pitch-angle moments of $f$, and DF-weighted pitch angle moments:
\begin{align}
\label{eqn:pitch.angle.x} \langle X \rangle_{\mu} &\equiv \frac{1}{4\pi}\,\int\,d\mu\,d\phi\,X, \\ 
\label{eqn:pitch.angle.f} \bar{f}_{n} &\equiv \langle \mu^{n}\,f \rangle_{\mu}, \\ 
\label{eqn:pitch.angle.mu} \mun{n} &\equiv \frac{\langle \mu^{n}\,f \rangle_{\mu}}{\langle f \rangle_{\mu}} = \frac{\bar{f}_{n}}{\bar{f}_{0}}. 
\end{align}

\section{Derivation of the CR Transport Moments Equations}
\label{sec:derivation}

\subsection{Ordering in $\mathcal{O}(u/c)$}

\subsubsection{General Moments Equations}
\label{sec:ordering:general}

Let us first discuss the general case before considering to the specific isotropic and anisotropic limits. 
Begin from Eq.~\ref{eqn:focused.cr}, and take the ``0th moment'' equation (average Eq.~\ref{eqn:focused.cr} over $\mu$). Integrating by parts, we have for a general gyrotropic DF:
\begin{align}
\label{eqn:zero.moment.full} 
\frac{1}{c}\,D_{t} \bar{f}_{0} &
+ \nabla \cdot(\beta\,\bhat\,\bar{f}_{1}) 
- \bar{f}_{0} \nabla \cdot \boldsymbol{\beta}_{u}  \\ 
\nonumber &
+ p\,\frac{\partial }{\partial p}\,\left[ \frac{1-3\,\mutwo}{2}\,\left(\bhat\bhat:\nabla\boldsymbol{\beta}_{u} \right) 
- \frac{1- \mutwo}{2}\,\nabla\cdot \boldsymbol{\beta}_{u} \right] \,\bar{f}_{0} \\
\nonumber &
+ \frac{3\,\mutwo-1}{2}\,\left[ \nabla\cdot\boldsymbol{\beta}_{u} 
- 3\,\left(\bhat\bhat:\nabla\boldsymbol{\beta}_{u} \right) \right]\,\bar{f}_{0} \\ 
\nonumber & 
- \frac{\bhat\cdot{\bf a}}{\beta\,c^{2}}\,\left[ 2\,\bar{f}_{1} 
+  p\,\frac{\partial \bar{f}_{1}}{\partial p}\right]
 =  \left\langle \frac{1}{c} \frac{\partial f}{\partial t}{\Bigr|}_{\rm coll} \right\rangle_{\mu},
\end{align}
Assuming $\mathcal{O}(\beta)\sim\mathcal{O}(1)$ and defining some gradient wavenumber $k \sim 1/\ell_{\rm grad}\sim \mathcal{O}(\nabla)$,  Eq.~\ref{eqn:zero.moment.full} has a collection of ``adiabatic'' terms $\mathcal{O}(\bar{f}_{0}\,\nabla\boldsymbol{\beta}_{u}) \sim \mathcal{O}(\bar{f}_{0}\,k\,u/c)$, acceleration terms $\mathcal{O}[\bar{f}_{1}\,{\bf a}/c^{2}]$, and a flux term $\mathcal{O}(k\,\bar{f}_{1})$. In the free-streaming limit, $\mathcal{O}(\bar{f}_{1}) \sim \mathcal{O}(\bar{f}_{0})$, so the adiabatic terms are $\mathcal{O}(u/c)$ smaller than the flux term, but in the strong-scattering/isotropic limits $\bar{f}_{1}$ can vanish (the bulk CR drift/streaming speed can be $\lesssim \mathcal{O}(u)$), so we need to keep the $\mathcal{O}(\nabla\boldsymbol{\beta}_{u})$ terms as they can be leading-order in some limits. 

Now consider the acceleration term: note $\mathcal{O}({\bf a}/c^{2}) \sim \mathcal{O}(\nabla P_{\rm eff,\,gas} / \rho\,c^{2})$ where $P_{\rm eff,\,gas}\sim \rho\,c_{\rm eff,\,gas}^{2}$ is the effective pressure exerting forces on the gas, and $c_{\rm eff,\,gas}$ is some effective sound speed so $\mathcal{O}(c_{\rm eff,\,gas}) \sim \mathcal{O}(u)$.\footnote{Note, even in the strong-coupling limit, if CR pressure dominates the forces on the gas, so $P_{\rm eff,\,gas} \rightarrow P_{\rm cr} \sim e_{\rm cr} \sim \gamma\,n_{\rm cr}\,m\,c^{2}$, we have $\mathcal{O}(\nabla P_{\rm eff,\,gas}/\rho\,c^{2}) \sim \mathcal{O}(k\,P_{\rm cr} / \rho\,c^{2}) \sim \mathcal{O}(k\,n_{\rm cr}/n_{\rm gas})$, i.e.\ this scales as the ratio of the number of CRs to non-relativistic particles, which is also extremely small for any limits we consider where we could treat the gas in the MHD limit.} So we have $\mathcal{O}({\bf a}/c^{2}) \sim \mathcal{O}(k\,u^{2}/c^{2})$, which is always at least one order in $\mathcal{O}(u/c)$ smaller than the other terms above and therefore should be dropped.

Next, take the ``1st moment'' equation by multiplying Eq.~\ref{eqn:focused.cr} by $\mu$ and averaging over $\mu$. This gives:
\begin{align}
\label{eqn:first.moment.full} \frac{1}{c}&\,D_{t} \bar{f}_{1}  
+ {\beta}\,\bhat \cdot \nabla \left( \mutwo\,\bar{f}_{0} \right) 
+ \left(\frac{3\,\mutwo - 1}{2}\right) \beta\,\bar{f}_{0}\,\nabla \cdot \bhat \\
\nonumber & 
+ \left[3\,\bhat\bhat:\nabla \boldsymbol{\beta}_{u} - 2\,\nabla \cdot \boldsymbol{\beta}_{u} \right]\,\bar{f}_{1} + \frac{1}{2}\left[ \bhat\bhat:\nabla\boldsymbol{\beta}_{u} - \nabla\cdot \boldsymbol{\beta}_{u} \right]\,p\,\frac{\partial \bar{f}_{1}}{\partial p} \\
\nonumber & 
- \left[6\,\bhat\bhat:\nabla \boldsymbol{\beta}_{u} - 2\,\nabla \cdot \boldsymbol{\beta}_{u} \right]\,\bar{f}_{3} - \frac{1}{2}\left[ 3\,\bhat\bhat:\nabla\boldsymbol{\beta}_{u} - \nabla\cdot \boldsymbol{\beta}_{u} \right]\,p\,\frac{\partial \bar{f}_{3}}{\partial p} \\
\nonumber & 
+ \frac{\bhat\cdot{\bf a}}{\beta\,c^{2}}\,\left[\left(1-3\,\mutwo \right)\,\bar{f}_{0}  
- p\,\frac{\partial \bar{f}_{2}}{\partial p} \right] 
=  \left\langle \frac{\mu}{c} \frac{\partial f}{\partial t}{\Bigr|}_{\rm coll} \right\rangle_{\mu} 
\end{align}
Going term by term, after the time derivative we first have ``flux'' and ``focusing'' terms which scale as $\mathcal{O}(k\,\bar{f}_{0})$ and $\mathcal{O}(k\,\bar{f}_{2})$; because  $\mathcal{O}(k\,\bar{f}_{0})\sim \mathcal{O}(k\,\bar{f}_{2})$ (at least in the isotropic limit), we cannot drop one of these relative to the other. Next we have a large number of ``adiabatic terms''  $\mathcal{O}(\nabla \boldsymbol{\beta} \,\bar{f}_{1}) \sim \mathcal{O}(k\,\bar{f}_{1}\,u/c)$; but these are always $\mathcal{O}(u/c)$ smaller than the flux/focusing terms $\mathcal{O}(k\,\bar{f}_{0})$, both in the free-streaming limit (where $\mathcal{O}(\bar{f}_{1}) \sim \mathcal{O}(\bar{f}_{0})$) by $\mathcal{O}(u/c)$, and in the isotropic limit by $\mathcal{O}[(\bar{f}_{1}/\bar{f}_{0})\,(u/c)] \ll \mathcal{O}(u/c)$.\footnote{Like the analogous radiation-hydrodynamics case, it is important here that we began from the co-moving focused transport equation, so $\bar{f}_{1}$ is comoving, and the dropped $\mathcal{O}(\nabla\boldsymbol{\beta}_{u})$ terms in the flux equations are those {\em outside} the operator $D_{t}$. If $\bar{f}_{1}$ were the ``lab-frame'' moment, leading-order $\mathcal{O}(\nabla\boldsymbol{\beta}_{u}\,\bar{f}_{1})$ terms in Eq.~\ref{eqn:first.moment.full} would appear outside the Eulerian derivatives $\partial_{t} \bar{f}_{1}$.} Next a similar set of terms appears in $\mathcal{O}(\nabla \boldsymbol{\beta} \,\bar{f}_{3})$, but since $\mathcal{O}(\bar{f}_{3}) \lesssim \mathcal{O}(\bar{f}_{1})$ (or more formally since $\bar{f}_{3}$ is bounded like $\bar{f}_{1}$ with $|\bar{f}_{3}| \le |\bar{f}_{0}|$) and we dropped the terms in $\mathcal{O}(\nabla \boldsymbol{\beta} \, \bar{f}_{1})$, we should drop the $\mathcal{O}(\nabla \boldsymbol{\beta} \,\bar{f}_{3})$ terms as well. Finally we have the acceleration terms $\mathcal{O}(\bar{f}_{0}\,|{\bf a}|/c^{2})$; given the order of $|{\bf a}|$ noted above, we immediately see this is $\mathcal{O}(u^{2}/c^{2})$ smaller than the leading terms. 

We can also obtain this hierarchy from the various integral equations. Multiplying Eq.~\ref{eqn:zero.moment.full} by $4\pi\,p^{2}\,dp\,E(p)$ and Eq.~\ref{eqn:first.moment.full} by $4\pi\,p^{2}\,dp\,E(p)\,v$ and integrating, we obtain the CR total energy and energy flux equations:
\begin{align}
\label{eqn:e.eqn.tmp}  \frac{1}{c}\,D_{t} e &+ \nabla \cdot\left( \frac{F_{e}}{c}\, \bhat \right) + \mathbb{P}:\,\nabla\boldsymbol{\beta}_{u} + \frac{F_{e}}{c}\,\frac{2\,\bhat\cdot{\bf a}}{c^{2}} =  \frac{1}{c} \frac{\partial e}{\partial t}{\Bigr|}_{\rm coll}, \\ 
\nonumber \frac{1}{c}\,D_{t}  \frac{F_{e}}{c}  &+ \bhat \cdot (\nabla \cdot \mathbb{P})
+ \frac{F_{e}}{c}\, \bhat\bhat:\nabla\boldsymbol{\beta}_{u}  
+ (e+\mathbb{P}:\bhat\bhat)\,\frac{\bhat\cdot{\bf a}}{c^{2}} = \frac{1}{c^{2}} \frac{\partial F_{e}}{\partial t}{\Bigr|}_{\rm coll} .
\end{align}
These are directly analogous to the comoving equations of radiation hydrodynamics (RHD; \citealt{mihalas:1984oup..book.....M}, Eqs.~95.87-95.88), with each featuring the co-moving time-derivative term ($D_{t}$), flux term ($\nabla\cdot(\bhat F)$ or $\bhat \cdot \nabla P$), velocity-gradient terms ($\propto \nabla\boldsymbol{\beta}_{u}$), acceleration term ($\propto {\bf a}$), and collisional/scattering terms. In RHD, it is well-established that in any relevant limit (free-streaming/unconfined, with $\nu\rightarrow 0$; or static/dynamical diffusion or strong-scattering, with $F_{e} \sim (c^{2}/\nu)\,\nabla e$; or advection, with $F_{e} \sim v_{\rm stream}\,e$; whether the gas or relativistic particle pressure dominates ${\bf a}$):  (1) the acceleration terms are always smaller by $\mathcal{O}(u^{2}/c^{2})$ compared to the dominant terms; and (2) the velocity gradient $\nabla\boldsymbol{\beta}_{u}$ terms in the {\em flux} ($\bar{f}_{1}$) equation are smaller by $\mathcal{O}(u/c)$, but must be retained in the {\em energy} ($\bar{f}_{0}$) equation to recover the correct behavior in the strong-scattering limit.


If we now return to Eq.~\ref{eqn:zero.moment.full} and keep only leading-order terms in in $\mathcal{O}(u/c)$, we have (after some algebra to simplify):
\begin{align}
\nonumber \frac{1}{c}\,D_{t} \bar{f}_{0} &
+ \nabla \cdot(\beta\,\bar{f}_{1}\,\bhat) 
- \frac{1}{p^{2}}\,\frac{\partial}{\partial p}\left[ p^{3}\,\bar{f}_{0}\,\mathbb{D}:\nabla\boldsymbol{\beta}_{u} \right]
=  \left\langle \frac{1}{c} \frac{\partial f}{\partial t}{\Bigr|}_{\rm coll} \right\rangle_{\mu}, \\
\nonumber \frac{1}{c}\,D_{t} \bar{f}_{1} & 
+ \nabla \cdot (\beta\,\bar{f}_{2}\,\bhat)  -\chi\,\beta\,\bar{f}_{0}\,\nabla \cdot \bhat 
=  \left\langle \frac{\mu}{c} \frac{\partial f}{\partial t}{\Bigr|}_{\rm coll} \right\rangle_{\mu}, \\
\label{eqn:zero.moment.leading} \mathbb{D} &\equiv \chi\,\mathbb{I} + (1-3\,\chi)\,\bhat\bhat \ \ \ \ \ , \ \ \ \ \ \chi \equiv \frac{1-\mutwo}{2}.
\end{align}

\subsubsection{Scattering Terms}
\label{sec:scattering}

Enormous controversy still surrounds the behavior of the CR scattering terms, and this is the focus of much of the CR literature \citep[see e.g.][]{chandran00,yan.lazarian.02,yan.lazarian.04:cr.scattering.fast.modes,yan.lazarian.2008:cr.propagation.with.streaming,Zwei13,zweibel:cr.feedback.review,zank:2014.book,bai:2015.mhd.pic,bai:2019.cr.pic.streaming,lazarian:2016.cr.wave.damping,holcolmb.spitkovsky:saturation.gri.sims,
2019MNRAS.tmp.2249V}. Our derivation here, on the other hand, is almost entirely focused on the collisionless CR transport terms (those outside $\partial_{t} f |_{\rm coll}$). However to write down a sensible galactic CR transport equation, we must make some assumption about scattering. So we will briefly consider these, in an intentionally simplified manner.

We begin from the usual quasi-linear theory (QLT) slab scalings \citep{schlickeiser:89.cr.transport.scattering.eqns}:
\begin{align}
\label{eqn:scattering} \frac{\partial f}{\partial t}{\Bigr|}_{\rm sc} = & 
\frac{\partial}{\partial\mu}\left( D_{\mu\mu} \frac{\partial f}{\partial \mu} + D_{\mu p} \frac{\partial f}{\partial p}  \right) \\
\nonumber &+ \frac{1}{p^{2}} \frac{\partial }{\partial p}\left[ p^{2}\,\left( D_{\mu p} \frac{\partial f}{\partial \mu} + D_{p p}\,\frac{\partial f}{\partial p} \right) \right], \\ 
\nonumber D_{\mu\mu} &= \frac{(1-\mu^{2})}{2}\,\left[ \left( 1-\mu\,\frac{v_{A}}{v} \right)^{2}\,\nu_{+} + \left( 1+\mu\,\frac{v_{A}}{v} \right)^{2}\,\nu_{-} \right]. \\ 
\nonumber D_{\mu p} &= \frac{(1-\mu^{2})}{2}\,\frac{p\,v_{A}}{v}\,\left[ \left( 1-\mu\,\frac{v_{A}}{v} \right)\,\nu_{+} -  \left( 1+\mu\,\frac{v_{A}}{v} \right)\,\nu_{-} \right], \\ 
\nonumber D_{p p} &= \frac{(1-\mu^{2})}{2}\,\frac{p^{2}\,v^{2}_{A}}{v^{2}}\,\left[ \nu_{+} + \nu_{-} \right],
\end{align}
where $v_{A}$ is the appropriate \Alf\ speed and $\nu_{\pm}(\mu)$ are the scattering rates from forward and backward-propagating waves \citep{1975MNRAS.172..557S}. 
Taking the appropriate moments and assuming $\mathcal{O}(v_{A}) \sim \mathcal{O}(u)$ gives:
\begin{align}
\nonumber \left\langle \frac{\partial f}{\partial t}{\Bigr|}_{\rm sc} \right\rangle_{\mu} &=  \frac{1}{p^{2}}\frac{\partial }{\partial p}\left[ p^{2}\,\left( S\,\bar{f}_{0}  
+ \bar{D}_{\mu p}\,\bar{f}_{1}
+ \bar{D}_{p p}\, \frac{\partial \bar{f}_{0}}{\partial p} 
\right)\right]  + \mathcal{O}\left( \frac{u^{2}}{v^{2}} \right), 
\\
\left\langle \mu\,\frac{\partial f}{\partial t}{\Bigr|}_{\rm sc} \right\rangle_{\mu} &= 
- \bar{D}_{\mu\mu,\,\mu}\,\bar{f}_{1} - \bar{D}_{\mu p,\,\mu}\,\frac{\partial \bar{f}_{0}}{\partial p}
 + \mathcal{O}\left( \frac{u^{2}}{v^{2}} \right) ,
\end{align}
where
\begin{align}
\bar{D}_{p p} &\equiv (\partial_{p} f)^{-1}\,\langle D_{pp}\,\partial_{p} f \rangle_{\mu} \approx \chi\,\frac{p^{2}\,v^{2}_{A}}{v^{2}}\,\bar{\nu}, \\\nonumber \bar{D}_{\mu p} &\equiv  \bar{f}_{1}^{-1}\,\langle D_{\mu p}\,\partial_{\mu} f \rangle_{\mu} \approx \frac{p\,\bar{v}_{A}}{v}\,\bar{\nu}, \\
 \nonumber \bar{D}_{\mu\mu,\,\mu} &\equiv - \bar{f}_{1}^{-1}\,\langle \mu\,\partial_{\mu} D_{\mu\mu} \partial_{\mu} f \rangle_{\mu} \approx \bar{\nu}, \\ 
\nonumber \bar{D}_{\mu p,\,\mu} &\equiv (\partial_{p} f)^{-1}\,\langle \mu\,D_{\mu p}\,\partial_{p} f \rangle_{\mu} \approx \chi\,\frac{p\,\bar{v}_{A}}{v}\,\bar{\nu}, \\ 
\nonumber \bar{v}_{A} &\equiv v_{A}\,\left( \frac{\bar{\nu}_{+} - \bar{\nu}_{-}}{\bar{\nu}_{+} + \bar{\nu}_{-}} \right) \ \ \ \ \ , \ \ \ \ \  \bar{\nu} \equiv \bar{\nu}_{+} + \bar{\nu}_{-}.
\end{align}
Note we have defined $\bar{\nu}$ and $\bar{v}_{A}$ for convenience, with $\bar{\nu}_{\pm}$ representing the appropriate $\mu$-averages. For completeness, the $\partial_{t} f|_{\rm coll}$ term should also include a term $p^{-2}\,\partial_{p}\,[p^{2}\,S\,\bar{f}_{0}]$ representing continuous external momentum loss/gain processes (e.g.\ radiative losses), and some $j$ representing injection or catastrophic losses. 

\subsubsection{Focused Transport Equation to Leading Order}
\label{sec:leading:general}

With \S~\ref{sec:ordering:general} in mind, we now return to the focused transport Eq.~\ref{eqn:focused.cr} to obtain a simplified form valid to $\mathcal{O}(u/c)$. First dropping just the (always higher-order) acceleration terms, after some tedious algebra we can write Eq.~\ref{eqn:focused.cr} as:
\begin{align}
\label{eqn:focused.cr.noaccel}  \frac{1}{c}&\,D_{t} f  
+ \nabla \cdot ( \mu\,\beta\,f\,\bhat )
- \frac{1}{p^{2}}\,\frac{\partial}{\partial p}\left[ p^{3}\,f\,\mathbb{D}:\nabla\boldsymbol{\beta}_{u} \right] \\ 
\nonumber & + \frac{\partial }{\partial \mu}\left[ \chi\,\left\{ \beta\,\nabla\cdot\bhat  + \mu\,(\mathbb{I} - 3\,\bhat\bhat):\nabla\boldsymbol{\beta}_{u}  \right\} \, f  \right]  
= \frac{1}{c} \frac{\partial f}{\partial t}{\Bigr|}_{\rm coll}.
\end{align}
Based on the arguments above in \S~\ref{sec:ordering:general}, we see that the $\mu\,(\mathbb{I}-3\,\bhat\bhat):\nabla\boldsymbol{\beta}_{u}$ term inside $\partial_{\mu}[\chi\{...\}]$ is smaller by $\mathcal{O}(u/c)$ than the others in all relevant regimes and can also be dropped. Specifically, this term produced only terms in the $D_{t}\bar{f}_{0}$ and $D_{t} \bar{f}_{1}$ equations which we argued were smaller by $\mathcal{O}(u/c)$ and should be dropped in those equations. But we can see this directly as well: in all relevant regimes, $\mu\,(\mathbb{I}-3\,\bhat\bhat):\nabla\boldsymbol{\beta}_{u}$ is smaller by $\mathcal{O}(u/c)$ compared to the focusing term $\beta\,\nabla\cdot\bhat$ inside $\partial_{\mu}[\chi\{...\}]$. Even if $\nabla \cdot \bhat=0$, the $\mu\,(\mathbb{I}-3\,\bhat\bhat):\nabla\boldsymbol{\beta}_{u}$ is still always smaller by $\mathcal{O}(u/c)$ compared to the flux-of-flux term (outside $\partial_{\mu}$), so it can be safely dropped here. 
Re-adding the leading-order scattering terms from \S~\ref{sec:scattering}, and keeping only the remaining (leading-order) terms in $\mathcal{O}(u/c)$ in each power of $\partial_{t,\,{\bf x},\,{\bf p}}$, $\nu$, etc, we have:
\begin{align}
\label{eqn:df.general}  \frac{1}{c}\,D_{t} f  &+  \nabla \cdot (\mu\,\beta\,f\,\bhat) =   \\
\nonumber & \frac{\partial}{\partial\mu}\left[ \chi\,\left\{ -f\, \beta\,\nabla\cdot\bhat + \frac{\nu}{c}\,\left( \frac{\partial f}{\partial \mu} + \frac{\bar{v}_{A}}{v} p\,\frac{\partial f}{\partial p} \right)  \right\} \right] + \\
\nonumber & \frac{1}{p^{2}} \frac{\partial }{\partial p}\left[ p^{3}\,\left\{ (\mathbb{D}:\nabla \boldsymbol{\beta}_{u})\, f + \frac{\nu\,\chi}{c}\,\left( \frac{\bar{v}_{A}}{v}\, \frac{\partial f}{\partial \mu} + \frac{v_{A}^{2}}{v^{2}}\,p\,\frac{\partial f}{\partial p} \right) \right\} \right], 
\end{align}
where  $\chi = (1-\mu^{2})/2$, and $\nu_{\pm}(\mu)$ are a function of  $\mu$. We note that all expansions and discussion used to derive Eq.~\ref{eqn:df.general} rely only on our $\mathcal{O}(u/c)$ expansion, and the derivation can, if desired, be carried out without needing to first follow the moments expansion in our \S~\ref{sec:ordering:general}.

\subsection{The Close-to-Isotropic-DF Case}
\label{sec:isotropic}

We now consider an example of a specific form for the CR DF that is nearly-isotropic in $\mu$. The derivation here will closely follow \citet{thomas.pfrommer.18:alfven.reg.cr.transport}, to whom we refer for more details. By assumption, if $f$ is close-to-isotropic in $\mu$, it can be expanded in pitch angle moments as $f(\mu) \approx \bar{f}_{0} + 3\,\mu\,\bar{f}_{1} + \mathcal{O}(|f_{1}|^{2}/|f_{0}|^{2} \ll 1)$, which implies $\bar{f}_{2} \approx \bar{f}_{0}/3$ or $\mutwo=1/3$ (and $\bar{f}_{3}\approx 3\,\bar{f}_{1}/5$). 
With this assumption the pressure tensor becomes isotropic: $\mathbb{P} = P_{0}\,\mathbb{I}$ (i.e.\ $\mathbb{D} = \mathbb{I}/3$) where $P_{0}=\int d^{3}{\bf p}\,f\,p\,v/3$ ($=\beta^{2}\,e/3$ integrated in a narrow interval of $p$). Either directly using this form for $f$ and taking the zeroth and first $\mu$ moment-averages of Eq.~\ref{eqn:focused.cr}, or simply inserting the above for $\mutwo$ in Eqs.~\ref{eqn:zero.moment.full}-\ref{eqn:first.moment.full}, we can immediately verify that these give consistent expressions, and the ordering in $\mathcal{O}(u/c)$ is the same as \S~\ref{sec:ordering:general}. 
For the leading-order terms, we have:
\begin{align}
\label{eqn:f0.eqn.iso.leading} \frac{1}{c}\,D_{t} \bar{f}_{0} &+ \nabla \cdot(\beta\,\bhat\,\bar{f}_{1}) - \bar{f}_{0} \nabla \cdot \boldsymbol{\beta}_{u} + ... \\ 
\nonumber &+ \left[ \frac{1-3\,\mutwo}{2}\,\left(\bhat\bhat:\nabla\boldsymbol{\beta}_{u} \right) - \frac{1- \mutwo}{2}\,\nabla\cdot \boldsymbol{\beta}_{u} \right] \,p\,\frac{\partial \bar{f}_{0}}{\partial p} \\
\nonumber & =  \left\langle \frac{1}{c} \frac{\partial f}{\partial t}{\Bigr|}_{\rm coll} \right\rangle_{\mu}, \\
\label{eqn:f1.eqn.iso.leading} \frac{1}{c}\,D_{t} \bar{f}_{1} &+ {\beta}\,\bhat \cdot \nabla (\mutwo\,\bar{f}_{0}) 
+ ... =  \left\langle \frac{\mu}{c} \frac{\partial f}{\partial t}{\Bigr|}_{\rm coll} \right\rangle_{\mu},
\end{align}
where $...$ denotes the dropped terms, and we write out $\mutwo$ (instead of inserting $1/3$) for reference below.
For the scattering terms, we obtain to leading order in $\mathcal{O}(u/c)$: $\bar{D}_{\mu\mu,\,\mu}=\bar{\nu}$, $\bar{D}_{\mu p}=(p\,\bar{v}_{A}/v)\,\bar{\nu}$, $\bar{D}_{\mu p,\,\mu}= (1/3)\, (p\,\bar{v}_{A}/v)\,\bar{\nu}$, $\bar{D}_{pp} = (1/3)\, (p\,v_{A}/v)^{2}\,\bar{\nu}$.


\subsection{The Maximally-Anisotropic-DF Case}
\label{sec:anisotropic}

Next, consider the opposite limit of the maximally anisotropic DF $f(\mu) = \bar{f}_{0}\,\delta(\mu - \mu_{0})$ -- i.e.\ all CRs at a given $({\bf x},\,p,\,s,\,...)$ have identical pitch angle, and $\bar{f}_{n}=\mun{n}\,\bar{f}_{0} = \mu_{0}^{n}\,\bar{f}_{0}$. Our ordering above in $\mathcal{O}(u/v)$ is not sensitive to this, so keeping only the terms to leading order, the moments of Eq.~\ref{eqn:focused.cr} become, 
\begin{align}
\label{eqn:f0.eqn.delta} 
\frac{1}{c}\,D_{t} \bar{f}_{0} &+ \nabla \cdot(\beta\,\bhat\,\bar{f}_{1}) - \bar{f}_{0} \nabla \cdot \boldsymbol{\beta}_{u} + ...  \\ 
\nonumber &+ \left[ \frac{1-3\,\mutwo}{2}\,\left(\bhat\bhat:\nabla\boldsymbol{\beta}_{u} \right) - \frac{1- \mutwo}{2}\,\nabla\cdot \boldsymbol{\beta}_{u} \right] \,p\,\frac{\partial \bar{f}_{0}}{\partial p} \\
\nonumber &+ \frac{3\,\mutwo-1}{2}\,\left[ \nabla\cdot\boldsymbol{\beta}_{u} - 3\,\left(\bhat\bhat:\nabla\boldsymbol{\beta}_{u} \right) \right]\,\bar{f}_{0} 
 =  \left\langle \frac{1}{c} \frac{\partial f}{\partial t}{\Bigr|}_{\rm coll} \right\rangle_{\mu},
\end{align}
and (again being careful regarding $\mu$ commutation),
\begin{align}
\label{eqn:f1.eqn.delta} \frac{1}{c}\,D_{t} \bar{f}_{1} +& {\beta}\,\bhat \cdot \nabla \left( \mutwo\,\bar{f}_{0} \right) + ... \\
\nonumber & + \left(\frac{3\,\mutwo - 1}{2}\right) \beta\,\bar{f}_{0}\,\nabla \cdot \bhat  + ... =  \left\langle \frac{\mu}{c} \frac{\partial f}{\partial t}{\Bigr|}_{\rm coll} \right\rangle_{\mu},
\end{align}
where $...$ denotes the dropped terms of sub-leading order in $\mathcal{O}(u/c)$. If $\mu_{0}$ is independent of $p$ (or we integrate over a narrow range of $p$), the pressure tensor is $\mathbb{P}_{\rm cr} = 3\,P_{0}\,\mathbb{D} = \mathbb{P}_{\rm iso} + \mathbb{P}_{\rm aniso}$ with 
\begin{align}
\label{eqn:pressure.tensor.delta} \mathbb{D} &= \left(\frac{1-\mutwo}{2}\right)\,\mathbb{I} + \left(\frac{3\,\mutwo-1}{2}\right)\,\bhat\bhat. 
\end{align}
Defining the mean scattering coefficients so that $\bar{\nu}_{\pm} = \nu_{\pm}(\mu=\mu_{0})$ because $f\propto \delta(\mu-\mu_{0})$,  we obtain to leading $\mathcal{O}(u/c)$: $\bar{D}_{\mu\mu,\,\mu}=\bar{\nu}$, $\bar{D}_{\mu p}=(p\,\bar{v}_{A}/v)\,\bar{\nu}$, $\bar{D}_{\mu p,\,\mu}= ([1-\mutwo]/2)\, (p\,\bar{v}_{A}/v)\,\bar{\nu}$, $\bar{D}_{pp} = ([1-\mutwo]/2)\, (p\,v_{A}/v)^{2}\,\bar{\nu}$.

Written this way, we verify an important connection to Eqs.~\ref{eqn:f0.eqn.iso.leading}-\ref{eqn:f1.eqn.iso.leading}: at this order, the equations differ {\em only} in the addition of terms with the pre-factor $(3\,\mutwo - 1)$, which vanish identically with the nearly-isotropic-DF closure $\mutwo = 1/3$. Likewise, the pressure tensor and these expressions for the $\bar{D}$ coefficients reduce to exactly their near-isotropic-DF values when $\mutwo=1/3$. Thus Eq.~\ref{eqn:zero.moment.leading} or Eqs.~\ref{eqn:f0.eqn.delta}-\ref{eqn:pressure.tensor.delta} are valid in {\em both} the nearly-isotropic-DF and maximally-anisotropic-DF cases, for appropriate choice of $\mutwo$.

\subsection{Co-Moving Expressions to Leading Order}
\label{sec:moment.expressions}

\subsubsection{General Expressions \&\ Closure Relation}
\label{sec:closures}

After some re-arrangement we can now write a series of expressions valid in both the nearly-isotropic-DF and maximally-anisotropic-DF limits:
\begin{align}
\label{eqn:f0.eqn.final} 
\frac{1}{c}\,D_{t} \bar{f}_{0} + & \nabla \cdot  (\beta\,\bhat\,\bar{f}_{1})  
-  \mathbb{D}:\nabla\boldsymbol{\beta}_{u} \left[ 3\,\bar{f}_{0} +  \,p\,\frac{\partial \bar{f}_{0}}{\partial p} \right] \\
\nonumber & \ \ \ =  
\frac{1}{c\,p^{2}}\frac{\partial }{\partial p}\left[ p^{2}\,\left( S\,\bar{f}_{0}  
+ \tilde{D}_{p \mu}\,\bar{f}_{1}
+ \tilde{D}_{p p}\, \frac{\partial \bar{f}_{0}}{\partial p} 
\right)\right]  + \frac{j_{0}}{c}, \\
\label{eqn:f1.eqn.final} 
\frac{1}{c}\,D_{t} \bar{f}_{1} + & 
\beta\,\momentgrad(\bar{f}_{0}) = - \frac{1}{c}\left[ \tilde{D}_{\mu\mu}\,\bar{f}_{1} + \tilde{D}_{\mu p}\,\frac{\partial \bar{f}_{0}}{\partial p} \right]
+ \frac{j_{1}}{c}, \\
\label{eqn:D.coefficient.defn} \tilde{D}_{p p}  = \chi\,&\frac{p^{2}\,v_{A}^{2}}{v^{2}}\,\bar{\nu} 
\  , \  \
\tilde{D}_{p \mu} = \frac{p\,\bar{v}_{A}}{v}\,\bar{\nu} 
\  , \ \
\tilde{D}_{\mu\mu} = \bar{\nu} 
\  , \  \
\tilde{D}_{\mu p} = \chi\,\frac{p\,\bar{v}_{A}}{v}\,\bar{\nu}.
\end{align}
We have added the terms $S$, which represents continuous (e.g.\ radiative) losses, and $j$, which represents injection or catastrophic losses. We also define the operator $\momentgrad(q)$ and Eddington tensor $\mathbb{D}$ in terms of the variable $\chi$:
\begin{align}
\label{eqn:gradop.defn} \momentgrad(q) &\equiv \bhat \cdot \nabla  \left( [1-2\,\chi]\,q \right)  + (1-3\,\chi)\,q\,\nabla \cdot \bhat,  \\
\nonumber &\ \ \ = \nabla \cdot \left( \mutwo\,q\,\bhat \right) - \chi\,q\,\nabla \cdot \bhat = \bhat\cdot\left[ \nabla \cdot \left( \mathbb{D}\,q \right) \right],
\\ 
\label{eqn:eddington.tensor.defn} \mathbb{D} &\equiv \chi\,\mathbb{I} + \left( 1-3\,\chi \right)\,\bhat\bhat,  \\
\label{eqn:chi.defn} \chi &\equiv \frac{1-\mutwo}{2} = \frac{1}{2}\,\left[ 1 - \frac{\bar{f}_{2}}{\bar{f}_{0}} \right].
\end{align}

Provided some expression for scattering rates and $\mutwo \equiv \bar{f}_{2} / \bar{f}_{0}$, the above form a complete system of equations for $(\bar{f}_{0},\,\bar{f}_{1})$. But we do not have a general equation for $\bar{f}_{2}$: we have the usual moments hierarchy problem, requiring some closure relation. Without solving for the entire $f(\mu,\,\phi,\,...)$, by analogy to the M1 closure(s) in RHD we can define an {\em approximate} closure $\mutwo \approx \closurefunction(\muone)$, which (with Eqs.~\ref{eqn:f0.eqn.final}-\ref{eqn:f1.eqn.final}) accurately captures both the isotropic-DF and maximally-anisotropic-DF limits (note that $\muone\equiv \bar{f}_{1}/\bar{f}_{0}$). The function $\closurefunction$ should satisfy the following: (1) in the nearly-isotropic-DF case, by definition, $|\muone| \ll 1$ and $\mutwo = 1/3 + \mathcal{O}(\muone^{2})$; (2) in the free-streaming case with $f\rightarrow \delta(\mu \pm 1)$ (maximally-anisotropic-DF case), $\mutwo = \muone^{2}$, with $\mathcal{O}( \bar{f}_{n}) \sim \mathcal{O}(\bar{f}_{0})$; and (3) the DF should be {\em realizable}, meaning that an $f(\mu)$ exists which is finite and non-negative for all $-1\le \mu\le 1$ with the given $\muone$ and $\mutwo$. 

A natural choice satisfying the above is the popular RHD closure from \citet{levermore:1984.FLD.M1}, which is the unique $\closurefunction$ if there exists any frame in which (after Lorentz boosting) the DF is isotropic:
\begin{align}
\label{eqn:closure} \mutwo &\approx \closurefunction\left( \muone \right) =  \frac{3+4\,\muone^{2}}{5 + 2\,(4-3\,\muone^{2})^{1/2}}. 
\end{align} 
This is not the only possible closure, however. For example, \citet{minerbo:1978.eddington.factors} note that if the DF satisfies a maximum entropy principle,  
\begin{equation}
    \label{eqn:closure.minerbo} \closurefunction = \frac{1}{3} + \frac{2}{15}\muone^{2}(3-|\muone| + 3\,\muone^{2}).
    \end{equation}
    Various other choices are reviewed in \citet{murchikova:m1.neutrino.transport.closures}. 
We stress that while  the closure relation Eq.~\ref{eqn:closure} (or Eq.~\ref{eqn:closure.minerbo}) is an approximation, Eqs.~\ref{eqn:f0.eqn.final}-\ref{eqn:ek.specific} are exact (to lowest order in $u/c$) for {\em any} DF, provided the ``correct'' $\mutwo$ and $\bar{\nu}$. So one can easily imagine constructing more complicated or exact closure relations, analogous to ``variable Eddington tensor'' methods in RHD, to assign the correct values of $\mutwo$.

\subsubsection{CR Number \&\ Energy Equations}
\label{sec:number.energy.equations}

We can now obtain equations for $(q,\,F_{q})$ by multiplying Eqs.~\ref{eqn:f0.eqn.final}-\ref{eqn:f1.eqn.final} by $4\pi\,p^{2}\,\psi_{q}\,dp$ and integrating. First, it is helpful to consider the equations integrated over an infinitesimal range of $p$, e.g.\ $\Delta n \equiv (dn/dp)\,\Delta p$. This gives:
\begin{align}
\label{eqn:n.specific} D_{t}\left( n^{\prime} \right) + \nabla \cdot \left( F_{n}^{\prime}\,\bhat \right) &= S_{n}^{\prime}, \\ 
\nonumber D_{t}\left( F_{n}^{\prime} \right) + c^{2}\,\momentgrad \left( \beta^{2}\,n^{\prime} \right) &= 
-\bar{\nu}\,\left[ F_{n}^{\prime} - 3\,\chi\,\bar{v}_{A}\,n^{\prime} \right] + S_{F_{n}}^{\prime},
\end{align}
where $n^{\prime} \equiv dn/dp = 4\pi\,p^{2}\,\bar{f}_{0}$, $F_{n}^{\prime}\equiv dF_{n}/dp = 4\pi\,p^{2}\,v\,\bar{f}_{1}$, $S_{n}^{\prime} \equiv 4\pi\,p^{2}\,j_{0}$, $S_{F_{n}}^{\prime} \equiv 4\pi\,p^{2}\,v\,j_{1}$. For total energy $e$ we have:
\begin{align}
\label{eqn:e.specific} D_{t}\left( e^{\prime} \right) + \nabla \cdot \left( F_{e}^{\prime}\,\bhat \right) &= S_{e}^{\prime} + \tilde{S}^{\prime}_{\rm sc} - \mathbb{P}^{\prime}:\nabla{\bf u}, \\ 
\nonumber D_{t}\left( F_{e}^{\prime} \right) + c^{2}\,\momentgrad \left( \beta^{2}\,e^{\prime} \right) &= 
-\bar{\nu}\,\left[ F_{e}^{\prime} - 3\,\chi\,\bar{v}_{A}\,(e^{\prime}+P_{0}^{\prime}) \right] + S_{F_{e}}^{\prime},
\end{align}
with $e^{\prime} \equiv de/dp=4\pi\,p^{2}\,E(p)\,\bar{f}_{0}$, $F_{e}^{\prime} \equiv d F_{e} /dp=4\pi\,p^{2}\,E(p)\,v\,\bar{f}_{1}$, $S^{\prime}_{e}\equiv4\pi\,p^{2}\,(E(p)\,j_{0}-S\,v)$, $S_{F_{e}}^{\prime} \equiv 4\pi\,p^{2}\,E(p)\,v\,j_{1}$, $P_{0}^{\prime} \equiv dP_{0}/dp = 4\pi\,p^{2}\,(p\,v/3)\,\bar{f}_{0}$, $\mathbb{P}^{\prime} \equiv 3\,P_{0}^{\prime}\,\mathbb{D}$, and 
\begin{align}
\tilde{S}_{\rm sc}^{\prime} &\equiv - \frac{\bar{\nu}}{c^{2}}\,\left[ \bar{v}_{A}\,F^{\prime}_{e} - 3\,\chi\,{v}^{2}_{A}\,\left( e^{\prime} + P_{0}^{\prime} \right)  \right] \\ 
\nonumber &=-  \frac{\bar{\nu}}{c^{2}}\,\left[ \frac{\gamma}{\gamma-1}\, \bar{v}_{A}\,F_{\epsilon}^{\prime} - 3\,\chi\,{v}^{2}_{A}\,\left( \frac{\gamma}{\gamma-1}\,\epsilon^{\prime} + P_{0}^{\prime} \right)  \right].
\end{align}
Then for kinetic energy $\epsilon$ we obtain:
\begin{align}
\label{eqn:ek.specific} D_{t}\left( \epsilon^{\prime} \right) + \nabla \cdot \left( F_{\epsilon}^{\prime}\,\bhat \right) &= S_{\epsilon}^{\prime} + \tilde{S}_{\rm sc}^{\prime} - \mathbb{P}^{\prime}:\nabla{\bf u}, \\ 
\nonumber D_{t}\left( F_{\epsilon}^{\prime} \right) + c^{2}\,\momentgrad \left( \beta^{2}\,\epsilon^{\prime} \right) &= 
-\bar{\nu}\,\left[ F_{\epsilon}^{\prime} - 3\,\chi\,\bar{v}_{A}\,(\epsilon^{\prime}+P_{0}^{\prime}) \right] + S_{F_{\epsilon}}^{\prime},
\end{align}
with $\epsilon^{\prime} \equiv d\epsilon/dp = 4\pi\,p^{2}\,T(p)\,\bar{f}_{0}$, $F_{\epsilon}^{\prime} \equiv d F_{e} /dp=4\pi\,p^{2}\,T(p)\,v\,\bar{f}_{1}$, $S^{\prime}_{\epsilon}\equiv4\pi\,p^{2}\,(T(p)\,j_{0}-S\,v)$, $S_{F_{\epsilon}}^{\prime} \equiv 4\pi\,p^{2}\,T(p)\,v\,j_{1}$. It is useful to note the relations: 
\begin{align}
P_{0}^{\prime} & = \frac{\beta^{2}\,e^{\prime}}{3} = (\gamma_{\rm eos}-1)\,\epsilon^{\prime} = \frac{1+\gamma^{-1}}{3}\,\epsilon^{\prime}, \\ 
\mathbb{P}^{\prime} &\equiv 3\,P_{0}^{\prime}\,\mathbb{D} = \beta^{2}\,e^{\prime}\,\mathbb{D} = 3\,P_{0}^{\prime}\,\left[ \chi\,\mathbb{I} + (1-3\,\chi)\,\bhat\bhat \right], \\ 
\muone &\equiv \frac{\bar{f}_{1}}{\bar{f}_{0}} = \frac{F_{q}}{q\,v} \ \ \ \ \ , \ \ \ \ \ \mutwo \approx \closurefunction\left( \muone \right),
\end{align}
i.e.\ the ``effective adiabatic index'' relating CR pressure and kinetic energy density is $\gamma_{\rm eos} = (4 + \gamma^{-1})/3$  at a given Lorentz factor $\gamma$. One uses $\muone=F_{q}/q\,v$ to determine the closure values of $\mutwo$ or $\chi$.

Note every term in the ``macroscopic'' equations for $q$ has a simple interpretation and correspondence with a term in Eqs.~\ref{eqn:f0.eqn.final}-\ref{eqn:f1.eqn.final} for $f$. The $D_{t}\bar{f}_{0,\,1} \rightarrow D_{t} (q,\,F_{q})$ term is the comoving conservative derivative; $\nabla \cdot (\beta\,\bar{f}_{1}\,\bhat) \rightarrow \nabla \cdot ({\bf F}_{q})$ is the normal flux; $\mathbb{D}:\nabla\boldsymbol{\beta}_{u} \rightarrow \mathbb{P}:\nabla {\bf u}$ is the ``adiabatic'' term (for $\mutwo=1/3$, $\mathbb{P}:\nabla{\bf u} \rightarrow P_{0} \nabla\cdot {\bf u}$) related in detail to the non-inertial frame (akin to the analogous RHD term); $S$ and $j$ represent loss/gain processes in number and momentum space (e.g.\ radiative/catastrophic losses, injection); $\beta\momentgrad(\bar{f}_{0}) \rightarrow \momentgrad(\beta^{2}\,q)$ is the ``flux of flux'' (flux source) term; $D_{\mu\mu}\bar{f}_{1} \rightarrow \bar{\nu}\,F$ the scattering term in the flux equation; $D_{\mu p}\,\partial_{p}\,\bar{f}_{0} \rightarrow \chi\,\bar{v}_{A}\,(q+...)$ is the ``streaming'' term if the scattering is asymmetric; and the $D_{p\mu}$ and $D_{pp}$ terms give rise to the gyro-resonant loss or diffusive re-acceleration terms $\tilde{S}_{\rm sc}$ (discussed below).

Taking the diffusive limit ($\mutwo \rightarrow 1/3$, $D_{t} F^{\prime}_{q} \rightarrow 0$), we immediately see that the parallel (anisotropic) spatial diffusivity\footnote{If we assume a scattering rate that scales with CR speed as $\bar{\nu} \sim (\beta\,c) / r_{0}$ for some characteristic scattering scale $r_{0}$ (e.g.\ for Bohm diffusion, $r_{0}$ is the gyro radius), then we obtain the common ansatz $\kappa(p) \sim \beta\,c\,r_{0}$.} at a given $p$ is $\kappa_{\|}(p) \equiv (\beta\,c)^{2} / (3\,\bar{\nu})$.

\subsubsection{Spectrally-Integrated Expressions}
\label{sec:spectrum.integrated.expressions}

Integrating Eqs.~\ref{eqn:n.specific}-\ref{eqn:ek.specific} over all CR momenta gives equations for the spectrally-integrated CR number and energy, for example: 
\begin{align}
D_{t} n + \nabla \cdot \left( F_{n} \, \bhat \right) &= S_{n}, \\ 
\nonumber D_{t} F_{n} + c^{2}\,\int dp\,\momentgrad\left( \beta^{2}\,n^{\prime} \right) &= S_{F_{n}}^{\prime}-\int dp\, \bar{\nu}\,\left[ F_{n}^{\prime} - 3\,\chi\,\bar{v}_{A}\,n^{\prime} \right].
\end{align}
Although $\int dp\,q^{\prime} = q$ is trivial, this immediately introduces practical difficulties in terms like $\int dp\,\momentgrad(\beta^{2}\,q^{\prime})$ and $\int dp\,\bar{\nu}\,[F_{q}^{\prime} - 3\,\chi\,\bar{v}_{A}\,(q^{\prime}+...)]$ in the flux, and $\int dp\,[ \tilde{S}_{\rm sc}^{\prime} - \mathbb{P}^{\prime}:\nabla{\bf u}]$ in the energy equations. The issue is that even if we know the form of $\bar{\nu}_{\pm}(p)$, we cannot write these equations in terms of a single ``effective'' $\chi$, $\bar{\nu}$, $\bar{v}_{A}$, $\beta$, $\gamma$, etc, because the ``weights'' (combination of $p$-dependent factors in the integrals) in each part of each term are different. Moreover, even if we specified an initial spectral shape ($\bar{f}_{0}(p)$ and $\bar{f}_{1}(p)$) to calculate some effective values, the $p$-dependence would immediately alter the spectrum and change those values. 

If one wishes to adopt the spectrally-integrated equations in practical applications, therefore, one must impose a universal (fixed) spectral shape. In that limit, the CR total energy is the meaningful quantity to evolve, since a ``fixed-spectrum'' CR number equation will not conserve energy or momentum. We can further simplify by noting that most of the total CR energy is in particles with $\beta \approx 1$ and $E(p) \sim T(p)$, giving: 
\begin{align}
\nonumber D_{t} e  + \nabla \cdot ( F_{e}\,\bhat ) &\approx S_{e} - \mathbb{P}_{e}:\nabla{\bf u}
-\frac{\bar{\nu}_{e}}{c^{2}}\left[ \bar{v}_{A}^{e}\,F_{e} - 3\,\chi_{e}\,{v}^{2}_{A}\,\left( e + P_{0} \right)  \right], \\
\label{eqn:e.total}  D_{t} F_{e}  + c^{2}\,\momentgrad_{e} \left( 3\,P_{0} \right) &\approx 
-\bar{\nu}_{e}\,\left[ F_{e} - 3\,\chi_{e}\,\bar{v}_{A}^{e}\,(e+P_{0}) \right] + S_{F_{e}}.
\end{align}
Here $P_{0} \approx e/3$; $\momentgrad_{e}(3\,P_{0}) = \bhat \cdot ( \nabla \cdot \mathbb{P}_{e} ) = \bhat\cdot \nabla([1-2\,\chi_{e}]\,3\,P_{0}) + (1-3\,\chi_{e})\,3\,P_{0}\,\nabla \cdot \bhat$; and $\mathbb{P}_{e} \equiv 3\,P_{0}\,\mathbb{D}_{e} = 3\,P_{0}\,[ \chi_{e}\,\mathbb{I} + (1-3\,\chi_{e})\,\bhat\bhat]$; with $\chi_{e}$, $\bar{v}_{A}^{e}$, and $\bar{\nu}_{e}$ understood to be the appropriate ``spectrally-averaged'' values.\footnote{For completeness, we note that the ``0th moment'' spectrally-integrated CR energy equation arises from Eq.~\ref{eqn:e.total} taking the strong-scattering (isotropic-DF, $\mutwo\rightarrow 1/3$), flux-steady-state ($D_{t} F_{e} \rightarrow 0$) limit, so $F_{e} \rightarrow \bar{v}_{A}^{e}\,(e + P_{0}) - (c^{2}/\bar{\nu}_{e})\,\bhat\cdot \nabla P_{0}$.
}

\subsection{The Gas Equations \&\ Conservation}
\label{sec:gas.eqns}

As discussed in \citet{Zwei13,zweibel:cr.feedback.review} and \citet{thomas.pfrommer.18:alfven.reg.cr.transport}, the CRs can exchange momentum with the (non-relativistic) gas and magnetic fields\footnote{Since we are working in the limit where the CR gyro radii are small, and obviously the non-relativistic ion+electron gyro radii are much smaller still, the MHD assumption that the non-relativistic ion gyro radii are vanishingly small compared to resolved scales is reasonable.} primarily via two effects: (1) scattering, and (2) Lorentz forces. If we note that the CR momentum density is $\int d^{3}{\bf p}\,{\bf p}\,f = (1/c^{2})\,{\bf F}_{e}$ (using ${\bf p} = E(p)\,{\bf v}/c^{2}$), then it is immediately clear how to account for (1): we simply add an equal-and-opposite momentum flux to the gas momentum equation to match the scattering ($\bar{\nu}$) term in Eq.~\ref{eqn:e.specific}, i.e.\ $D_{t}(\rho\,{\bf u}) + ... = +(1/c^{2})\,\bhat\,\int dp\,\bar{\nu}\,[F_{e}^{\prime} - 3\,\chi\,\bar{v}_{A}\,(e^{\prime}+P^{\prime}_{0})]$. 

Deriving the Lorentz term (2) requires re-visiting the CR momentum equation before gyro-averaging. In generality (making no assumption about the form of $f$) for a non-relativistic background, the comoving Vlasov equation for $f$ is ${\rm d}_{t} f + \nabla_{\bf x}\cdot({\bf v}\,f) + \nabla_{\bf p}\cdot({\bf F} \, f) = {\rm d}_{t} f |_{\rm coll}$, where $\nabla_{\bf x,\,p}$ denote gradients in position and momentum space, respectively, and ${\bf F}$ is the external force term. Here ${\bf F} = {\bf F}_{\rm Lorentz} + \mathcal{O}(u/c)$ with ${\bf F}_{\rm Lorentz} = (q/c)\,({\bf v} \times {\bf B})$ in this frame.\footnote{We neglect other exchange terms such as e.g.\ the gravity of the CRs, secondary transfer of momentum from scattering of beamed CR radiation, etc, as these are several orders-of-magnitude smaller.} Now, take the momentum density by multiplying by ${\bf p}$ and integrating over $d^{3}{\bf p}$. Integrating by parts and using various identities, note: 
$\int d^{3}{\bf p}\,{\bf p}\, \nabla_{\bf p} \cdot ({\bf F}\, f) = 
- \int d^{3}{\bf p}\,f\,\{ \nabla_{\bf p} \cdot ({\bf F}\,{\bf p}) \} = -\int d^{3}{\bf p}\,f\,[ {\bf p}\,(\nabla_{\bf p}\cdot {\bf F}) + ({\bf F} \cdot \nabla_{\bf p})\,{\bf p}] = -\int d^{3}{\bf p}\,f\,{\bf F}_{\rm Lorentz}$.
\footnote{In this last step, we have used the fact that ${\bf F} \approx {\bf F}_{\rm Lorentz}$ can be written as ${\bf F} = {\bf p} \times {\bf Q}$ where ${\bf Q} = {\bf Q}(p)$ depends only on the magnitude (but not direction) of ${\bf p}$ and external/constant properties, so $\nabla_{\bf p} \cdot {\bf F} = \nabla_{\bf p} \cdot ({\bf p} \times {\bf Q}[p]) = (\nabla_{\bf p} \times {\bf p}) \cdot {\bf Q} - {\bf p} \cdot ( \nabla_{\bf p} \times {\bf Q}(p) ) = 0$, and $({\bf F} \cdot \nabla_{\bf p})\,{\bf p} = {\bf F}$.} Now separate this into parallel and perpendicular components by projecting with $\bhat\bhat$ and $(\mathbb{I}-\bhat\bhat)$, respectively. Because $\bhat\cdot{\bf F}_{\rm Lorentz}=0$, the parallel equation becomes $\int d^{3}{\bf p}\, \bhat \,({\bf p}\,\cdot\bhat)\,(D_{t} f - f\,\nabla_{\bf x} \cdot {\bf u}) + \bhat\bhat\cdot {\bf p}\,{\bf v} \cdot \nabla_{\bf x} f+ ... = \bhat\,(\bhat \cdot {\rm d}_{t}{\bf f}_{\rm coll})$. Recalling that ${\bf p}\cdot \bhat = e\,v\,\mu/c^{2}$, this is immediately recognizable as $\bhat\,(1/c^{2})\,D_{t} F_{e} + .... = -\bar{\nu}(...)$, i.e.\ our Eq.~\ref{eqn:e.specific} for $(1/c^{2})\,D_{t} F_{e}$, multiplied by $\bhat$. Since the terms on the left-hand side of this parallel equation represent free transport and relativistic corrections (coordinate-transformation terms), with no ${\bf F}$ term appearing, the scattering term represents the only parallel momentum exchange with the gas -- i.e.\ we have  re-derived the scattering term (1), which was derived 
more heuristically above from momentum-conservation arguments. 

Now consider the perpendicular component. Averaged over the ``macroscopic'' spatial/time scales ($\ell_{\rm macro}$, $t_{\rm macro}$) much larger than the gyro radius/time ($r_{g}$, $\Omega_{g}$), the first term $D_{t}{\bf F}_{e,\,\bot} = \langle \int d^{3}{\bf p}\,(\mathbb{I}-\bhat\bhat)\,{\bf p}\,f \rangle_{\Omega}$ must vanish, because there can be no coherent flux of CRs perpendicular to the field (more precisely, this term must be smaller than the dominant terms by $\mathcal{O}(r_{g} / \ell_{\rm macro})$). The second term (the $\nabla_{\bf x}$ term) does not vanish, but gives: 
$(\mathbb{I}-\bhat\bhat)\cdot  \int d^{3}{\bf p}\,{\bf p}\,({\bf v} \cdot \nabla_{\bf x})\,f 
= (\mathbb{I}-\bhat\bhat)\cdot \{ \nabla_{\bf x} \cdot [ \int d^{3}{\bf p}\,{\bf p}\,{\bf v}\,f ]\} 
=\nabla_{\bot} \cdot \mathbb{P}$.\footnote{We define the parallel and perpendicular tensor divergence as $\nabla_{\|}  \cdot \mathbb{P} \equiv \bhat\bhat\cdot (\nabla \cdot \mathbb{P}) $ and $\nabla_{\bot} \cdot \mathbb{P} \equiv (\mathbb{I}-\bhat\bhat)\cdot ( \nabla \cdot \mathbb{P} )$.} 
The third term $(\mathbb{I}-\bhat\bhat) \cdot \langle -\int d^{3}{\bf p}\,({\bf F}_{\rm Lorentz}\,f) \rangle_{\Omega} = -\langle \int d^{3}{\bf p}\,({\bf F}_{\rm Lorentz}\,f) \rangle_{\Omega} = -\langle \int d^{3}{\bf p}\,(q/c)\,({\bf v} \times {\bf B})\,f \rangle_{\Omega}= -(1/c)\,\langle {\bf j}_{\rm cr} \times {\bf B} \rangle_{\Omega} = -{\bf f}_{\rm Lorentz}^{\rm cr}$ represents the {\em total} Lorentz force per unit volume on CRs ${\bf f}_{\rm Lorentz}^{\rm cr}$. The scattering term in the perpendicular direction $(\mathbb{I}-\bhat\bhat)$ is negligible compared to the Lorentz forces by $\mathcal{O}(r_{g}/\ell_{\rm mfp})$ (where $\ell_{\rm mfp} \sim 3\,c/\bar{\nu} \sim \mathcal{O}(\ell_{\rm macro})$), so force balance requires 
${\bf f}_{\rm Lorentz}^{\rm cr} = \nabla_{\bot} \cdot \mathbb{P}\,\{1  + \mathcal{O}(r_{g}/\ell_{\rm macro}) \}$. 
The Lorentz force on CRs redirecting ${\bf v}$ requires an equal-and-opposite force on gas,\footnote{Equivalently, we can insert ${\bf j}_{\rm cr}$ in Ampere's law to obtain $\nabla \times {\bf B}  = ({\bf j}_{\rm gas} + {\bf j}_{\rm cr})/c$, and use this to calculate the ``back-reaction'' force $=-{\bf f}_{\rm Lorentz}^{\rm cr}$ on gas.} giving 
$D_{t}(\rho\,{\bf u}) + ... = - {\bf f}_{\rm Lorentz}^{\rm cr} = -\nabla_{\bot} \cdot \mathbb{P}\, \{1 + \mathcal{O}(r_{g}/\ell_{\rm macro})\}$.\footnote{It may appear inconsistent with our assumption of a gyrotropic CR distribution elsewhere to show 
$\langle {\bf j}_{\rm cr} \times {\bf B} \rangle / c \approx \nabla_{\bot} \cdot \mathbb{P} \ne \mathbf{0}$, 
since for a perfectly gyrotropic distribution ${\bf j}_{\rm cr} \times {\bf B} = \mathbf{0}$ exactly. Physically, one can think of this as the perpendicular CR pressure gradient inducing a very small non-gyrotropic perturbation to compensate. The fractional deviation from perfectly-gyrotropic orbits can be estimated as 
$\sim \langle {\bf j} \times {\bf B} / c \rangle / | {\bf j} \times {\bf B}/c |_{\rm max} 
\sim | \nabla_{\bot} \cdot \mathbb{P}^{\prime} | / (n^{\prime}\,q\,v\,|{\bf B}|/c) 
\sim | \nabla P_{0}^{\prime} | / (n^{\prime}\,p\,v/r_{g}) 
\sim (n^{\prime}\,p\,v/\ell_{\rm grad} )/(n^{\prime}\,p\,v/r_{g}) 
 \sim r_{g} / \ell_{\rm grad}$ 
 where $\ell_{\rm grad} \sim P^{\prime}_{0} / |\nabla P^{\prime}_{0}| \sim \mathcal{O}(\ell_{\rm macro})$. So in all other expressions derived in this paper, this correction is sub-dominant by $\mathcal{O}(r_{g} / \ell_{\rm macro})$ and can be safely neglected. However in the back-reaction force on the gas, this term remains finite and leading order even as $(r_{g}/\ell_{\rm macro}) \rightarrow 0$.}

This has a simple interpretation: spatial differences in the collisionless CR pressure tensor (non-zero $\nabla \cdot \mathbb{P}$) source a net CR current (mean $\langle {\bf v} \rangle$ or net flux ${\bf F}_{e}$). The parallel momentum current is $\bhat\,F_{e}$, which is resisted only by scattering (exchanging momentum with gas). The perpendicular current, on the other hand, is immediately redirected by Lorentz forces, exerting an equal-and-opposite force on the gas.
The gas momentum equation becomes: 
\begin{align}
\nonumber D_{t}(\rho\,{\bf u}) + ... = \sum_{s} \int & 4\pi\,p^{2}\,dp\, {\Bigl\{} -\left(\mathbb{I}-\bhat\bhat \right)\cdot \left[ \nabla \cdot  \left( \mathbb{D}\,p\,v\,\bar{f}_{0} \right) \right] \\ 
\label{eqn:gas.momentum.diff} & + \bhat\,\left[ \tilde{D}_{\mu\mu} \bar{f}_{1}\,p  + \tilde{D}_{\mu p} p^{2}\,\frac{\partial \bar{f}_{0} }{\partial p} \right]  {\Bigr\}}  
\end{align}
or
\begin{align}
\label{eqn:gas.momentum} D_{t}(\rho\,{\bf u}) + ... =& \sum_{s} \int dp\, \left[ \bhat\,\frac{\bar{\nu}}{c^{2}}\,\left[ F^{\prime}_{e} - 3\,\chi\,\bar{v}_{A}\,(e^{\prime} + P^{\prime}_{0}) \right] -\nabla_{\bot} \cdot \mathbb{P}^{\prime} \right],
\end{align}
where the $...$ refers to all the non-CR terms, and the sum and integral refer to the summation over all CR species \&\ integration over all momenta. Noting $\bhat\,\momentgrad(\beta^{2}\,e^{\prime}) = \nabla_{\|} \cdot \mathbb{P}^{\prime}$, it is often convenient to rewrite this as:
\begin{align}
\label{eqn:gas.momentum.riemann} D_{t}(\rho\,{\bf u}) &+ ... + \nabla \cdot  \mathbb{P}  = -\frac{1}{c^{2}}\,\bhat\,D_{t}\,F_{e}  \\
\nonumber &=  \bhat\,\sum_{s} \int dp\,{\Bigl \{} \momentgrad(\beta^{2}\,e^{\prime}) +  \frac{\bar{\nu}}{c^{2}}\,\left[ F^{\prime}_{e} - 3\,\chi\,\bar{v}_{A}\,(e^{\prime} + P^{\prime}_{0}) \right] {\Bigr \}}. 
\end{align}
This has the form of a hyperbolic pressure gradient term $\nabla \cdot \mathbb{P}$ that can be included in a Riemann solver, plus a ``source term'' (the right hand side) which vanishes identically when the energy flux equation is in local steady-state.

In the total gas+radiation energy equation, the behavior is straightforward: the kinetic energy terms simply follow the momentum equation: $D_{t} e_{\rm gas} = ... {\bf u}\cdot D_{t}(\rho\,{\bf u})\,|_{\rm cr}$ (where $D_{t}(\rho\,{\bf u})\,|_{\rm cr}$ collects the terms on the right-hand side of Eq.~\ref{eqn:gas.momentum}), and the thermal+magnetic+radiation terms see the source terms $D_{t} e_{\rm gas+rad} + ... = -\sum_{s}\,\int dp\, [ \tilde{S}^{\prime}_{\rm sc} + S^{\prime}_{e}]$, so
\begin{align}
\label{eqn:gas.energy} D_{t} e_{\rm gas+rad} + ... = {\bf u} \cdot \left[ D_{t}(\rho\,{\bf u})\,|_{\rm cr} \right] - \sum_{s} \int dp\, \left[ \tilde{S}^{\prime}_{\rm sc} + S^{\prime}_{e} \right].
\end{align}
Physically, the source/sink $S_{e}^{\prime}$ term corresponds to either energy lost to CR acceleration at injection, or thermalized or radiated away  from various loss processes (thus determining how much goes into thermal vs.\ radiation energy). The kinetic terms reflect work done and, in flux steady state, behave like an adiabatic ``PdV'' term balancing the $\mathbb{P}:\nabla{\bf u}$ term in the CR energy equation. The scattering term $\tilde{S}_{\rm sc}$ corresponds to energy loss/gain from scattering with micro-scale (gyro-resonant) magnetic fluctuations. By definition for the applications of interest, these are unresolved, and have rapid thermalization times, so this can be treated as part of the gas thermal/internal energy budget, although one could also evolve them explicitly as in e.g.\ \citealt{Zwei13,thomas.pfrommer.18:alfven.reg.cr.transport}.

As discussed at length in \citet{mihalas:1984oup..book.....M} in the RHD context and \citet{thomas.pfrommer.18:alfven.reg.cr.transport} for the CR limit, there are subtle ambiguities related to exact, separate energy and momentum conservation if we include the CR inertia at this order in $\mathcal{O}(u/c)$. These are related  to  the definition of frame, the consistency of other terms of higher $\mathcal{O}(u/c)$, and the fact that  non-relativistic MHD drops terms of higher order  in $\mathcal{O}(u/c)$. For example, including the CR inertia, the momentum change includes terms $D_{t}{\bf F}_{e}/c^{2} = \bhat\,D_{t}\,F_{e}/c^{2} + (F_{e}/c^{2})\,D_{t}\bhat$, where the latter term becomes (for ideal MHD) $(F_{e}/c)\,(\mathbb{I}-\bhat\bhat)\,(\bhat\cdot\nabla)\,\boldsymbol{\beta}_{u}$, which is $\mathcal{O}(u/c)$ smaller than all the retained terms in the flux equation. These could be added to maintain manifest conservation if desired, but are not well-posed, as they relate to higher-order terms dropped in both the CR and MHD equations. However, one can immediately verify that in the flux-steady-state or Newtonian ($c\rightarrow \infty$) limits, as assumed in MHD, manifest conservation in the lab and comoving frames is recovered.

\section{Explicit Pitch-Angle Evolution Methods}
\label{sec:direct.methods}

\subsection{DF Equation in Finite-Volume Form}
\label{sec:finite.volume}

Although we have focused on developing the $\mu$-moments equations, there may be occasions where one wishes to directly evolve the pitch-angle distribution, as in our exact solution cases below in \S~\ref{sec:closure.list}. This can be done explicitly by integrating on a phase-space grid that includes the $\mu$ dimension explicitly, similar to e.g.\ direct ray integration methods for RHD like those in e.g.\ \citealt{jiang:2014.rhd.solver.local}. This is actually simpler for CRs as compared to RHD, because we retain the gyrotropic assumption so can still integrate out the $\phi$ dimension. For these applications, it is useful to take the focused transport equation in Eq.~\ref{eqn:df.general}, which incorporates the scattering terms (Eq.~\ref{eqn:scattering}) and carefully retains only leading-order terms in $\mathcal{O}(u/c)$. This can be conveniently written as:
\begin{align}
\label{eqn:df.general.alt}  D_{t} f  &+  \nabla \cdot (\mu\,v\,f\,\bhat) =   \\
\nonumber & \frac{\partial}{\partial\mu}\left[ \chi\,\left\{ -f\, v\,\nabla\cdot\bhat + \nu\,\left( \frac{\partial f}{\partial \mu} + \frac{\bar{v}_{A}}{v} p\,\frac{\partial f}{\partial p} \right)  \right\} \right] + \\
\nonumber & \frac{1}{p^{2}} \frac{\partial }{\partial p}\left[ p^{3}\,\left\{ (\mathbb{D}:\nabla {\bf u})\, f + \nu\,\chi\,\left( \frac{\bar{v}_{A}}{v}\, \frac{\partial f}{\partial \mu} + \frac{v_{A}^{2}}{v^{2}}\,p\,\frac{\partial f}{\partial p} \right) \right\} \right], 
\end{align}
where now terms like $\chi = (1-\mu^{2})/2$, $\mathbb{D}=\chi\,\mathbb{I} + (1-3\,\chi)\,\bhat\bhat$, $\nu\equiv\nu_{+}(\mu) + \nu_{-}(\mu)$ refer to each value of $\mu$ (without averaging).\footnote{It is also often useful to write Eq.~\ref{eqn:df.general.alt} in terms of the one-dimensional DF such that $dn = dp\,d\mu\,f_{\rm 1D}$ as opposed to $dn=d^{3}{\bf p}\,f = p^{2}\,dp\,d\mu\,d\phi\,f$ defined above. This gives
\begin{align}
 D_{t}& f_{\rm 1D}  +  \nabla \cdot (\mu\,v\,f_{\rm 1D}\,\bhat)  =  \\
\nonumber  & \frac{\partial}{\partial\mu}\left[ \chi\,\left\{ -f_{\rm 1D}\, v\,\nabla\cdot\bhat + \nu\,\left( \frac{\partial f_{\rm 1D}}{\partial \mu} - \frac{\bar{v}_{A}}{v} \left[ 2\,f_{\rm 1D} - p\,\frac{\partial f_{\rm 1D}}{\partial p} \right] \right)  \right\} \right] + \\
\nonumber & \frac{\partial }{\partial p}\left[ p\,\left\{ (\mathbb{D}:\nabla {\bf u})\, f_{\rm 1D} + \nu\,\chi\,\left( \frac{\bar{v}_{A}}{v}\, \frac{\partial f_{\rm 1D}}{\partial \mu} - \frac{v_{A}^{2}}{v^{2}}\,\left[ 2\,f_{\rm 1D} - p\,\frac{\partial f_{\rm 1D}}{\partial p} \right] \right) \right\} \right]. 
\end{align}} There is a one-to-one correspondence between each term in Eq.~\ref{eqn:df.general.alt} and their pitch-angle-averaged equivalents in $\bar{f}_{0}$, $\bar{f}_{1}$ (Eq.~\ref{eqn:f0.eqn.final}-\ref{eqn:f1.eqn.final}).

Eq.~\ref{eqn:df.general.alt} is straightforward to implement numerically using standard finite-volume methods: the time-evolution $D_{t} f$ of the comoving $f$ can be operator split into three terms representing (1) translation/flux in position-space (the $\nabla \cdot (...)$ advection term) at fixed $\mu$ and $p$; (2) translation/flux in pitch-angle space (the $\partial_{\mu}(...)$ terms) at fixed ${\bf x}$ and $p$; (3) translation/flux in rigidity/energy space (the $\partial_{p}(...)$ terms) at fixed ${\bf x}$ and $\mu$. Each reduces to a finite-volume problem in the ${\bf x}$, $\mu$, $p$ space, and (2)-(3) being local in position space allows them to be integrated efficiently; the major overhead is the higher dimensionality of the problem causing (potentially excessive) computation. For an example where e.g.\ the $p$ terms are integrated in a finite-volume fashion in $p$-space, see \citet{girichidis:cr.spectral.scheme}.

\subsection{Equations for the Mean Evolution of a CR ``Group''}
\label{sec:monte.carlo}

It is instructive to consider the gyro-averaged evolution equations for the mean state of a CR ``wave packet'' or ``group'' with instantaneous state ${\bf U}(t) = \langle{\bf U}\rangle(t) \equiv ({\bf x},\,\mu,\,p,\,s)[t] = (\langle{\bf x}\rangle,\,\langle\mu\rangle,\,\langle p\rangle,\,\langle s \rangle)$. This is obtained by taking $p^{2}\,f({\bf x},\,\mu,\,p,\,t,\,s) \rightarrow \delta({\bf x}-\langle{\bf x}\rangle[t],\,\mu-\langle \mu \rangle[t],\,p-\langle p\rangle[t],\,s-\langle s\rangle[t],\,t)$ in the general DF Eq.~\ref{eqn:df.general.alt}, and then multiplying Eq.~\ref{eqn:df.general.alt} by ${\bf U}$ and integrating over ${\bf x},\,\mu,\,p,\,s$ to obtain $\dot{\langle {\bf U} \rangle}$, the rate-of-change of the state vector along the path of the group. The ``species equation'' for $s$ trivially evaluates to $\dot{\langle s\rangle} = 0$, since we have not included explicit spallation or other species-changing processes. The position equation is simply $\dot{\langle {\bf x} \rangle} = {\bf u} + \langle \mu\rangle\,\langle v\rangle\,\bhat$, i.e.\ translation with the gas velocity and along the field. The pitch angle and momentum equations are non-trivial, however. For $\mu$, we have
\begin{align}
\nonumber \dot{\langle \mu \rangle} &= \langle\chi\rangle\,\langle v\rangle \,\nabla\cdot\bhat  + \langle\chi\rangle\frac{\partial \nu}{\partial \mu} \\ 
\nonumber &  \ \ \ \ \ \ \ \ \ \  - \nu\,\left[\langle \mu \rangle - \langle \chi \rangle\,\frac{\bar{v}_{A}}{\langle v\rangle}\,\left(2 + \langle \beta \rangle^{2} + \frac{\partial \ln{\delta\nu}}{\partial \ln{p}} \right)\right] 
\\
\label{eqn:individual.mu}  &\approx \langle \chi \rangle \,\langle v \rangle  \,\nabla\cdot\bhat  - \nu\,\left[ \langle \mu \rangle - \langle \chi \rangle\,\frac{\bar{v}_{A}}{\langle v \rangle}\,(2+\langle \beta \rangle^{2}) \right],
\end{align}
where $\delta\nu\equiv\nu_{+}-\nu_{-}$, $\nu=\nu(\langle {\bf U} \rangle)$, and the $\approx$ makes the grey approximation for $\nu$ (which slightly changes the pre-factors but none of the behaviors). We can understand the physics of each term in Eq.~\ref{eqn:individual.mu}: (1) The term $\propto \langle v \rangle\,\nabla \cdot \bhat$ is the ``focusing'' term, corresponding to the $\nabla\cdot\bhat$ terms in $\mathcal{G}(q)$ in the flux equations; (2) $D_{\mu\mu}\,\partial_{\mu} f \rightarrow \nu\,\partial_{\mu} f \rightarrow \nu\,\langle \mu \rangle$ is the normal scattering term ($\sim \nu\,F_{q}$ in the flux equations), which acts like a ``drag'' term on the mean $\langle \mu \rangle$ -- but note, because this an equation just for $\langle \mu \rangle$, the diffusive behavior (which would increase $\langle \mu^{2} \rangle$ if we started from a $\delta$-function DF) does not appear here; (3) The term $D_{\mu p} \rightarrow \nu\,(\bar{v}_{A}/v)\,p\,\partial_{p}\,f \rightarrow \nu\,\langle \chi \rangle\,\bar{v}_{A}/\langle v\rangle$ gives rise to trans-\Alf{ic} CR streaming, appearing as the $\chi\,\bar{v}_{A}\,q$ terms in the flux equations, and giving a mean $\langle\mu\rangle \rightarrow \bar{v}_{A}/\langle v\rangle$, i.e.\ streaming at $\sim \bar{v}_{A}$, in the strong-scattering ($\nu\rightarrow\infty$) limit.

For the momentum equation:
\begin{align}
\nonumber \frac{\dot{\langle p \rangle}}{\langle p \rangle} &= -\left( \langle \mathbb{D} \rangle:\nabla{\bf u}  \right) 
- \frac{v_{A}}{\langle v\rangle}\,\left[\langle \mu \rangle\,\delta\nu - \langle\chi\rangle\,\frac{\partial \delta\nu}{\partial \mu} \right] \\
\nonumber & \ \ \ \ \ \ \ \ \ \ + 2\,\nu\,\langle\chi\rangle\,\frac{v_{A}^{2}}{\langle v\rangle^{2}}\left[ 1+\langle \beta \rangle^{2} +  \frac{\langle p \rangle}{2\,\nu}\frac{\partial \nu}{\partial p} \right]  \\
\label{eqn:individual.momentum} &\approx -\left( \langle\mathbb{D}\rangle:\nabla{\bf u}  \right) - \nu\,\left[ \langle\mu\rangle \frac{\bar{v}_{A}}{\langle v \rangle} -\langle\chi\rangle\,\frac{v_{A}^{2}}{\langle v\rangle^{2}}\,(2 + 2\,\langle\beta\rangle^{2}) \right],
\end{align}
where again $\approx$ indicates the grey approximation. Again, the terms can be understood as follows: (1) $\langle \mathbb{D} \rangle:\nabla{\bf u}$ is the ``adiabatic'' term (immediately analogous to the term in the energy equations); (2) $D_{p \mu} \rightarrow \nu\,\chi\,\bar{v}_{A}\,\partial_{\mu} f \rightarrow \nu\,\bar{v}_{A}\,\langle\mu\rangle$ is the streaming/gyro-resonant loss term ($\propto \nu\,\bar{v}_{A}\,F_{e}$ in $\tilde{S}_{\rm sc}$ in the energy equations); (3) $D_{pp} \rightarrow \nu\,v_{A}^{2}\,\chi\,p\,\partial_{p} f \rightarrow \nu\,\langle\chi \rangle\,v_{A}^{2}/\langle v\rangle^{2}$ is the turbulent/diffusive re-acceleration term. 

If desired, these equations can be directly integrated as well, in Monte Carlo-type methods where each explicitly-evolved CR ``super-particle'' represents the gyro-averaged behavior of an ensemble of CRs with a $\delta$-function DF, but this would require adding some stochastic scattering terms to capture the diffusive/second-derivative behavior (i.e.\ the change in $\langle \mu^{2} \rangle$ or non-$\delta$-function behavior of $f$ as it evolves away from an initial $\delta$-function).

\begin{figure*}
	\includegraphics[width=1\textwidth]{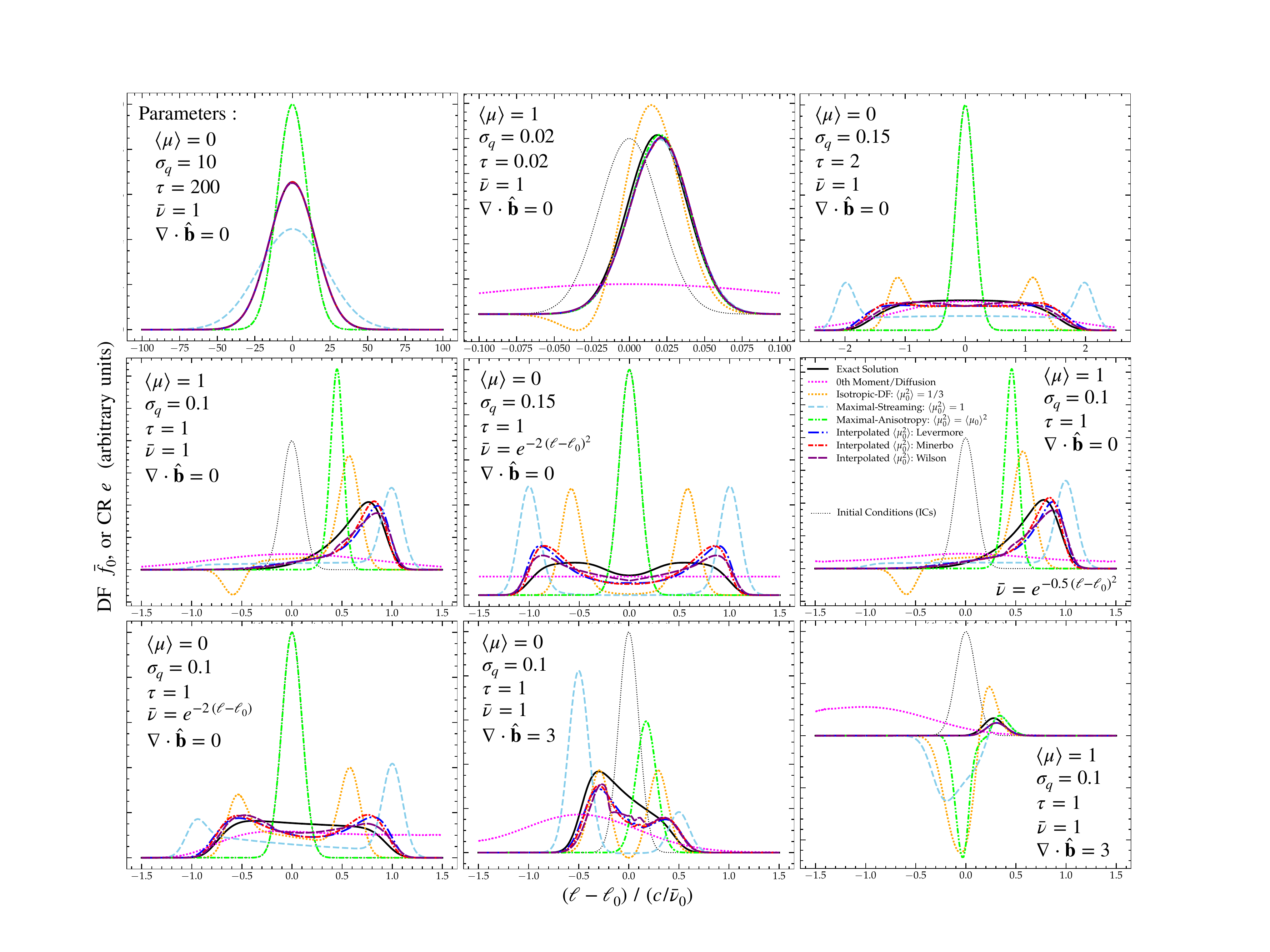}
	\vspace{-0.5cm}
	\caption{Idealized test problems from \S~\ref{sec:examples} comparing different closure assumptions (\S~\ref{sec:closure.list}) for the Boltzmann/Vlasov moments hierarchy vs.\ exact solutions. We simplify to ``pure transport'' problems in a stationary background where the ICs are specified by the initial pitch-angle DF ($\muone=0$ corresponding to an isotropic DF $f=\bar{f}_{0}$, $\muone=1$ to a free-streaming DF with $f=\bar{f}_{0}\,\delta(\mu-1)$), width of the (initially-Gaussian) CR number or $\bar{f}_{0} \propto \exp{\{ -(\ell-\ell_{0})^{2}/2\,\sigma_{q}^{2}\}}$, scattering coefficient $\bar{\nu}(\ell)$ and field divergence $\nabla\cdot \bhat$. We plot the value of the $\mu$-integrated DF $\bar{f}_{0}$ or its moments ($n$, $e$) versus spatial coordinate along a field line $\ell$, in units of scattering time $1/\bar{\nu}_{0}$ and length $c/\bar{\nu}_{0}$, at plotted time $\tau = \bar{\nu}_{0}\,t$. Exact solutions evolve the entire pitch-angle-resolved DF $f(\mu)$ explicitly. The ``interpolated'' closures evolve the first two CR $\mu$-moments equations, differing in the exact form of $\mutwo=\closurefunction(\muone)$ used to close the moments hierarchy; they give very similar results and qualitatively reproduce the exact solution behavior (albeit imperfectly) in all problems while retaining positive-definite $\bar{f}_{0}$. The ``isotropic-DF,'' ``maximal-streaming'' and ``maximal-anisotropy'' closures adopt $\mutwo=1/3$, $=1$, $=\muone^{2}$ (appropriate for isotropic or free-streaming or $\delta$-function DFs) respectively; these can give qualitatively incorrect behavior and produce solutions with negative $\bar{f}_{0}$ (negative energy/particle number) in some circumstances. ``0th-Moment/Diffusion'' refers to the common diffusion closure at 0th order by assuming flux-steady-state and strong-scattering; this preserves positive-definite behavior but produces qualitatively wrong behaviors and super-luminal CR transport in many problems.
	\label{fig:closure.compare}}
\end{figure*}

\section{Example Problems \&\ Illustrative Behaviors}
\label{sec:examples}

\subsection{Setup \&\ Closures Considered}
\label{sec:closure.list}

We now consider some extremely-simplified test problems to illustrate how solutions of the CR transport equations differ depending on the closure. In that spirit, we take ultra-relativistic CRs ($\beta\rightarrow1$) in a gas medium with negligible fluid motion (${\bf u} \rightarrow 0$), $\bhat=\bhat({\bf x})$ and $\bar{\nu}=\bar{\nu}({\bf x})$ independent of time, uniform $\rho$, with symmetric scattering and weak fields ($\bar{v}_{A} \rightarrow 0$, $v_{A} \rightarrow0$) no sources/sinks/other losses, and sufficiently low CR density such that the CR forces on gas are negligible (i.e.\ ``pure CR transport''). We will make the problem dimensionless by defining $f \rightarrow f/f_{i}$ ($e\rightarrow e/e_{i}$), $\tau\rightarrow \bar{\nu}_{0}\,t$, ${\bf x}\rightarrow {\bf x}\,\bar{\nu}_{0}/c$, for some reference $f_{i}$ (or $e_{i}$) and $\bar{\nu}_{0}$, and define the path-length $\ell$ integrated along a field line $\ell = \int_{{\bf x}_{0}}^{{\bf x}_{f}} d{\bf x} \cdot \bhat$ (so $\bhat\cdot\nabla X \rightarrow \partial_{\ell} X$). With these simplifications, the equations are effectively one-dimensional in $\ell$ and are identical for any moments pair $(q,\,F_{q}) = (\bar{f}_{0},\,\bar{f}_{1})$, $(n,\,F_{n})$, $(e,\,F_{e})$, etc: $\partial_{\tau} q = - \nabla \cdot (F_{q}\,\bhat)$ and $\partial_{\tau} F_{q} + \momentgrad(q) = -\bar{\nu}\,F_{q}$. 

For initial conditions (ICs), we take $q$ to be a Gaussian with $q(\tau=0) = \exp{\{-(\ell-\ell_{0})^{2} / 2\,\sigma_{q}^{2} \}}$ for arbitrary $\ell_{0}$. For the same $q(\tau=0)$, we will consider (1) isotropic ICs, where $\muone|_{\tau=0}=F_{q}/q|_{\tau=0} = 0$, and (2) ``streaming'' ICs, where $\muone|_{\tau=0}=F_{q}/q|_{\tau=0} = 1$.

We will compare the following closure assumptions. Except for the 0th-Moment/Diffusion and Exact Solution cases, all adopt the two-moment expansion, but make different assumptions about the closure assumption for $\mutwo$ or $\bar{f}_{2}$.
\begin{enumerate}

\item{\label{closure:diffusion} \bf 0th-Moment/Diffusion Approximation:} Assume the isotropic-DF limit ($\chi\rightarrow1/3$) and Newtonian+strong-scattering limits ($D_{\tau} F_{q}\rightarrow 0$),  so we obtain the single diffusion equation: $\partial_{\tau} q = \nabla \cdot [ (3\,\bar{\nu})^{-1}\,\bhat\bhat\cdot \nabla q]$. 

\item{\label{closure:isotropic} \bf Isotropic-DF:} Assume $\mutwo=1/3$ ($\chi=1/3$) always, appropriate for an isotropic DF, so $\momentgrad(q) \rightarrow (1/3)\,\bhat \cdot \nabla q$.

\item{\label{closure:streaming} \bf Maximal-Streaming:} Assume $\mutwo=1$ ($\chi=0$) always, appropriate for the fastest-possible-streaming DF, $f \propto \delta(\mu \pm 1$), so $\momentgrad(q) \rightarrow \nabla \cdot (q\,\bhat)$. 

\item{\label{closure:anisotropic} \bf Maximal-Anisotropy:} Assume the DF corresponds to a $\delta$-function with the given $\muone = \bar{f}_{1}/\bar{f}_{0} = F_{q}/q\,v$, so $\mutwo = \muone^{2}$ always.

\item{\label{closure:levermore} \bf Interpolated $\mutwo$: Levermore:} This adopts the proposed scaling $\mutwo = \closurefunction(\muone) = (3+4\,\muone^{2}) / (5 + 2\,\sqrt{4 - 3\,\muone^{2}})$ from \citet{levermore:1984.FLD.M1}, which interpolates between the isotropic-DF and anisotropic-DF limits and represents the exact closure for any DF which can be made isotropic under some Lorentz transformation.

\item{\label{closure:minerbo} \bf Interpolated $\mutwo$: Minerbo:} Adopt $\mutwo = \closurefunction(\muone) = (1/3) + (2\,\muone^{2}/15)\,(3 - |\muone|+3\,\muone^{2})$ from \citet{minerbo:1978.eddington.factors}, which similarly interpolates between limits but is exact for a DF satisfying a classical maximum-entropy principle.

\item{\label{closure:wilson} \bf Interpolated $\mutwo$: Wilson:} Adopt $\mutwo = \closurefunction(\muone) = (1 - |\muone| + 3\,\muone^{2})/3$, from \citet{wilson:1975.m1.closure}, which is realizable but represents an ad-hoc interpolation function between isotropic and anisotropic limits.

\item{\label{closure:exact} \bf Exact Solution:} We compare these to the results of {\em directly} integrating the focused CR transport equation for $f(\mu)$ explicitly as a function of $\mu$ and ${\bf x}$ per Eq.~\ref{eqn:df.general} (\S~\ref{sec:direct.methods}), using a grid of $\sim 1000$ elements in the $\mu$ dimension at each spatial position. For the isotropic IC we initialize an isotropic DF $f(\mu)$, for the streaming IC we initialize $f(\mu) \propto \delta(\mu-1)$, and for simplicity we assume isotropic scattering $\nu=\bar{\nu}$. 

\end{enumerate}

Note we have also considered other closures such as the Kershaw function $\closurefunction(\muone)=(1+2\,\muone^{2})/3$ or \citet{1992A&A...256..452J} functions $\closurefunction(\muone)=(1+\alpha_{0}\,\muone^{\alpha_{1}}+(2-\alpha_{0})\,\muone^{\alpha_{2}})$ with various $(\alpha_{0},\,\alpha_{1},\,\alpha_{2})$ suggested therein, but these generally perform more poorly than the other interpolated closures considered above.

\subsection{1D Pure-Propagation in A Homogenous Medium}
\label{sec:prop.line}

Take $\bhat=\hat{z}=$\,constant, $\bar{\nu}=\bar{\nu}_{0}=\,$constant, so the transport equations simplify to $\partial_{\tau} q = -\partial_{\ell} F_{q}$ and $\partial_{\tau} F_{q} + \partial_{\ell}(\mutwo\,q) = - F_{q}$. The problem is one-dimensional and the solutions depend only on the ICs and closure $\mutwo$, which we vary and compare in Fig.~\ref{fig:closure.compare}.

First ({\em top-left} panel), consider a case which is well-described by the isotropic, diffusive limit: ICs with $\sigma_{q} = 10$, $F_{q}(\tau=0)=0$, evolved to $\tau=200$. The CRs begin isotropic, and, recalling that $\ell=1$ corresponds in these units to the scattering mean-free-path (MFP) $=c/\bar{\nu}$, all of the gradient length and time scales even in the ICs are much larger than the CR scattering MFP. Indeed, the 0th Moment \ref{closure:diffusion}, Isotropic-DF \ref{closure:isotropic}, and all the interpolated closures \ref{closure:levermore}-\ref{closure:wilson} give nearly identical results here in excellent agreement with the Exact solution \ref{closure:exact}, as they should. The Maximal-Anisotropy closure \ref{closure:anisotropic} fails catastrophically: it assumes an initial $\muone=0$ corresponds to a pitch angle distribution with all CRs at $\mu=0$, so no flux can ever develop. The Maximal-Streaming closure \ref{closure:streaming} fails as well: although the flux equation approaches steady-state, the assumed $\mutwo=1$ means that the effective diffusion coefficient is $3\times$ larger than the correct value.

Second ({\em top-center} panel), consider a case which is close to free-streaming, a ``streaming'' IC with $\sigma_{q}=0.02$, $F_{q}(\tau=0)=q$, evolved to $\tau=0.02$, so the CRs are initially free-streaming and all scales are much shorter than the MFP. Now, the Maximally-Anisotropic \ref{closure:anisotropic}, Maximal-Streaming \ref{closure:streaming}, and interpolated \ref{closure:levermore}-\ref{closure:wilson} closures are very similar to the Exact solution \ref{closure:exact}. 0th Moment/Diffusion \ref{closure:diffusion} fails catastrophically as expected, since the system is not in the diffusive limit. The Isotropic-DF closure \ref{closure:isotropic} under-estimates the correct speed of propagation of the ``pulse,'' as expected,\footnote{Taking the derivative of the $q$ equation in \S~\ref{sec:prop.line} to combine it with the $F_{q}$ equation, we have $\partial^{2}_{\tau} F_{q} + \partial_{\tau} F_{q} =\partial_{\ell}(\mutwo\,\partial_{\ell} F_{q})$. If we enforce the isotropic-DF $\mutwo = 1/3$, then we see immediately that this reduces the maximum free-streaming speed from $c$ to $c/\sqrt{3}$.} but more problematically we see that $q$ (e.g.\ $\bar{f}_{0}$ or $e$ or $n$) has become negative in some places. This is the formally correct solution if we impose $\mutwo=1/3$ -- the issue stems from the fact that this closure violates the realizability constraint from \S~\ref{sec:closures}: there exists no positive-definite DF with $\muone=1$ (imposed by the ICs) and $\mutwo=1/3$ everywhere. 

Third, consider two intermediate cases.  
For an isotropic IC with $\sigma_{q}=0.15$ evolved to $\tau=2$ ({\em top-right} panel), the exact solution (for isotropic scattering; \ref{closure:exact}) is a symmetric flat-topped ``shelf'' moving outwards at speed intermediate between the isotropic and free-streaming cases, with diffusive ``tails.''\footnote{We stress that this is different from the ``streaming problem'' discussed extensively in e.g.\ \citet{sharma.2010:cosmic.ray.streaming.timestepping,jiang.oh:2018.cr.transport.m1.scheme,thomas.pfrommer.18:alfven.reg.cr.transport}, which also produces a ``flat shelf'' behavior. That problem effectively takes the assumptions here but further imposes (1) the strong-scattering limit with $\bar{\nu}$ very large so that $|\bar{v}_{A}| \gg (c\,|\nabla \bar{f}_{0}|)/(\nu\,\bar{f}_{0})$, (2) an isotropic-DF closure, and (3) non-zero $\bar{v}_{A}=\,$constant, so $F_{q} \rightarrow v_{\rm stream}\,q$ for some constant $v_{\rm stream}$. That is a less interesting problem for our purposes, however, since all of the interpolated closures here trivially reproduce the exact solution in this limit, and even a 0th-order closures can capture the relevant behavior provided careful numerical treatment \citep{sharma.2010:cosmic.ray.streaming.timestepping}.} None of the closures perfectly reproduces this, but the interpolated closures \ref{closure:levermore}-\ref{closure:wilson} are much closer to the exact solution and behave qualitatively similar to one another (and also rapidly converge to the exact solution as we evolve further in time). Maximal-Anisotropy \ref{closure:anisotropic} again fails catastrophically as it cannot propagate starting from $\muone=0$. Despite the IC being isotropic, the 0th Moment/Diffusion approximation \ref{closure:diffusion} also performs poorly (producing excessive ``tails'' and an incorrectly-peaked shape), as the strong-scattering/flux-steady-state assumption does not apply. Both the Isotropic-DF \ref{closure:isotropic} and Maximal-Streaming \ref{closure:streaming}, or any other closure with $\mutwo=\,$constant, produce two spurious ``peaks'' which propagate outwards with a low central density in between. 

For a streaming IC with $\sigma_{q}=0.1$ evolved to $\tau=1$ ({\em middle-left} panel), the interpolated closures \ref{closure:levermore}-\ref{closure:wilson} all resemble the exact solution \ref{closure:exact} (the peak propagates at the correct speed, with just a slightly modified shape). As expected the 0th-order/Diffusive closure \ref{closure:diffusion} fails totally. The Isotropic-DF closure \ref{closure:isotropic} again produces an unphysically negative $\bar{f}_{0}$, and under-estimates the pulse speed. The Maximal-Streaming \ref{closure:streaming} closure over-estimates the front speed but also produces an artifact of a ``shelf'' extending to $\ell < \ell_{0}$. Unlike the previous streaming IC, the Maximal-Anisotropy \ref{closure:anisotropic} closure now also under-estimates the propagation speed, as assuming $\mutwo=\muone^{2}$ suppresses the flux source term too rapidly when $\muone$ is not very close to $\pm 1$.

\subsection{1D Propagation With Variable Scattering}
\label{sec:1d.prop.variable.nu}

Now consider a spatially-variable $\bar{\nu} = \bar{\nu}_{0}\,g(\ell)$ (dimensionless equations $\partial_{\tau} q =- \partial_{\ell} F_{q}$, $\partial_{\tau} F_{q} + \partial_{\ell}(\mutwo\, q) = -g\,F_{q}$). First consider $g=\exp{\{-(\ell-\ell_{0})^{2}/(2\,\sigma_{g}^{2}) \}}$ with $\sigma_{g} \sim 0.1-10$, qualitatively akin to analytic models for Galactic CR transport with $\ell$ representing the height in the Galactic disk/halo, with both an isotropic ({\em middle-center} panel) and streaming ({\em middle-right} panel) IC. The effect here is primarily to exaggerate the differences already seen in \S~\ref{sec:prop.line}. Most notably, the 0th Moment/Diffusion approximation fails much more dramatically here, because $\nu \rightarrow 0$, causing the diffusivity $\kappa \rightarrow \infty$ at $|\ell - \ell_{0}| \gtrsim \sigma_{g}$. This leads to the PDF becoming almost perfectly flat and the diffusive ``tails'' travelling at $v \gg c$ (e.g.\ at the times plotted, we obtain fronts moving at $\gtrsim 10^{6}\,c$). 

Next, consider $g = \exp{\{ -2\,(\ell-\ell_{0}) \}}$ ({\em bottom-left} panel), where there is an asymmetric gradient across the injection region (akin to injection in any off-center location in a disk or galaxy). With the streaming IC (not shown) the differences between closures are similar to the case above. With an isotropic IC (bottom-left panel), the broken symmetry is important: at $\tau=1$, the exact solution predicts an asymmetric shelf from $-0.5 \lesssim \ell-\ell_{0} \lesssim 0.8$, with slightly higher density $f$ at $\ell<0$ (as CRs are being scattered more rapidly at $\ell < \ell_{0}$). The constant-$\mutwo$ closures \ref{closure:isotropic}, \ref{closure:streaming} fail to capture this: they again produce two peaks but these move with nearly-symmetric speed, and actually predict much larger amplitude of the peak in the $\ell > \ell_{0}$ direction (the opposite of the correct behavior). The 0th Moment \ref{closure:diffusion} case predicts essentially infinite transport speeds in the $+\ell$ direction. Interestingly, of the interpolated closures here the Wilson closure \ref{closure:wilson} best captures the correct asymmetry, suggesting this test can distinguish between more subtle variations.
\begin{figure}
	\includegraphics[width=1\columnwidth]{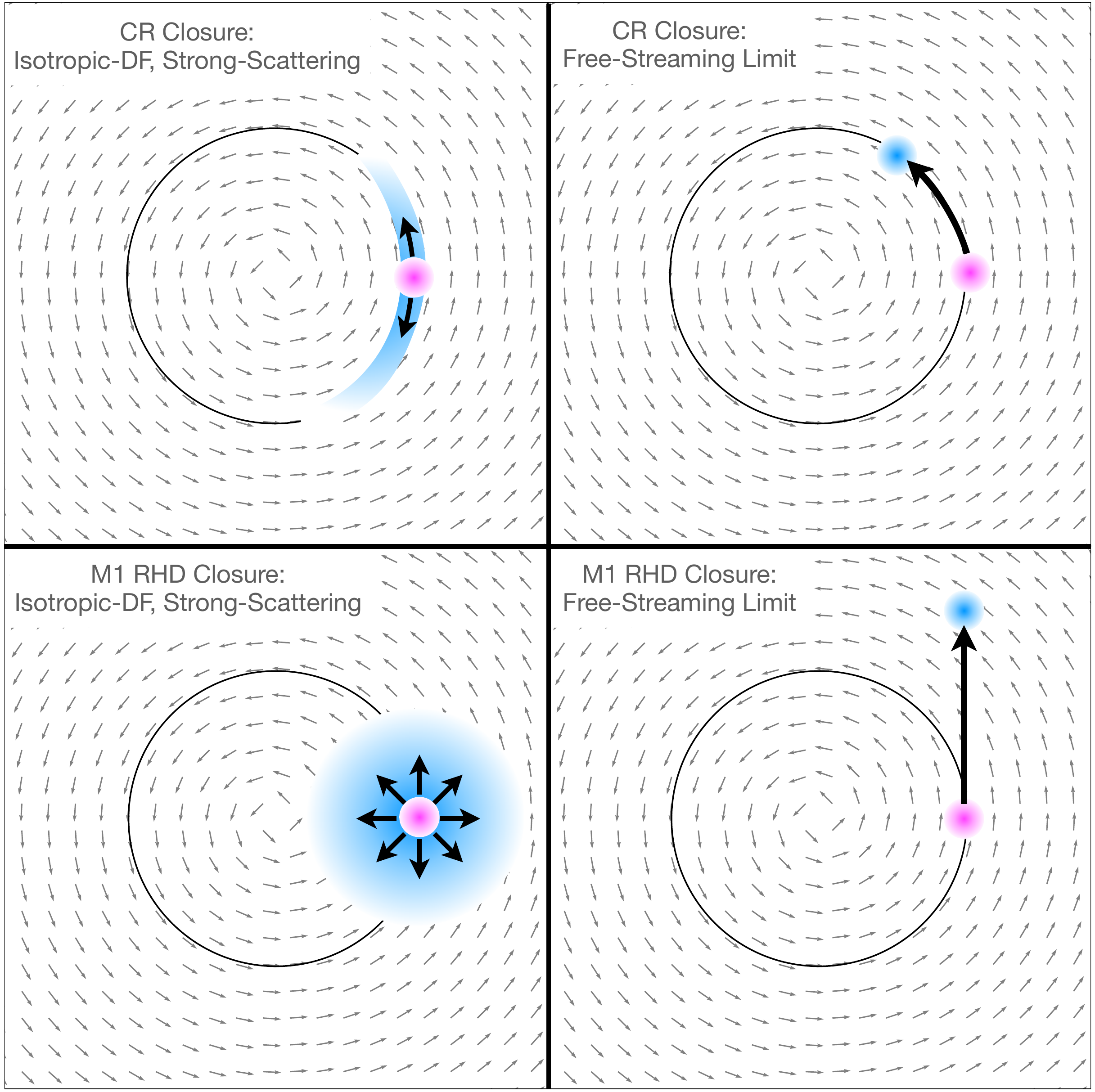}
	\vspace{-0.5cm}
	\caption{Cartoon illustrating the qualitative difference in behaviors between the CR closures proposed here ({\em top}) and the analogous M1 RHD closures ({\em bottom}), following the mathematical demonstration in \ref{sec:prop.m1}. Although the functional form of the pressure tensor $\mathbb{P}$ and its dependence on $\mutwo$, $\muone$ (the ``closure relation'') is seemingly identical if we equate $\bhat$ with the specific intensity direction $\hat{\bf n}$, Lorentz forces confining CRs give rise to fundamentally different anisotropic transport confined to fields. The figure illustrates this in a problem with purely cylindrical fields (arrows show the local direction $\bhat$, with an initial narrow Gaussian distribution of $f$ (magenta circle) injected at some position (black circle shows the closed field line along which this appears), and distribution at a later time in blue. In the isotropic-DF strong-scattering limit ({\em left}) the CR equations here reduce to spatially-anisotropic diffusion (despite the DF being isotropic in $\mu$) along the field line in both directions; in the RHD closure they reduce to globally-isotropic multi-dimensional diffusion. In the anisotropic-DF free-streaming limit ($\muone=1$, initially; {\em right}) the CR closure reduces to free-streaming ``around'' the field lines, while the RHD closure produces straight-line trajectories.
	\label{fig:m1.closure}}
\end{figure}

\subsection{Propagation With Bent Fields in A Simple Geometry}
\label{sec:prop.rings}

Now consider a variant of the ``diffusing ring'' in a cylindrical field geometry, with $\bar{\nu}=\,$constant and $\bhat=\hat{\boldsymbol{\phi}}$ purely azimuthal about some axis. This is a useful problem to illustrate the differences between the closure relation (even for ``pure transport'' in the ultra-relativistic limit) for CRs, derived here, and the analogous M1 closure relation for photons (RHD), as discussed in \S~\ref{sec:compare.M1}. 

\subsubsection{Comparison to the M1 RHD Closure}
\label{sec:prop.m1}

To illustrate the key behaviors, here we  explore mathematically the intuitive idea that CR streaming and diffusion is confined along field lines (unlike RHD). This is also sketched in Fig.~\ref{fig:m1.closure}. Take the Newtonian limit ($c \rightarrow \infty$) or flux steady-state $D_{t} F \rightarrow 0$, so we have $\partial_{\tau} q = -\nabla \cdot {\bf F}_{q}$ with ${\bf F}_{q} = -K\,{\bf g}_{\rm cr}$ where $K\sim c^{2}/\bar{\nu}$ is some effective diffusivity and ${\bf g}_{\rm cr} \equiv \bhat\,\momentgrad(q) = \bhat\bhat \cdot [\nabla \cdot (\mathbb{D}\,q)]$. For $q=e$, this becomes ${\bf g}_{\rm cr} = \bhat\bhat \cdot (\nabla \cdot \mathbb{P})$. Compare this to the RHD M1 closure, where the flux equation has the form $D_{t} {\bf F}_{\rm rad} + \nabla \cdot \mathbb{P}_{\rm rad}$ with $\mathbb{P}_{\rm rad} = \mathbb{D}_{\rm rad}\,e_{\rm rad}$, where $\mathbb{D}_{\rm rad} = \chi_{\rm rad}\,\mathbb{I} + (1-3\,\chi_{\rm rad})\,\hat{\bf n}\hat{\bf n}$ with $\chi_{\rm rad} \equiv (1-\langle \mu_{\rm rad}^{2} \rangle)/2$, identical to our definition for CRs if we identify $\bhat = \hat{\bf n}$ (the radiation flux direction). In flux steady-state, this gives ${\bf F}_{\rm rad} = -K_{\rm rad}\, {\bf g}_{\rm rad} $ with ${\bf g}_{\rm rad} = \nabla \cdot \mathbb{P}_{\rm rad}$.

Thus, even in flux-steady-state with identical effective diffusivities, we see that although the anisotropic $\mathbb{P}$ and $\mathbb{P}_{\rm rad}$ are similar, ${\bf F}_{\rm cr}$ fundamentally differs from ${\bf F}_{\rm rad}$ in that ${\bf F}_{\rm cr}$ is projected along $\bhat$. This leads to major qualitative differences in behaviors in both isotropic-DF and streaming limits. First take the isotropic-DF ($\mutwo \rightarrow 1/3$) case: ${\bf g}_{\rm cr} \rightarrow (1/3)\,\bhat\bhat \cdot \nabla e = (1/3)\,\hat{\boldsymbol{\phi}} \hat{\boldsymbol{\phi}} \cdot \nabla e$ and ${\bf g}_{\rm rad} \rightarrow (1/3)\,\nabla e$. So for CRs, even if the pitch-angle distribution is isotropic, we still have anisotropic diffusion with only parallel diffusion along the field lines allowed, owing to our assumption of a gyrotropic DF with small gyro radii. For RHD we obtain isotropic diffusion, and all information about the field lines is lost, because photons are not ``confined'' to field lines. Now consider the free-streaming limit: for CRs ${\bf g}_{\rm CR} \rightarrow \bhat\, \{ \nabla \cdot (e\,\bhat) \}$ while for RHD ${\bf g}_{\rm rad} \rightarrow \nabla \cdot(e\,\hat{\bf n}\hat{\bf n}) = \nabla \cdot(e\,\bhat\bhat)$. Now the difference is less obvious, as the RHD case is still anisotropic. But the ordering here produces totally different behavior: ${\bf g}_{\rm cr} \rightarrow \bhat \nabla \cdot (e\,\bhat) = \hat{\boldsymbol{\phi}} \hat{\boldsymbol{\phi}} \cdot \nabla e$ corresponds again to transport around an azimuthal ring (following $\bhat$),\footnote{For the cylindrical field $\bhat=\hat{\boldsymbol{\phi}}$, it is worth noting that $\nabla \cdot \hat{\boldsymbol{\phi}} = 0$, so ${\bf g}_{\rm cr} \rightarrow \hat{\boldsymbol{\phi}}\hat{\boldsymbol{\phi}} \cdot \nabla ([1-2\,\chi]\,e)$ generically and the free-streaming and isotropic-DF cases for CRs differ only in the $\chi$ factor in this test problem.} while ${\bf g}_{\rm rad} \rightarrow \nabla \cdot(e\,\bhat\bhat) = (\hat{\boldsymbol{\phi}} \cdot \nabla)\,(e \,\hat{\boldsymbol{\phi}}) = \hat{\boldsymbol{\phi}} \hat{\boldsymbol{\phi}}\cdot \nabla e + e\,(\hat{\boldsymbol{\phi}} \cdot \nabla) \hat{\boldsymbol{\phi}} = \hat{\boldsymbol{\phi}} \hat{\boldsymbol{\phi}}\cdot \nabla e - (e/r)\,\hat{r}$ produces a radially-propagating flux. Notably, while ${\bf g}_{\rm CR}$ saturates once $e\rightarrow e(r)$ becomes azimuthally-symmetric, the RHD solution in this limit actually corresponds to a ring which expands outwards at speed $\sim K/r$ \citep[see e.g.][]{hopkins:gizmo.diffusion}, because in the free-streaming limit there is nothing to ``bend'' the photon trajectories.

\subsubsection{Behavior of the CR Closures}
\label{sec:ring.closure.behavior}

Returning to the two-moment CR equations, noting for $\bhat=\hat{\boldsymbol{\phi}}$ that $\nabla \cdot \hat{\boldsymbol{\phi}}=0$, so $\nabla \cdot(q\,\bhat) = \hat{\boldsymbol{\phi}} \cdot \nabla q = \partial_{\ell} q$, we can write $\partial_{\tau} q = -\partial_{\ell} F_{q}$, $\partial_{\tau} F_{q} + \partial_{\ell}(\mutwo\,q) = -F_{q} $. But this is exactly identical to the equations with $\bhat=$\,constant in \S~\ref{sec:prop.line}, written in terms of the distance $\ell$ along the field line (so we have already shown the effects of different closures in Fig.~\ref{fig:closure.compare}). The only difference is (1) that this line is globally curved, but that can simply be considered an embedding/coordinate transformation; and (2) the circular nature of $\hat{\boldsymbol{\phi}}$ means that the boundaries for $e$, $f$ are periodic, whereas in \S~\ref{sec:prop.line} we implicitly considered open boundaries. In these simplified cases with ${\bf u}=0$, time-invariant background, $\nabla \cdot \bhat=0$, etc., any field geometry can be transformed into an equivalent 1D problem since CRs are confined along $\bhat$. The physical assumption that drives this behavior, fundamentally, is that the gyro radii of the CRs are much smaller than the radius of curvature of $\bhat$ smoothed on the scales of interest.

\subsection{Propagation in a Non-Trivial Field Geometry}
\label{sec:pop.divb}

Now consider a case with non zero ``focusing,'' $\nabla \cdot \bhat \ne 0$, for example a dipole field ${\bf B} \propto (1/r^{3})\,(2\,\cos{[\theta]}\,\hat{r} + \sin{[\theta]}\,\hat{\theta})$, which gives $\nabla \cdot \bhat = r^{-1}\,(3/\sqrt{2})\,(27\,\cos{[\theta]} + 5\,\cos{[3\,\theta]}) / (5 + 3\,\cos{[2\,\theta]})^{3/2}$. For $\bar{\nu}=$\,constant, let $\varpi \equiv (c/\bar{\nu})\,\nabla \cdot \bhat$, so our equations become $\partial_{\tau} q = -(\partial_{\ell} F_{q} + F_{q}\,\varpi)$ and $\partial_{\tau} F_{q} + \partial_{\ell}[(1-2\,\chi)\,q] + (1-3\,\chi)\,q\,\varpi = -F_{q}$. Since $\bhat$ is constant in time, we can write $\varpi=\varpi(\ell,\,{\bf x}_{0})$ as a function of length $\ell$ along some path following $\bhat$, and again the problem becomes one-dimensional along each field line. Mathematically, $\nabla\cdot\bhat$ acts like a source/sink term representing the (de)focusing of field lines (e.g.\ for a dipole, near the ``pole'' with $\theta\ll \pi/2$, $\nabla \cdot \bhat \approx 3/r$); but, we see the effect in the flux equation depends on the closure $\chi$. For simplicity, we take $\varpi=3$ to be constant over the interval calculated, and consider an isotropic IC ({\em bottom-middle} panel of Fig.~\ref{fig:closure.compare}) and streaming IC ({\em bottom-right} panel).

With an isotropic IC, we see that the isotropic-DF \ref{closure:isotropic} and Maximal-Anisotropy \ref{closure:anisotropic} cases fail completely to capture the correct anisotropy: in the $F_{q}$ equation, an isotropic-DF closure exactly eliminates the focusing term, and the maximal-anisotropy case produces propagation opposite the exact solution \ref{closure:exact}. Meanwhile, Maximal-Streaming \ref{closure:streaming} strongly over-estimates the anisotropy. The interpolated closures \ref{closure:levermore}-\ref{closure:wilson} at least capture the key qualitative behaviors.

With the streaming IC, the interpolated closures \ref{closure:levermore}-\ref{closure:wilson} are nearly identical and all behave qualitatively akin to the exact solution \ref{closure:exact}. Both constant-$\mutwo$ (isotropic or anisotropic) \ref{closure:isotropic}, \ref{closure:streaming}, and the Maximally-Anisotropic \ref{closure:anisotropic} cases produce negative DFs.\footnote{While technically closure \ref{closure:streaming} with $\mutwo=1$ is realizable for any $\muone$, this always represents a sum of $\delta$-functions with $\mu=\pm1$, which means even a local minimum in $f$ can have net ``outgoing'' flux in $\pm\ell$ directions, producing negative solutions. Meanwhile realizability for \ref{closure:anisotropic} fails as it attempts to interpolate through a position where $f\rightarrow 0$.} We also see the 0th Moment \ref{closure:diffusion} closure fail in a new manner: this closure cannot correctly treat the focusing term. For anisotropic diffusion ${\bf F}_{q} \propto -\bhat\bhat\cdot \nabla q$, as required for realistic CR dynamics, a non-zero $\nabla \cdot \bhat$ still appears as a source term in the $q$ equation, but the flux closure assumption \ref{closure:diffusion} means the focusing term in the flux is not included. The result  is that the front for \ref{closure:diffusion} actually propagates in the opposite direction to that of the correct solution.

\subsection{Summary}
\label{sec:prop.summary}

Just like the analogous RHD case, no two-moment closure can capture the exact behavior of full phase-space solutions for $f(\mu)$. However, the interpolated closures \ref{closure:levermore}-\ref{closure:wilson}  at least capture the qualitative behaviors of all terms in all test problems considered here. Constant-$\mutwo$ closures like assuming a near-isotropic-DF \ref{closure:isotropic} or a free-streaming-DF \ref{closure:streaming} or a maximally-anisotropic ($\delta$-function) DF \ref{closure:anisotropic} fail catastrophically on some problems and, most crucially, fail to ensure non-negative solutions for $f$ or $\bar{f}_{0}$ (e.g.\ CR number and energy density). While taking the 0th-Moment/Diffusion limit \ref{closure:diffusion} does ensure positive-definite solutions, it fails catastrophically in other ways: it drives CR transport in the incorrect direction in situations with strong focusing, streaming, or scattering-rate-gradients, and it produces super-luminal transport. 

Among the interpolated closures, the Levermore and Minerbo closures \ref{closure:levermore}-\ref{closure:minerbo} produce very similar results (not surprising since they give nearly-identical $\mutwo(\muone)$ functions). The Wilson closure \ref{closure:wilson} performs slightly more accurately with isotropic ICs, though it sometimes slightly under-estimates peak-amplitude in free-streaming ICs, which is expected as it gives $\mutwo$ slightly closer to the isotropic-DF $\mutwo=1/3$ at intermediate $\muone$.

Of course, real problems will be vastly more complex, with advection velocities ${\bf u}$ comparable to CR transport speeds, spatial-and-time variable versions of all quantities above, $\bar{\nu}$ dependent on $\mu$ as well as space and time, etc. We emphasize that many of the most important consequences of the proposed closures may only be evident in those scenarios. For example, if the ``adiabatic'' terms $\propto (\chi\,\mathbb{I}+[1-3\,\chi]\,\bhat\bhat):\nabla{\bf u}$, gyro-resonant losses $\propto \bar{v}_{A}\,F_{e}$, diffusive re-acceleration gains $\propto 3\,\chi\,v_{A}^{2}\,(e+P_{0})$, trans-\Alf{ic} or CR ``streaming'' speed $\propto 3\,\chi\,\bar{v}_{A}$ are important, these depend quite strongly on $\chi$ and therefore on the closure (with re-acceleration and \Alf{ic} streaming behaviors vanishing entirely in the anisotropic limit). Likewise, simulations where the CR forces on gas are important will be sensitive to the closure relation because the shape and anisotropic form of $\mathbb{P}$ depend explicitly on the closure relation.

\section{Relation to Other CR \&\ Radiation Transport Formulations}
\label{sec:relations}

\subsection{Relation to Previous CR Moments Formulations}
\label{sec:prev.moments}

Recently, \citet{jiang.oh:2018.cr.transport.m1.scheme,chan:2018.cosmicray.fire.gammaray,thomas.pfrommer.18:alfven.reg.cr.transport,hopkins:cr.transport.constraints.from.galaxies} have explored two-moment formulations of the CR energy transport equation ($q=e$). Those in \citet{chan:2018.cosmicray.fire.gammaray,hopkins:cr.transport.constraints.from.galaxies} and \citet{jiang.oh:2018.cr.transport.m1.scheme} were heuristically motivated by two-moment treatments of RHD but the authors did not attempt to derive a set of equations consistent with the actual DF equation for CRs
(nor appropriate closure, etc). \citet{thomas.pfrommer.18:alfven.reg.cr.transport} (here TP) did attempt such a derivation for the nearly-isotropic-DF case, and indeed \S~\ref{sec:isotropic} mostly follows their more detailed and comprehensive discussion. It is therefore worth noting how the work here extends their formulation. The major differences here are: 
{\bf (1)} We derive moments equations for the DF $f$ itself as well as integrals like CR number/total energy/kinetic energy $n$, $e$, $\epsilon$, while TP primarily focused on just $e$.
{\bf (2)} Our equations are valid for arbitrary CR $\gamma$, while TP considered only the ultra-relativistic ($\gamma \gg 1$, $\beta\approx 1$ case). 
{\bf (3)} We develop the equations for the entire CR spectrum $f(p)$ or $e^{\prime}(p)$, while TP focused on the spectrally-integrated expressions.
{\bf (4)} Our equations are agnostic to the specific scattering model (this physics is not our focus), while TP focused in detail on  deriving specific expressions for $\bar{\nu}_{\pm}$ due to CR scattering from \Alf\ waves within the context of  CR self-confinement scenarios. 
{\bf (5)} Most importantly, TP focused exclusively on the nearly-isotropic-DF case and enforced the strong-scattering closure $\mutwo = 1/3$; we derive a more general set of expressions that allow for anisotropic DFs and CR pressure, and can approximately capture the CR free-streaming limit. 

Most earlier CR transport models in galaxy simulations  adopted a ``zeroth-moment'' or pure-diffusion approximation, evolving e.g.\ the spectrally-integrated $e$ with ${\bf F}_{e} = \boldsymbol{\kappa}\,\nabla P_{0}$. The anisotropic version of this, with $\boldsymbol{\kappa} = \kappa_{\|}\,\bhat\bhat$, of course arises if we take the isotropic-DF, strong-scattering, Newtonian ($c\rightarrow \infty$, so flux-steady-state always applies) limit. Although simpler, this can give a number of unphysical behaviors, as discussed above. This can be mitigated by adopting a flux-limited-diffusion-type approximation, replacing $\kappa_{\|}\,\bhat\bhat\cdot \nabla e \rightarrow \phi_{\rm lim}\,\kappa_{\|}\,\bhat\bhat\cdot\nabla e$ with $\phi_{\rm lim} \equiv {\rm MIN}[1,\, \beta\,e\,c / | \kappa_{\|}\,\bhat\bhat\cdot \nabla e |]$, but as we have shown, there are qualitative phenomena this closure still fails to capture.

\subsection{Relation to the Isotropic FP Equation}
\label{sec:isotropic.fp}

By far the most popular form of the CR transport equations adopted in Galactic models of CR transport that do not attempt to explicitly follow galactic dynamics -- e.g.\ GALPROP \citep{strong:2001.galprop} or DRAGON \citep{evoli:dragon2.cr.prop} -- is the isotropic Fokker-Planck equation: 
\begin{align}
\label{eqn:iso.fp} \frac{\partial f}{\partial t} &= \nabla \cdot(\bar{D}_{x x} \nabla f) + \frac{1}{p^{2}}\,\frac{\partial }{\partial p}\left[ p^{2}\,\left( S\,f + \bar{D}_{pp}\,\frac{\partial f}{\partial p} \right) \right] + j.
\end{align}
If fluid velocities are included (these are often dropped), they are taken to add the terms $-{\bf u}\cdot \nabla f + (1/3)\,(\nabla \cdot {\bf u})\,p\,\partial_{p}\,f$ to the right-hand side of Eq.~\ref{eqn:iso.fp}.

This equation arises from our Eqs.~\ref{eqn:f0.eqn.final}-\ref{eqn:f1.eqn.final}, if we make the following assumptions: 
{\bf (1)} assume an isotropic-DF closure, so $\mutwo \rightarrow 1/3$, $\mathbb{D}\rightarrow \mathbb{I}/3$, $\momentgrad(q) \rightarrow \bhat\cdot\nabla q/3$, etc.; 
{\bf (2)} assume the Newtonian limit ($c\rightarrow \infty$) or the infinite-strong-scattering ($\bar{\nu}\rightarrow \infty$) limit in the CR flux or first $\mu$-moment $\bar{f}_{1}$ equation (Eq.~\ref{eqn:f1.eqn.final}), so that the CR flux reaches its local equilibrium value instantaneously, with $D_{t} \bar{f}_{1}\rightarrow 0$; 
{\bf (3)} assume that the scattering is {\em also} exactly isotropic with respect to pitch angle, so that $\bar{\nu}_{+}=\bar{\nu}_{-}$ (to $\mathcal{O}(u/c)$) and $\bar{v}_{A}\rightarrow 0$; this causes the $D_{\mu p}$ and $D_{p \mu}$ terms to vanish; {
\bf (4)} take the resulting anisotropic spatial diffusion term $\nabla \cdot(\beta\,\bhat\,\bar{f}_{1}) \rightarrow \nabla \cdot ( \bar{D}_{\|}\,\bhat\bhat \cdot \nabla f_{0})$ with $\bar{D}_{\|} = (\beta\,c)^{2}/(3\,\bar{\nu})$, and assume that the magnetic field direction $\bhat$ is isotropically random or ``tangled'' on scales of the mean free path (below some averaging scale), allowing it to be approximated as an isotropic diffusion $\nabla \cdot (\bar{D}_{x x}\,\nabla \bar{f}_{0})$ with $\bar{D}_{x x} \equiv \bar{D}_{\|}/3$ (which produces the commonly-assumed relation for this limit $D_{xx}\,D_{pp} = p^{2}\,v^{2}_{A}/9$); 
and 
{\bf (5)} drop the terms involving the fluid velocities ${\bf u}$ (sometimes called ``convective'' terms). 

The major limitations of Eq.~\ref{eqn:iso.fp} are therefore that it cannot capture anisotropy in the DF $f(\mu)$, anisotropy in the scattering rates $\nu_{\pm}(\mu)$, or anisotropy in the field geometry $\bhat$ (each of which is independent). It also cannot  correctly describe the free-streaming/weak-scattering or out-of-flux-equilibrium limit (e.g. $D_{t} F \ne 0$, relevant just after injection, or when $\bhat$ changes direction rapidly, or when $\bar{\nu}$ varies spatially or temporally). Finally, depending on the form adopted, it ignores or treats less accurately the fluid velocity and comoving-vs-inertial frame terms.

\subsection{Relation to the M1 RHD Equations} 
\label{sec:compare.M1}

Our derivation of the CR moment equations \&\ closure from the focused transport equation closely parallels the derivation of the radiation moments and M1 closure from the specific intensity equation in e.g.\ \citet{levermore:1984.FLD.M1,mihalas:1984oup..book.....M} and others, and indeed there are many similarities. However there are some important differences. The physics, of course, is completely distinct, and the detailed form of the scattering and collisional/loss terms totally different. Most obviously, radiation is always in the ultra-relativistic limit, so properties like $\beta\rightarrow1$ and $\epsilon\rightarrow e$ are always satisfied in RHD. Nonetheless, even for ``free'' transport of ultra-relativistic CRs, important differences arise  from two key effects: (1) the CRs are gyrotropic and feel Lorentz forces, and there is a scale hierarchy imposed by the assumption that the gyro radius is much smaller than resolved scales; (2) the ``preferred direction'' is $\bhat$ (not the solid angle vector $\hat{\bf n}$ in RHD), which can change direction and responds to the gas physics.

As a result, a number of terms appear which do not have an RHD analog, including (1) the $\tilde{S}_{\rm sc}$ terms and $\bar{v}_{A}$ terms that introduce the \Alf\ frame; (2) the perpendicular pressure forces in the gas hydro equation (which relate to Lorentz forces and therefore do not vanish even with weak parallel scattering), and (3) various geometric terms that alter the directions of key transport behaviors. For the latter, mathematically we see that the non-commutation of $\bhat$ and $\nabla$ results in the flux equation having the form $\bhat\,D_{t} F$ instead of $D_{t}{\bf F}$. Terms such as $\momentgrad (q) = (1-2\,\chi)\,\bhat \cdot \nabla q + (1-3\,\chi)\,q\, \nabla \cdot \bhat$ have fundamentally non-hyperbolic components and do not have the same form as their RHD analog, which can be written $D_{t}{\bf F} =-\nabla \cdot \mathbb{P}+...$. We could only do this if $\bhat$ and $\chi$ were uniform everywhere. The consequences of this are plainly illustrated in \S~\ref{sec:prop.m1} -- it produces qualitatively different behaviors.

Like the M1 case in RHD, there are still cases where our ``interpolated'' closure (Eq.~\ref{eqn:closure}) fails. For example, it cannot capture the ``intersecting rays'' problem, where $\muone = 0$ not because of an isotropic distribution (as the proposed closure in Eq.~\ref{eqn:closure} assumes), but because $f(\mu) = (1/2)\,\bar{f}_{0}\,( \delta(\mu-1) + \delta(\mu+1) )$. If $\bar{\nu} \rightarrow 0$, the closures predict that two free-streaming rays will ``collide'' and then diffuse out, rather than pass one another truly collisionlessly. More complicated closure schemes for $\mutwo$ can be devised to address this. It is less clear, however, whether this is as much a problem for CRs as for radiation, since the CRs are not truly collisionless ``test particles'' as they stream, in the way photons are. In fact, in this particular situation the CRs would be unstable to two-stream instabilities, so ``collide then diffuse'' may indeed be a more accurate description of their true dynamics. Fully kinetic CR models that do not assume even CR gyrotropy (as assumed from the start of our derivations) are needed to properly address such physics. 

Related to this, an important physical difference is that the M1-RHD closure imposes the assumption that the DF is symmetric about the flux direction $\hat{\bf F}$ {\em ad-hoc}, without any particular physical motivation. This can be violated rather severely on all spatial scales, e.g.\ if rays intersect at oblique angles. Here, the gyrotropic CR assumption is much more well-motivated, and has a well-defined scale length (the gyro scale) providing a formal scale-separation hierarchy.

\subsection{Hybrid Schemes \&\ a Note on the ``Gyro-Resonant Loss'' \&\ ``Re-Acceleration'' Terms}
\label{sec:hybrid}

Recently hybrid schemes have been proposed that evolve $f({\bf x},\,{\bf p})$ in large-scale simulations by directly evolving $\bar{f}_{0}({\bf p}\,|\,{\bf x})$ in momentum-space at each cell position ${\bf x}$, while using a zeroth or first-moment expansion scheme for the spatial terms \citep[e.g.][]{girichidis:cr.spectral.scheme}. These are straightforward to generalize to the methods here, by evolving $\bar{f}_{0}$, $\bar{f}_{1}$ according to Eqs.~\ref{eqn:f0.eqn.final}-\ref{eqn:f1.eqn.final}. In these approaches, the equations for $q$ or $\bar{f}_{0}$  can be operator-split into a hyperbolic spatial transport step $D_{t} \bar{f}_{0}+\nabla\cdot(v\,\bar{f}_{1}\,\bhat)=0$ and a momentum-space step where all the source and sink terms (including e.g.\ the ``adiabatic'' term $\mathbb{P}:\nabla {\bf u}$, $\tilde{S}_{\rm sc}$, and $S_{q}$) are evolved following Eq.~\ref{eqn:f0.eqn.final}.

In this spirit, recall from \S~\ref{sec:monte.carlo} that we can derive from the momentum-space translation/diffusion terms (including the adiabatic and $p^{-2}\,\partial_{p} p^{2} (\bar{D}_{p\mu}\bar{f}_{1} + \bar{D}_{pp} \partial_{p} \bar{f}_{0})$ terms) a mean rate-of-change $\dot{\langle p\rangle}$ of the CR momentum or energy (of a CR ``group'' with the same initial $p$; see Eq.~\ref{eqn:individual.momentum}). Pitch-angle averaging Eq.~\ref{eqn:individual.momentum}, using $\muone = F_{q}/q$, gives $\dot{\langle p \rangle}/p = -(\mathbb{D}:\nabla{\bf u}) - (\nu/v^{2})\,[\bar{v}_{A}\,F_{q}/q - 2\,\chi\,v_{A}^{2}\,(1+\beta^{2})]$. If we take $q=e^{\prime}$, and use various identities in \S~\ref{sec:number.energy.equations} to replace $\beta$, we can rewrite this as:
\begin{align}
\frac{\dot{\langle p \rangle}}{p} &= -\mathbb{D}:\nabla{\bf u} - \frac{\bar{\nu}}{c^{2}}\,\frac{1}{3\,P^{\prime}_{0}} \left[ \bar{v}_{A}\,F^{\prime}_{e} - 2\,\chi\,{v}^{2}_{A}\,\left( e^{\prime} + 3\,P^{\prime}_{0} \right)  \right] + ...
\end{align}
The first (adiabatic) term immediately reduces to the familiar $\dot{\langle p\rangle}=-(1/3)\,(\nabla\cdot{\bf u})\,p$ expression if we assume an isotropic-DF closure. The second (scattering) term closely resembles $\tilde{S}^{\prime}_{\rm sc}$, and indeed in the ultra-relativistic limit where $E\propto p$ (and $P_{0}^{\prime} = e^{\prime}/3$) it becomes exactly $\tilde{S}^{\prime}_{\rm sc} / e^{\prime}$ (i.e.\ the rate of change of energy and momentum become identical). In this term the first ($\propto \bar{v}_{A}\,F$) part stems from $D_{p \mu}$, while the second ($\propto v_{A}^{2}\,e$) stems from $D_{p p}$. The ``...'' term refers to other collisional terms (e.g.\ radiative losses).

In self-confinement scenarios where the scattering waves are excited by gyro-resonant instabilities sourced by the CR flux, waves are excited only in the direction of $\hat{\bf F}_{e}^{\prime}$, so we generically expect\footnote{As discussed in \citet{hopkins:cr.transport.constraints.from.galaxies}, if one somehow did have $\bar{\nu} \sim \bar{\nu}_{-}$ on micro-scales, the timescale for the $\bar{\nu}_{\pm}$ to come into the equilibrium state with $\bar{v}_{A} \rightarrow v_{A}\,\hat{\bf F}_{e}^{\prime}\cdot \bhat$ is much smaller than resolved timescales in galaxy-scale simulations.} an extreme forward/backward difference with $\bar{\nu}_{+} \gg \bar{\nu}_{-}$ or $\bar{\nu}_{+} \ll \bar{\nu}_{-}$, corresponding to whichever points in the direction of $F_{e}^{\prime}$. This gives $\bar{v}_{A} = v_{A}\,\hat{\bf F}_{e}^{\prime}\cdot \bhat = \pm v_{A}$. While the scattering term in $\dot{\langle p \rangle}$ can be positive if the CRs are streaming sub-\Alf{ic}ally ($|F_{e}^{\prime}| \lesssim v_{A}\,e^{\prime}$), it is generically negative, and if the CR energy (Eq.~\ref{eqn:e.specific}) is in flux-steady-state ($D_{t} F_{e}^{\prime} \rightarrow 0$) in the strong-scattering or isotropic-DF limit, it takes the negative-definite value $\dot{\langle p \rangle}/p \rightarrow -(v_{A}\,|\bhat\cdot\nabla P_{0}|/3\,P_{0}) - \bar{\nu}\,(v_{A}/\gamma\,\beta\,c)^{2}$. In this limit, this represents the CR energy loss to gyro-resonant instabilities -- the ``streaming loss'' or ``gyro-resonant loss'' term \citep{Wien13,wiener:cr.supersonic.streaming.deriv,Rusz17,thomas.pfrommer.18:alfven.reg.cr.transport}.\footnote{In these studies the CRs were taken to be ultra-relativistic so the gyro-resonant losses simply become $-v_{A}\,|\bhat\cdot\nabla P_{0}|/3\,P_{0}$.}

In extrinsic turbulence scenarios, if the turbulence and scattering rates are perfectly isotropic in the \Alf\ frame, then $\bar{v}_{A}=0$ ($\bar{\nu}_{+}=\bar{\nu}_{-}$), so the $D_{p \mu}$ or $F$ term above vanishes and the scattering term becomes positive-definite with $\dot{\langle p  \rangle}/p \rightarrow \bar{\nu}\,(v_{A}/v)^{2}\,2\,\chi\,(1+\beta^{2}) \sim v_{A}^{2}/D_{x x}$. 
This is the ``turbulent'' or ``diffusive'' re-acceleration term. However, note that in the anisotropic-DF case ($\chi\rightarrow 0$) this vanishes; even very weakly anisotropic-scattering (unless $\nu_{+}(\mu)=\nu_{-}(\mu)$ cancel to high precision $|\nu_{+} - \nu_{-}|/|\nu_{+} + \nu_{-}| \ll |v_{A}\,(e+P)/F| \sim v_{A}/v_{\rm eff}$) the $\bar{\nu}\,F$ or $D_{p \mu}$ (loss) term will usually dominate. 

In any case, the preceding discussion makes it clear that our derived scalings include both the ``gyro-resonant'' or ``streaming'' loss and ``turbulent/diffusive reacceleration'' terms, in a more general form.

\subsection{Where and When Are These Differences Most Important?}
\label{sec:importance}

It is helpful to ask ``under what conditions will the predictions from the more accurate expressions herein differ most dramatically from the predictions of simpler, less-accurate (e.g. isotropic Fokker-Plank, zeroth-moment/diffusion, or isotropic-DF) CR transport expressions?'' Examination of the relevant equations and our tests in Fig.~\ref{fig:closure.compare} suggest this will typically be most important when the CR scattering mean free time ($\sim \bar{\nu}^{-1}$) or path ($\ell_{\rm MFP}\sim c/\bar{\nu}$, since we must consider the full range of $\mu$) become larger than some other scales of interest or relevance for CR transport (e.g.\ the gradient scale-lengths for $\bhat$, $\ell_{\bhat}\equiv|\bhat|/|\nabla\cdot\bhat|$, or $\bar{\nu}$, $\ell_{\bar{\nu}}\equiv\bar{\nu}/|\nabla\bar{\nu}|$, or background quantities such as the gas density or pressure if CR-gas interactions are of interest). As shown in Fig.~\ref{fig:closure.compare}, this is true {\em even if} the CR DF is close-to-isotropic. And although the scattering time $\bar{\nu}^{-1}$ is generally short, the scattering length can be quite large: if we take state-of-the-art empirical estimates of $\bar{\nu}$ in the Solar neighborhood/LISM \citep[e.g.][converting from an isotropic diffusivity to $\bar{\nu}$]{evoli:dragon2.cr.prop,2018AdSpR..62.2731A,chan:2018.cosmicray.fire.gammaray,hopkins:cr.transport.constraints.from.galaxies,delaTorre:2021.dragon2.methods.new.model.comparison}, we obtain $\ell_{\rm MFP} \sim 10\,{\rm pc}\,R_{\rm GV}^{0.5}$, where $R_{\rm GV}$ is the CR rigidity in GV. 

In phenomenological models where $\bar{\nu}$ is constant, $\ell_{\bar{\nu}}\rightarrow\infty$ by definition, so the effects of the expressions here will generally be more modest. However, for $\sim1-10\,$GV CRs, $\ell_{\bhat}$ (essentially the \Alf\ scale of ISM turbulence) can be comparable to $\ell_{\rm MFP}$, and for $\gtrsim 10\,$GV CRs, $\ell_{\rm MFP}$ can begin to exceed the Galactic disk scale-height. So propagation models over these scales, especially for high-energy CRs and/or models where the CR-gas coupling is important (e.g.\ models of CR-driven winds where the ``launching'' occurs from the disk) could be sensitive to the more detailed CR transport expressions here.

Much more dramatically, in physically-motivated models where the scattering rates $\nu$ are set by some competition between damping and driving either by gyro-resonant instabilities (self-confinement models) or extrinsic turbulence, then $\bar{\nu}$ can be a strong function of quantities such as the neutral fraction or gas temperature or local Mach numbers (see e.g.\ \citealt{yan.lazarian.04:cr.scattering.fast.modes,zweibel:cr.feedback.review}, or the review in \citealt{hopkins:cr.transport.constraints.from.galaxies}), which can vary on vastly smaller scales (the skin depth of phase transitions or shock widths, orders-of-magnitude smaller than $\ell_{\rm MFP}$). These rapid changes can be tightly associated with phenomena such as CR ``bottlenecks'' (as CRs propagate across phase transitions) or the CR ``staircase'' which arises in self-confinement models of CR-driven outflows, all of which have been the subject of considerable recent study using variations of the simpler CR transport expressions that may not accurately represent the exact solutions in this regime \citep[e.g.][]{bustard:2020.crs.multiphase.ism.accel.confinement,winner:2020.cr.harps.transport,huang.davis:2021.cr.staircase.in.outflows,quataert:2021.cr.outflows.diffusion.staircase,tsung:2021.cr.outflows.staircase}. In these regimes, the bulk CR behavior could differ substantially with the more accurate expressions proposed herein (\S~\ref{sec:closures}).

Finally, if $\nu(\mu)$ itself is strongly anisotropic, then an approach which evolves the pitch-angle DF, as in \S~\ref{sec:finite.volume}, becomes crucial to obtaining accurate results.

\section{The Reduced-Speed-of-Light (RSOL) Approximation}
\label{sec:rsol}

Explicitly integrating Eqs.~\ref{eqn:f0.eqn.final}-\ref{eqn:ek.specific} imposes a Courant-type timestep limiter $\Delta t \le C\,\Delta x/c$ in Lagrangian codes (or $\Delta t \le C\,\Delta x/(c+u)$ in Eulerian codes). While this is generally less onerous at high resolution than the quadratic condition imposed by ``pure diffusion'' or ``zeroth moment'' schemes (where $\partial_{t} f \propto \kappa\,\nabla^{2}\,f$, imposing $\Delta t \le C\,\Delta x^{2}/\kappa$), it is still often numerically prohibitive because $c$ is much faster than any other signal speed in the problem. By analogy to RHD, we can therefore adopt a ``reduced speed of light'' or RSOL approximation, as in many previous CR studies \citep{jiang.oh:2018.cr.transport.m1.scheme,su:2018.stellar.fb.fails.to.solve.cooling.flow,su:turb.crs.quench,ji:fire.cr.cgm,chan:2018.cosmicray.fire.gammaray,hopkins:cr.mhd.fire2,hopkins:2020.cr.outflows.to.mpc.scales,hopkins:cr.transport.constraints.from.galaxies,hopkins:2020.cr.transport.model.fx.galform,buck:2020.cosmic.ray.low.coeff.high.Egamma}. However, in those studies, the CR transport equations were developed ad-hoc, as described above. Here we develop two viable RSOL formulations, and describe the terms where additional corrections are needed. 

\subsection{Alternative (Viable) Formulations}
\label{sec:rsol.viable.eqns}

Per the preceding derivations, we can generically write the spatial transport terms in the CR moment equations for $(\bar{f}_{0}$, $\bar{f}_{1})$\footnote{Note Eqs.~\ref{eqn:f0.eqn.final}-\ref{eqn:f1.eqn.final} can be written $c^{-1}\,D_{t} \bar{f}_{0} + \nabla \cdot(\beta\,\bar{f}_{1}\,\bhat) = (...)$, $c^{-1}\,D_{t} (\beta\,\bar{f}_{1}) +  \beta^{2}\,\momentgrad(\bar{f}_{0}) = \beta\,(...)$, matching the form in Eq.~\ref{eqn:generic.transport} for $(q,\,F_{q})=(\bar{f}_{0},\,v\,\bar{f}_{1})$.} or $(q,\,F_{q})$ with $q=(n,\,e,\,\epsilon)$ for some species and energy interval as:
\begin{align}
\label{eqn:generic.transport} 
\frac{1}{c}\,{\rm D}_{t} q + \nabla\cdot\left( \frac{F_{q}}{c}\,\bhat \right) &= \frac{1}{c}\,\Seff_{q}(...,\,c), \\ 
\nonumber \frac{1}{c}\,{\rm D}_{t} \left( \frac{F_{q}}{c} \right) + \beta^{2}\,\momentgrad(q) &= \frac{1}{c}\,\Seff_{F_{q}}(...,\,c) 
\end{align}
(we collect all of the non-transport terms such as scattering and sources/sinks in $\Seff$). 

When using the RSOL approximation, it is important to be careful which values of $c$ are replaced with the RSOL $\tilde{c}$. We wrote these equations in the form $c^{-1}\,D_{t} q = ...$ because then (just like in radiation hydrodynamics; see \citealt{2013ApJS..206...21S}, and references therein) the RSOL replaces {\em only} the value[s] of $c$ associated with the $D_{t}$ term.\footnote{Because our moments are defined in the comoving frame, we associate $\tilde{c}$ with $D_{t}$, as opposed to $\partial_{t}$, which is more appropriate when the salient quantities are defined in the lab frame.} There are then two choices of viable scheme, first:
\begin{align}
\label{eqn:rsol.1} 
\frac{1}{\tilde{c}}\,{\rm D}_{t} q + \nabla\cdot\left( \frac{F_{q}}{c}\,\bhat \right) &= \frac{1}{c}\,\Seff_{q}(...,\,c), \\ 
\nonumber \frac{1}{\tilde{c}}\,{\rm D}_{t} \left( \frac{F_{q}}{c} \right) + \beta^{2}\,\momentgrad(q) &= \frac{1}{c}\,\Seff_{F_{q}}(...,\,c) ,
\end{align}
or alternatively 
\begin{align}
\label{eqn:rsol.2} 
\frac{1}{c}\,{\rm D}_{t} q + \nabla\cdot\left( \frac{F_{q}}{c}\,\bhat \right) &= \frac{\Psi}{c}\,\Seff_{q}(...,\,c), \\ 
\nonumber \frac{1}{\tilde{c}}\,{\rm D}_{t} \left( \frac{F_{q}}{\tilde{c}} \right) + \beta^{2}\,\momentgrad(q) &= \frac{1}{c}\,\Seff_{F_{q}}(...,\,c) .
\end{align}
The formulation in Eq.~\ref{eqn:rsol.1} is exactly equivalent to replacing $c^{-1}\,D_{t} f \rightarrow \tilde{c}^{-1}\,D_{t} f$ in the original focused transport Eq.~\ref{eqn:focused.cr},\footnote{Consider the free-streaming limit of the focused transport Eq.~\ref{eqn:focused.cr}, with negligible scattering in a homogeneous medium: $c^{-1}\,D_{t} f + \nabla \cdot (\mu\,\beta\,f\,\bhat) = 0$. This is pure advection with $v=\beta\,\mu\,c$; taking $c\rightarrow \tilde{c}$ correspondingly reduces the maximum bulk/free-streaming advection speed from $\beta\,c$ to $\beta\,\tilde{c}$.} then following our derivations identically. It is also the more common scheme in RHD. The formulation in Eq.~\ref{eqn:rsol.2} associates $\tilde{c}$ {\em only} with the flux equation, instead, and introduces the function $\Psi \equiv {\rm MIN}[1,\,|F_{q}|/F_{\rm true}]$ with $F_{\rm true} \approx {\rm MIN}[q\,\beta\,c,\,|F_{q}(\tilde{c}\rightarrow \infty)|]$, as justified below.\footnote{\citet{jiang.oh:2018.cr.transport.m1.scheme,chan:2018.cosmicray.fire.gammaray,hopkins:cr.mhd.fire2} used a formulation similar to Eq.~\ref{eqn:rsol.2}, but set $\Psi=1$, which as we argue below leads to significantly slower convergence with respect to $\tilde{c}/c$.} 

These share the most important features: (1) the maximum signal speed for free-streaming is reduced to $\beta\,\tilde{c}$, meaning that the stable Courant timestep condition becomes $\Delta t \propto \Delta x/(\beta\,\tilde{c})$, allowing much larger timesteps (the reason to introduce the RSOL); (2) both exactly recover the true Eq.~\ref{eqn:generic.transport} as $\tilde{c} \rightarrow c$; (3) both converge exactly to the true ($\tilde{c}=c$) solutions for $q$, $F_{q}$, $\Seff_{q}$, in local steady-state (when $D_{t} \rightarrow 0$).

\subsection{Out of Equilibrium Behaviors and Timescales}
\label{sec:rsol.limits}

The differences between the schemes come when $\tilde{c}\ll c$ out of steady-state. Define $\Gamma\equiv c/\tilde{c}$ and consider some key timescales: the flux-convergence timescale $\Delta t_{\rm F}$, the loss/injection timescale $\Delta t_{\rm in/loss}$, and the CR transport/escape timescale $\Delta t_{\rm esc}$. First assume $\Seff_{F}$ is dominated by a scattering term $\sim -\nu\,F/c^{2}$: with $\Gamma=1$ (Eq.~\ref{eqn:generic.transport}), the flux equation should converge to steady state ($D_{t}\rightarrow 0$) on a scattering time $\Delta t_{\rm F}^{\rm true} \sim \nu^{-1}$.
For Eq.~\ref{eqn:rsol.1}, $\Delta t_{\rm F}^{(\ref{eqn:rsol.1})} \sim \Gamma\,\nu^{-1} \sim \Gamma\,\Delta t_{\rm F}^{\rm true}$; for Eq.~\ref{eqn:rsol.2}, $\Delta t_{\rm F}^{(\ref{eqn:rsol.2})} \sim \Gamma^{2}\,\nu^{-1}\sim \Gamma^{2}\,\Delta t_{\rm F}^{\rm true}$.
Now assume in the number/energy equation $\Seff_{q} \sim \pm q/(c\,\tau)$, for some loss or production/injection processes. These processes reach equilibrium in $\Delta t_{\rm in/loss}^{\rm true} \sim \tau$ for Eq.~\ref{eqn:generic.transport}. For Eq.~\ref{eqn:rsol.1}, $\Delta t_{\rm in/loss}^{(\ref{eqn:rsol.1})} \sim \Gamma\,\Delta t_{\rm in/loss}^{\rm true}$, and for Eq.~\ref{eqn:rsol.2} $\Delta t_{\rm in/loss}^{(\ref{eqn:rsol.2})} \sim \Psi^{-1}\,\Delta t_{\rm in/loss}^{\rm true}$. The CR transport/escape time $\Delta t_{\rm esc}\sim L/v_{\rm eff}$ to some distance $L$ is given by the effective transport speed $v_{\rm eff}$ (writing $D_{t} q + \nabla(v_{\rm eff}\,q) = ...$): for Eq.~\ref{eqn:rsol.1}, $\Delta t_{\rm esc}^{(\ref{eqn:rsol.1})} \sim \Gamma\,L\,q/F$; for Eq.~\ref{eqn:rsol.2}, $\Delta t_{\rm esc}^{(\ref{eqn:rsol.2})} \sim L\,q/F$. But $F$ depends on whether the flux equation has reached steady state. First consider case {\bf (a)}, where $\Delta t \gg \Delta t_{\rm F}$ and $v_{\rm eff} \lesssim \tilde{c}$, so both Eq.~\ref{eqn:rsol.1} and Eq.~\ref{eqn:rsol.2} have $F\rightarrow F_{\rm true}$, and therefore $\Delta t_{\rm esc}^{(\ref{eqn:rsol.1})} \rightarrow \Gamma\,\Delta t_{\rm esc}^{\rm true}$, $\Delta t_{\rm esc}^{(\ref{eqn:rsol.2})} \rightarrow \Delta t_{\rm esc}^{\rm true}$. In case {\bf (b)}, $\Delta t \ll \Delta t_{\rm F}$, or equivalently the system is free-streaming/unconfined; thus the true $v_{\rm eff} \gg \tilde{c}$ and Eqs.~\ref{eqn:rsol.1}-\ref{eqn:rsol.2} have $v_{\rm eff} \rightarrow \tilde{c}$, giving $\Delta t_{\rm esc}^{(\ref{eqn:rsol.1})}\sim \Delta t_{\rm esc}^{(\ref{eqn:rsol.2})} \sim L/\tilde{c} \sim \Gamma\,\Delta t_{\rm esc}^{\rm true}$. 

The quantities of interest in CR models -- e.g. CR number densities of a given species at a given energy, primary-to-secondary or radioactive-to-stable ratios, etc. -- are set by the appropriate ratios of injection/loss/escape timescales  (for a given galactic background). Since injection and non-transport (e.g. collisional) losses scale together in $\Delta t_{\rm in/loss}$ in both Eq.~\ref{eqn:rsol.1} \&\ Eq.~\ref{eqn:rsol.2}, their ratio (and therefore scalings that depend on balancing injection and non-escape losses) is insensitive to $\tilde{c}$. For Eq.~\ref{eqn:rsol.1}, in all limits, the {\em ratio} $\Delta t_{\rm in/loss} / \Delta t_{\rm esc}$ is also equal to its ``true'' ($\tilde{c}=c$) value, as both scale identically with $\Gamma$. For Eq.~\ref{eqn:rsol.2}, however, this is only true if $\Psi \rightarrow 1$ in case {\bf (a)} and $\Psi \rightarrow \Gamma^{-1}$ (or more generically $\Psi \rightarrow F/F^{\rm true}$) in case {\bf (b)}. 

\subsection{(Dis)Advantages of Each Formulation}
\label{sec:rsol.advantages}

This leads us to the major (dis)advantages of each method. The formulation of Eq.~\ref{eqn:rsol.1} ``uniformly'' slows down CR transport: it is essentially equivalent to a uniform rescaling of time, as seen by the CRs, by a factor $\tilde{c}/c$. This has the advantage that although the time $\Delta t$ to reach equilibrium  in  $q$ and $F_{q}$ is increased, in both the free-streaming and confined limits (equivalent to the  optically thin and thick limits in the RHD literature where these were first derived), the system reaches the ``correct'' number/energy density and losses/production at the same distance $\Delta x$ from any source. Also, the flux equation converges more rapidly than Eq.~\ref{eqn:rsol.2} ($\Delta t_{\rm F}^{(\ref{eqn:rsol.1})} \ll \Delta t_{\rm F}^{(\ref{eqn:rsol.2})}$), although all terms in the number/energy equation (transport and production/loss) converge more slowly ($\Delta t_{\rm in/loss}^{(\ref{eqn:rsol.1})} \gg \Delta t_{\rm in/loss}^{(\ref{eqn:rsol.2})}$, $\Delta t_{\rm transport}^{(\ref{eqn:rsol.1})} \gg \Delta t_{\rm transport}^{(\ref{eqn:rsol.2})}$). The problem this can create is that the timescales $\Delta t_{\rm in/loss}^{(\ref{eqn:rsol.1})}$, $\Delta t_{\rm transport}^{(\ref{eqn:rsol.1})}$ can potentially become {\em so} long, for computationally tractable RSOL values $\tilde{c}$, that the system never actually reaches that $\Delta x$ or steady-state. This is most acute in the circum/inter-galactic medium (CGM/IGM) around galaxies, where many have argued CRs may be most important \citep{Boot13,wiener:cr.supersonic.streaming.deriv,Buts18,2020arXiv200804915B,hopkins:2020.cr.outflows.to.mpc.scales,ji:fire.cr.cgm,ji:20.virial.shocks.suppressed.cr.dominated.halos}. Consider that even for rapid diffusion (diffusivity $\kappa \sim \kappa_{30}\,10^{30}\,{\rm cm^{2}\,s^{-1}}$), at $L\sim L_{30}\,30\,$kpc from a galaxy $\Delta t_{\rm esc}^{(\ref{eqn:rsol.1})} \sim \Gamma\,L^{2}/\kappa \sim 100\,{\rm Gyr}\,(\tilde{c}/1000\,{\rm km\,s^{-1}})^{-1}\,L_{30}^{2}\,\kappa_{30}^{-1}$. In other words, we require $\tilde{c} \gg 10^{4}\,{\rm km\,s^{-1}}$ for the CRs to ``reach'' the CGM in less than a Hubble time in the formulation of Eq.~\ref{eqn:rsol.1}. Similarly, we need very large $\tilde{c}$ to ensure $\Delta t_{\rm in/loss}^{(\ref{eqn:rsol.1})}$ is not much longer than galaxy dynamical times (which would risk converging to the wrong equilibrium).

The formulation of Eq.~\ref{eqn:rsol.2} avoids this, by converging in the number/energy (loss and transport) equations much more rapidly (on the ``correct'' timescale, independent of $\tilde{c}$, on large scales). It converges in the flux equation more slowly, but this is still rapid in absolute terms, as e.g.\ $\Delta t_{\rm F}^{(\ref{eqn:rsol.2})} \sim 3\,{\rm Myr}\,\kappa_{30}\,(\tilde{c}/1000\,{\rm cm^{2}\,s^{-1}})^{-2}$. The problem with Eq.~\ref{eqn:rsol.2} is that we can find ourselves in case {\bf (b)}, and potentially in the sub-case where $\Delta t_{\rm F}^{(\ref{eqn:rsol.2})}$ is larger than one of $\Delta t_{\rm in/loss}^{\rm true}$ or $\Delta t_{\rm esc}^{\rm true}$ -- the limit where capturing the correct behavior with $\tilde{c} \ll c$ {\em requires} including the $\Psi$ term with $\Psi \rightarrow F/F^{\rm true}$. Motivated by the above and treatments of the flux-limiter in flux-limited RHD with an RSOL, we therefore suggest the interpolation function $\Psi = {\rm MIN}[1,\,|F_{q}|/F_{\rm true}]$, where $F_{\rm true}={\rm MIN}[e^{\prime}\,\beta\,c,\,|\bar{v}_{A}\,(e^{\prime}+P_{0}^{\prime}) + \kappa_{\|}\nabla_{\|}e^{\prime} |]$ for $q=e^{\prime}$ (or $F_{\rm true}={\rm MIN}[n^{\prime}\,\beta\,c,\,|\bar{v}_{A}\,n + \kappa_{\|}\,\nabla_{\|} n^{\prime}|]$ for $q=n^{\prime}$, etc.) is given by the value the flux would have in local steady-state ($D_{t} F_{q} \rightarrow 0$) for $\tilde{c}=c$ at the given energy. This ensures the correct behavior in both asymptotic limits discussed in \S~\ref{sec:rsol.limits}.

With this definition, one can verify that {\em both} formulations in Eq.~\ref{eqn:rsol.1} \&\ \ref{eqn:rsol.2} converge to identical solutions as $\tilde{c}$ increases. One would expect from the above that in the dense ISM, the formulation of Eq.~\ref{eqn:rsol.1} converges somewhat faster with respect to $\tilde{c}/c$ (i.e.\ one can obtain converged solutions with lower $\tilde{c}$, hence lower computational expense). But for the reasons above, in the CGM, the formulation of Eq.~\ref{eqn:rsol.2} converges at much lower values of $\tilde{c}$. Eq.~\ref{eqn:rsol.2} therefore has advantages for applications in, e.g.\ cosmological galaxy formation simulations, while the formulation in Eq.~\ref{eqn:rsol.1} potentially advantageous for transport around sources or in the ISM within galaxies.

\subsection{Which Speed of Light Enters the Closure Relation?}
\label{sec:rsol.closure}

Recall that for the closure relation Eq.~\ref{eqn:closure} that we proposed to estimate $\mutwo$, we used $\muone = \bar{f}_{1}/\bar{f}_{0} =  F_{q} / (\beta\,q\,c)$. For the formulation in Eq.~\ref{eqn:rsol.1},  the ``actual'' flux of $q$ is $(\tilde{c}/c)\,F_{q}$, so $F_{q}$ retains its usual meaning -- free streaming will still have $F_{q} = \beta\,q\,c$, so we can use this relation in unmodified form, $\muone = \bar{f}_{1}/\bar{f}_{0} = F_{q} / (\beta\,q\,c)$ (provided we follow all the definitions above). For the formulation in Eq.~\ref{eqn:rsol.2}, we need to be more careful: $F_{q}$ saturates at $\sim q\,\tilde{c}$, but this can occur even if the system approaches a near-isotropic DF, for sufficiently-large diffusivity. So in the closure relation, we require a function similar to the $\Psi$ term above; for example, taking $F_{q}/(\beta\,q\,c) \rightarrow F_{q} / {\rm MAX}[ \beta\,q\,\tilde{c},\, |F_{q}(\tilde{c} \rightarrow \infty)|]$.

\subsection{Rigidity-Dependent RSOL}
\label{sec:rsol.vs.energy}

Finally, we note that although the arguments above assume $\tilde{c}$ is constant in space and time, they {\em do not} require $\tilde{c}$ be the same for different CR species or energies. In calculations that evolve a set of CR species of energies binned in rigidity, for example, one can adopt a $\tilde{c}$ that increases for the highest-rigidity CRs (for example, as $\tilde{c} = \tilde{c}_{0}$ for $R<1$\,GV, and $\tilde{c}_{0}\,(R/{\rm GV})$ at larger values). Larger-rigidity CRs have larger $\kappa$ (e.g.\ larger $\Delta t_{\rm F}$) so require larger $\tilde{c}$ to converge. By sub-cycling the CR equations for the highest-rigidity values, faster convergence may be possible.

\subsection{Appearance in the Gas+Radiation (Momentum+Energy) Equations \&\ Conservation}
\label{sec:rsol.gas}

Just like with RHD \citep[see e.g.][]{2013ApJS..206...21S}, it is important that the RSOL appear {\em only} in the dynamical equations for the CRs, {\em not} in the terms that couple to the gas that are written in terms of physical quantities. Otherwise certain terms, like the parallel forces or CR thermal heating rates, would not, in fact, converge to equilibrium when $D_{t} \rightarrow 0$ and would be severely incorrect. Thus, for example, the form of the gas momentum Eq.~\ref{eqn:gas.momentum} as written remains identical. Likewise the gas heating terms have their ``normal'' values with respect to $e$, etc. One consequence of this, again identical to RHD, is that the formally conserved quantities with an RSOL are not total energy ($E_{\rm other}+E_{\rm cr}$) and momentum (${\bf P}_{\rm other} + c^{-2}\,{\bf F}_{\rm cr}$). Instead, for the formulation in Eq.~\ref{eqn:rsol.1}, they are ($E_{\rm other}+(c/\tilde{c})\,E_{\rm cr}$) and (${\bf P}_{\rm other} + (c\,\tilde{c})^{-1}\,{\bf F}_{\rm cr}$), while  for the formulation in Eq.~\ref{eqn:rsol.2}, they are ($E_{\rm other}+E_{\rm cr}$) and (${\bf P}_{\rm other} + \tilde{c}^{-2}\,{\bf F}_{\rm cr}$). This is important to note but introduces no conceptual difficulty, provided the definitions above are used.

\section{Summary}
\label{sec:conclusions}

Beginning from the focused  CR transport equation allowing for an arbitrary pitch-angle distribution, we have derived and tested a consistent set of moments equations for CR-MHD applications, analogous to widely used closures for RHD. We present equations for either e.g.\ the first two pitch-angle moments of the DF $f$ ($\langle f \rangle_{\mu}$, $\langle \mu\,f \rangle_{\mu}$), or corresponding integrated pairs like CR number density and its flux $(n,\,F_{n})$, total CR energy and flux $(e,\,F_{e})$, or CR kinetic energy and its flux $(\epsilon,\,F_{\epsilon})$. We present two different schemes to integrate these explicitly in simulations with a RSOL approximation, discuss their relative convergence properties and merits, and note some important terms missing from previous CR-RSOL implementations. The derived equations are summarized in Appendix~\ref{sec:equations}.

Our equations are valid for all relevant CR $\beta=v/c$ (not just the ultra-relativistic limit), and do not impose any assumption about the slope or form of $f(p)$. Unlike the Fokker-Planck or pure diffusion+streaming (zeroth-moment) formulations of the CR transport equations, the expressions here can handle both free-streaming/weak-coupling (arbitrarily large mean-free-path) and strong-scattering (static or dynamic diffusion or advective) limits, for both near-isotropic and arbitrarily anisotropic DFs, anisotropic forward/backward scattering, and anisotropic magnetic fields/global transport. The expressions are accurate to leading order in $\mathcal{O}(u/c)$ in all limits. The key assumptions are: (1) that the background fluid is non-relativistic, $|{\bf u}| \ll c$; and (2) the CRs have a gyrotropic DF, with gyro radii much smaller than resolved scales. 

It is easy to imagine extending this even further to include more complicated ``variable Eddington tensor'' formulations akin to RHD (representing arbitrary CR DFs), although the gyrotropic nature of CRs removes some of the ambiguities associated with RHD formulations. 
In this spirit we also present the relevant gyro-averaged equations for direct finite-volume phase-space integration of the pitch-angle distribution (following $f({\bf x},\,p,\,\mu,\,t,\,...)$ explicitly on a grid of ${\bf x},\,\mu,\,p$), 
as  there may be cases where the different formulations are beneficial.

Finally, it is worth commenting on a major practical difference  between RHD and CR-MHD applications: in many astrophysical RHD applications, the collisional/scattering terms (absorption and scattering coefficients) are reasonably well understood, and much of the debate in the literature has centered on methods to accurately handle the actual radiation transport. In contrast, in CR-MHD, the scattering terms -- and, as a consequence, the diffusion/streaming coefficients -- are enormously uncertain. This is true even of their qualitative form and dimensional scalings. Different state-of-the-art models for CR scattering rates $\nu$ differ by several orders of magnitude and often predict opposite dependence on properties like magnetic field or turbulence strength \citep[see the review in][]{hopkins:cr.transport.constraints.from.galaxies}. Real progress in predictions will require a better understanding of the form of the CR scattering rates, their dependence on pitch angle and local plasma/ISM properties, and developing new diagnostics to compare models  to observations. Nonetheless, the hope is that the calculations in this paper can aid in reducing some of the  better-understood uncertainties in CR transport. And we argue in \S~\ref{sec:importance} that there are many physically important situations, especially those which involve rapidly-varying CR scattering rates and/or CR ``bottlenecks,'' where the more accurate form of the equations herein may predict significantly different behaviors compared to more simplified and less-accurate expressions. Further, in numerical applications where an RSOL is adopted, it is crucial to adopt treatments that can correctly interpolate between different limits. Finally, the basic principles of the closure structure proposed here can be used to include additional information about scattering coefficients in the CR-moment framework. For example, if one wished to model a scattering rate $\bar{\nu} = \bar{\nu}[f(\mu)] \approx \bar{\nu}(\muone,\,\mutwo,\,...)$ that is a function of the CR pitch-angle distribution, the structure herein provides a well-defined way to retain and estimate some (though certainly not all) of this physics without having to evolve the entire pitch angle distribution function at each momentum and position.

\acknowledgments{Support for PFH was provided by NSF Research Grants 1911233 \&\ 20009234, NSF CAREER grant 1455342, NASA grants 80NSSC18K0562, HST-AR-15800.001-A. Numerical calculations were run on the Caltech compute cluster ``Wheeler,'' allocations FTA-Hopkins supported by the NSF and TACC, and NASA HEC SMD-16-7592. Support for JS  was provided by Rutherford Discovery Fellowship RDF-U001804 and Marsden Fund grant UOO1727, which are managed through the Royal Society Te Ap\=arangi. }

\datastatement{The data supporting this article are available on reasonable request to the corresponding author.} 

\bibliography{ms_extracted}

\begin{thebibliography}{}
\makeatletter
\relax
\def\mn@urlcharsother{\let\do\@makeother \do\$\do\&\do\#\do\^\do\_\do\%\do\~}
\def\mn@doi{\begingroup\mn@urlcharsother \@ifnextchar [ {\mn@doi@}
  {\mn@doi@[]}}
\def\mn@doi@[#1]#2{\def\@tempa{#1}\ifx\@tempa\@empty \href
  {http://dx.doi.org/#2} {doi:#2}\else \href {http://dx.doi.org/#2} {#1}\fi
  \endgroup}
\def\mn@eprint#1#2{\mn@eprint@#1:#2::\@nil}
\def\mn@eprint@arXiv#1{\href {http://arxiv.org/abs/#1} {{\tt arXiv:#1}}}
\def\mn@eprint@dblp#1{\href {http://dblp.uni-trier.de/rec/bibtex/#1.xml}
  {dblp:#1}}
\def\mn@eprint@#1:#2:#3:#4\@nil{\def\@tempa {#1}\def\@tempb {#2}\def\@tempc
  {#3}\ifx \@tempc \@empty \let \@tempc \@tempb \let \@tempb \@tempa \fi \ifx
  \@tempb \@empty \def\@tempb {arXiv}\fi \@ifundefined
  {mn@eprint@\@tempb}{\@tempb:\@tempc}{\expandafter \expandafter \csname
  mn@eprint@\@tempb\endcsname \expandafter{\@tempc}}}

\bibitem[\protect\citeauthoryear{{Amato} \& {Blasi}}{{Amato} \&
  {Blasi}}{2018}]{2018AdSpR..62.2731A}
{Amato} E.,  {Blasi} P.,  2018, \mn@doi [Advances in Space Research]
  {10.1016/j.asr.2017.04.019}, \href
  {http://adsabs.harvard.edu/abs/2018AdSpR..62.2731A} {62, 2731}

\bibitem[\protect\citeauthoryear{{Bai}, {Caprioli}, {Sironi}  \&
  {Spitkovsky}}{{Bai} et~al.}{2015}]{bai:2015.mhd.pic}
{Bai} X.-N.,  {Caprioli} D.,  {Sironi} L.,   {Spitkovsky} A.,  2015, \mn@doi
  [\apj] {10.1088/0004-637X/809/1/55}, \href
  {https://ui.adsabs.harvard.edu/abs/2015ApJ...809...55B} {809, 55}

\bibitem[\protect\citeauthoryear{{Bai}, {Ostriker}, {Plotnikov}  \&
  {Stone}}{{Bai} et~al.}{2019}]{bai:2019.cr.pic.streaming}
{Bai} X.-N.,  {Ostriker} E.~C.,  {Plotnikov} I.,   {Stone} J.~M.,  2019,
  \mn@doi [\apj] {10.3847/1538-4357/ab1648}, \href
  {https://ui.adsabs.harvard.edu/abs/2019ApJ...876...60B} {876, 60}

\bibitem[\protect\citeauthoryear{{Booth}, {Agertz}, {Kravtsov}  \&
  {Gnedin}}{{Booth} et~al.}{2013}]{Boot13}
{Booth} C.~M.,  {Agertz} O.,  {Kravtsov} A.~V.,   {Gnedin} N.~Y.,  2013,
  \mn@doi [\apjl] {10.1088/2041-8205/777/1/L16}, \href
  {http://adsabs.harvard.edu/abs/2013ApJ...777L..16B} {777, L16}

\bibitem[\protect\citeauthoryear{{Buck}, {Pfrommer}, {Pakmor}, {Grand}  \&
  {Springel}}{{Buck} et~al.}{2020}]{buck:2020.cosmic.ray.low.coeff.high.Egamma}
{Buck} T.,  {Pfrommer} C.,  {Pakmor} R.,  {Grand} R. J.~J.,   {Springel} V.,
  2020, \mn@doi [\mnras] {10.1093/mnras/staa1960}, \href
  {https://ui.adsabs.harvard.edu/abs/2020MNRAS.497.1712B} {497, 1712}

\bibitem[\protect\citeauthoryear{{Bustard} \& {Zweibel}}{{Bustard} \&
  {Zweibel}}{2020}]{bustard:2020.crs.multiphase.ism.accel.confinement}
{Bustard} C.,  {Zweibel} E.~G.,  2020, \mnras, in press, arXiv:2012.06585,
  \href {https://ui.adsabs.harvard.edu/abs/2020arXiv201206585B} {p.
  arXiv:2012.06585}

\bibitem[\protect\citeauthoryear{{Butsky} \& {Quinn}}{{Butsky} \&
  {Quinn}}{2018}]{Buts18}
{Butsky} I.~S.,  {Quinn} T.~R.,  2018, \mn@doi [\apj]
  {10.3847/1538-4357/aaeac2}, \href
  {https://ui.adsabs.harvard.edu/abs/2018ApJ...868..108B} {868, 108}

\bibitem[\protect\citeauthoryear{{Butsky}, {Fielding}, {Hayward}, {Hummels},
  {Quinn}  \& {Werk}}{{Butsky} et~al.}{2020}]{2020arXiv200804915B}
{Butsky} I.~S.,  {Fielding} D.~B.,  {Hayward} C.~C.,  {Hummels} C.~B.,  {Quinn}
  T.~R.,   {Werk} J.~K.,  2020, arXiv e-prints, \href
  {https://ui.adsabs.harvard.edu/abs/2020arXiv200804915B} {p. arXiv:2008.04915}

\bibitem[\protect\citeauthoryear{{Chan}, {Kere{\v{s}}}, {Hopkins}, {Quataert},
  {Su}, {Hayward}  \& {Faucher-Gigu{\`e}re}}{{Chan}
  et~al.}{2019}]{chan:2018.cosmicray.fire.gammaray}
{Chan} T.~K.,  {Kere{\v{s}}} D.,  {Hopkins} P.~F.,  {Quataert} E.,  {Su} K.~Y.,
   {Hayward} C.~C.,   {Faucher-Gigu{\`e}re} C.~A.,  2019, \mn@doi [\mnras]
  {10.1093/mnras/stz1895}, \href
  {https://ui.adsabs.harvard.edu/abs/2019MNRAS.488.3716C} {488, 3716}

\bibitem[\protect\citeauthoryear{{Chandran}}{{Chandran}}{2000}]{chandran00}
{Chandran} B. D.~G.,  2000, \mn@doi [\prl] {10.1103/PhysRevLett.85.4656}, \href
  {https://ui.adsabs.harvard.edu/abs/2000PhRvL..85.4656C} {85, 4656}

\bibitem[\protect\citeauthoryear{{Cummings} et~al.,}{{Cummings}
  et~al.}{2016}]{2016ApJ...831...18C}
{Cummings} A.~C.,  et~al., 2016, \mn@doi [\apj] {10.3847/0004-637X/831/1/18},
  \href {http://adsabs.harvard.edu/abs/2016ApJ...831...18C} {831, 18}

\bibitem[\protect\citeauthoryear{{Evoli}, {Gaggero}, {Vittino}, {Di Bernardo},
  {Di Mauro}, {Ligorini}, {Ullio}  \& {Grasso}}{{Evoli}
  et~al.}{2017}]{evoli:dragon2.cr.prop}
{Evoli} C.,  {Gaggero} D.,  {Vittino} A.,  {Di Bernardo} G.,  {Di Mauro} M.,
  {Ligorini} A.,  {Ullio} P.,   {Grasso} D.,  2017, \mn@doi [Journal of
  Cosmology and Astroparticle Physics] {10.1088/1475-7516/2017/02/015}, \href
  {http://adsabs.harvard.edu/abs/2017JCAP...02..015E} {2, 015}

\bibitem[\protect\citeauthoryear{{Girichidis}, {Naab}, {Hanasz}  \&
  {Walch}}{{Girichidis} et~al.}{2018}]{Giri18}
{Girichidis} P.,  {Naab} T.,  {Hanasz} M.,   {Walch} S.,  2018, \mn@doi
  [\mnras] {10.1093/mnras/sty1653}, \href
  {https://ui.adsabs.harvard.edu/abs/2018MNRAS.479.3042G} {479, 3042}

\bibitem[\protect\citeauthoryear{{Girichidis}, {Pfrommer}, {Hanasz}  \&
  {Naab}}{{Girichidis} et~al.}{2020}]{girichidis:cr.spectral.scheme}
{Girichidis} P.,  {Pfrommer} C.,  {Hanasz} M.,   {Naab} T.,  2020, \mn@doi
  [\mnras] {10.1093/mnras/stz2961}, \href
  {https://ui.adsabs.harvard.edu/abs/2020MNRAS.491..993G} {491, 993}

\bibitem[\protect\citeauthoryear{{Guo}, {Tian}  \& {Jin}}{{Guo}
  et~al.}{2016}]{2016ApJ...819...54G}
{Guo} Y.-Q.,  {Tian} Z.,   {Jin} C.,  2016, \mn@doi [\apj]
  {10.3847/0004-637X/819/1/54}, \href
  {http://adsabs.harvard.edu/abs/2016ApJ...819...54G} {819, 54}

\bibitem[\protect\citeauthoryear{{Hin Navin Tsung}, {Oh}  \& {Jiang}}{{Hin
  Navin Tsung} et~al.}{2021}]{tsung:2021.cr.outflows.staircase}
{Hin Navin Tsung} T.,  {Oh} S.~P.,   {Jiang} Y.-F.,  2021, arXiv e-prints,
  \href {https://ui.adsabs.harvard.edu/abs/2021arXiv210707543H} {p.
  arXiv:2107.07543}

\bibitem[\protect\citeauthoryear{{Holcomb} \& {Spitkovsky}}{{Holcomb} \&
  {Spitkovsky}}{2019}]{holcolmb.spitkovsky:saturation.gri.sims}
{Holcomb} C.,  {Spitkovsky} A.,  2019, \mn@doi [\apj]
  {10.3847/1538-4357/ab328a}, \href
  {https://ui.adsabs.harvard.edu/abs/2019ApJ...882....3H} {882, 3}

\bibitem[\protect\citeauthoryear{{Hopkins}}{{Hopkins}}{2017}]{hopkins:gizmo.diffusion}
{Hopkins} P.~F.,  2017, \mn@doi [\mnras] {10.1093/mnras/stw3306}, \href
  {http://adsabs.harvard.edu/abs/2017MNRAS.466.3387H} {466, 3387}

\bibitem[\protect\citeauthoryear{{Hopkins}, {Chan}, {Ji}, {Hummels}, {Keres},
  {Quataert}  \& {Faucher-Giguere}}{{Hopkins}
  et~al.}{2020a}]{hopkins:2020.cr.outflows.to.mpc.scales}
{Hopkins} P.~F.,  {Chan} T.~K.,  {Ji} S.,  {Hummels} C.,  {Keres} D.,
  {Quataert} E.,   {Faucher-Giguere} C.-A.,  2020a, \mnras, in press,
  arXiv:2002.02462, \href
  {https://ui.adsabs.harvard.edu/abs/2020arXiv200202462H} {p. arXiv:2002.02462}

\bibitem[\protect\citeauthoryear{{Hopkins}, {Squire}, {Chan}, {Quataert}, {Ji},
  {Keres}  \& {Faucher-Giguere}}{{Hopkins}
  et~al.}{2020b}]{hopkins:cr.transport.constraints.from.galaxies}
{Hopkins} P.~F.,  {Squire} J.,  {Chan} T.~K.,  {Quataert} E.,  {Ji} S.,
  {Keres} D.,   {Faucher-Giguere} C.-A.,  2020b, \mnras, in press,
  arXiv:2002.06211, \href
  {https://ui.adsabs.harvard.edu/abs/2020arXiv200206211H} {p. arXiv:2002.06211}

\bibitem[\protect\citeauthoryear{{Hopkins}, {Chan}, {Squire}, {Quataert}, {Ji},
  {Keres}  \& {Faucher-Giguere}}{{Hopkins}
  et~al.}{2020c}]{hopkins:2020.cr.transport.model.fx.galform}
{Hopkins} P.~F.,  {Chan} T.~K.,  {Squire} J.,  {Quataert} E.,  {Ji} S.,
  {Keres} D.,   {Faucher-Giguere} C.-A.,  2020c, \mnras, in press,
  arXiv:2004.02897, \href
  {https://ui.adsabs.harvard.edu/abs/2020arXiv200402897H} {p. arXiv:2004.02897}

\bibitem[\protect\citeauthoryear{{Hopkins} et~al.,}{{Hopkins}
  et~al.}{2020d}]{hopkins:cr.mhd.fire2}
{Hopkins} P.~F.,  et~al., 2020d, \mn@doi [\mnras] {10.1093/mnras/stz3321},
  \href {https://ui.adsabs.harvard.edu/abs/2020MNRAS.492.3465H} {492, 3465}

\bibitem[\protect\citeauthoryear{{Huang} \& {Davis}}{{Huang} \&
  {Davis}}{2021}]{huang.davis:2021.cr.staircase.in.outflows}
{Huang} X.,  {Davis} S.~W.,  2021, arXiv e-prints, \href
  {https://ui.adsabs.harvard.edu/abs/2021arXiv210511506H} {p. arXiv:2105.11506}

\bibitem[\protect\citeauthoryear{{Isenberg}}{{Isenberg}}{1997}]{1997JGR...102.4719I}
{Isenberg} P.~A.,  1997, \mn@doi [\jgr] {10.1029/96JA03671}, \href
  {https://ui.adsabs.harvard.edu/abs/1997JGR...102.4719I} {102, 4719}

\bibitem[\protect\citeauthoryear{{Janka}}{{Janka}}{1992}]{1992A&A...256..452J}
{Janka} H.~T.,  1992, \aap, \href
  {https://ui.adsabs.harvard.edu/abs/1992A&A...256..452J} {256, 452}

\bibitem[\protect\citeauthoryear{{Ji}, {Kere{\v{s}}}, {Chan}, {Stern},
  {Hummels}, {Hopkins}, {Quataert}  \& {Faucher-Gigu{\`e}re}}{{Ji}
  et~al.}{2020a}]{ji:20.virial.shocks.suppressed.cr.dominated.halos}
{Ji} S.,  {Kere{\v{s}}} D.,  {Chan} T.~K.,  {Stern} J.,  {Hummels} C.~B.,
  {Hopkins} P.~F.,  {Quataert} E.,   {Faucher-Gigu{\`e}re} C.-A.,  2020a,
  \mnras, submitted, arXiv:2011.04606, \href
  {https://ui.adsabs.harvard.edu/abs/2020arXiv201104706J} {p. arXiv:2011.04706}

\bibitem[\protect\citeauthoryear{{Ji} et~al.,}{{Ji}
  et~al.}{2020b}]{ji:fire.cr.cgm}
{Ji} S.,  et~al., 2020b, \mn@doi [\mnras] {10.1093/mnras/staa1849}, \href
  {https://ui.adsabs.harvard.edu/abs/2020MNRAS.496.4221J} {496, 4221}

\bibitem[\protect\citeauthoryear{{Jiang} \& {Oh}}{{Jiang} \&
  {Oh}}{2018}]{jiang.oh:2018.cr.transport.m1.scheme}
{Jiang} Y.-F.,  {Oh} S.~P.,  2018, \mn@doi [\apj] {10.3847/1538-4357/aaa6ce},
  \href {http://adsabs.harvard.edu/abs/2018ApJ...854....5J} {854, 5}

\bibitem[\protect\citeauthoryear{{Jiang}, {Stone}  \& {Davis}}{{Jiang}
  et~al.}{2014}]{jiang:2014.rhd.solver.local}
{Jiang} Y.-F.,  {Stone} J.~M.,   {Davis} S.~W.,  2014, \mn@doi [\apjs]
  {10.1088/0067-0049/213/1/7}, \href
  {http://adsabs.harvard.edu/abs/2014ApJS..213....7J} {213, 7}

\bibitem[\protect\citeauthoryear{{J{\'o}hannesson} et~al.,}{{J{\'o}hannesson}
  et~al.}{2016}]{2016ApJ...824...16J}
{J{\'o}hannesson} G.,  et~al., 2016, \mn@doi [\apj]
  {10.3847/0004-637X/824/1/16}, \href
  {http://adsabs.harvard.edu/abs/2016ApJ...824...16J} {824, 16}

\bibitem[\protect\citeauthoryear{{Korsmeier} \& {Cuoco}}{{Korsmeier} \&
  {Cuoco}}{2016}]{2016PhRvD..94l3019K}
{Korsmeier} M.,  {Cuoco} A.,  2016, \mn@doi [\prd]
  {10.1103/PhysRevD.94.123019}, \href
  {http://adsabs.harvard.edu/abs/2016PhRvD..94l3019K} {94, 123019}

\bibitem[\protect\citeauthoryear{Kulsrud}{Kulsrud}{1983}]{Kulsrud1983}
Kulsrud R.~M.,  1983, in Sagdeev R.~N.,  Rosenbluth M.~N.,  eds, , {Handbook of
  Plasma Physics}.
Princeton University

\bibitem[\protect\citeauthoryear{{Lazarian}}{{Lazarian}}{2016}]{lazarian:2016.cr.wave.damping}
{Lazarian} A.,  2016, \mn@doi [\apj] {10.3847/1538-4357/833/2/131}, \href
  {https://ui.adsabs.harvard.edu/\#abs/2016ApJ...833..131L} {833, 131}

\bibitem[\protect\citeauthoryear{{Levermore}}{{Levermore}}{1984}]{levermore:1984.FLD.M1}
{Levermore} C.~D.,  1984, \mn@doi [Journal of Quantitative Spectroscopy and
  Radiative Transfer] {10.1016/0022-4073(84)90112-2}, \href
  {http://adsabs.harvard.edu/abs/1984JQSRT..31..149L} {31, 149}

\bibitem[\protect\citeauthoryear{{Mao} \& {Ostriker}}{{Mao} \&
  {Ostriker}}{2018}]{Mao18}
{Mao} S.~A.,  {Ostriker} E.~C.,  2018, \mn@doi [\apj]
  {10.3847/1538-4357/aaa88e}, \href
  {http://adsabs.harvard.edu/abs/2018ApJ...854...89M} {854, 89}

\bibitem[\protect\citeauthoryear{{Mihalas} \& {Mihalas}}{{Mihalas} \&
  {Mihalas}}{1984}]{mihalas:1984oup..book.....M}
{Mihalas} D.,  {Mihalas} B.~W.,  1984, {Foundations of radiation
  hydrodynamics}.
New York, Oxford University Press, 731 p.

\bibitem[\protect\citeauthoryear{{Minerbo}}{{Minerbo}}{1978}]{minerbo:1978.eddington.factors}
{Minerbo} G.~N.,  1978, \mn@doi [\jqsrt] {10.1016/0022-4073(78)90024-9}, \href
  {https://ui.adsabs.harvard.edu/abs/1978JQSRT..20..541M} {20, 541}

\bibitem[\protect\citeauthoryear{{Murchikova}, {Abdikamalov}  \&
  {Urbatsch}}{{Murchikova}
  et~al.}{2017}]{murchikova:m1.neutrino.transport.closures}
{Murchikova} E.~M.,  {Abdikamalov} E.,   {Urbatsch} T.,  2017, \mn@doi [\mnras]
  {10.1093/mnras/stx986}, \href
  {https://ui.adsabs.harvard.edu/abs/2017MNRAS.469.1725M} {469, 1725}

\bibitem[\protect\citeauthoryear{{Pakmor}, {Pfrommer}, {Simpson}  \&
  {Springel}}{{Pakmor} et~al.}{2016}]{Pakm16}
{Pakmor} R.,  {Pfrommer} C.,  {Simpson} C.~M.,   {Springel} V.,  2016, \mn@doi
  [\apjl] {10.3847/2041-8205/824/2/L30}, \href
  {http://adsabs.harvard.edu/abs/2016ApJ...824L..30P} {824, L30}

\bibitem[\protect\citeauthoryear{{Quataert}, {Thompson}  \& {Jiang}}{{Quataert}
  et~al.}{2021}]{quataert:2021.cr.outflows.diffusion.staircase}
{Quataert} E.,  {Thompson} T.~A.,   {Jiang} Y.-F.,  2021, arXiv e-prints, \href
  {https://ui.adsabs.harvard.edu/abs/2021arXiv210205696Q} {p. arXiv:2102.05696}

\bibitem[\protect\citeauthoryear{{Ruszkowski}, {Yang}  \&
  {Zweibel}}{{Ruszkowski} et~al.}{2017}]{Rusz17}
{Ruszkowski} M.,  {Yang} H.-Y.~K.,   {Zweibel} E.,  2017, \mn@doi [\apj]
  {10.3847/1538-4357/834/2/208}, \href
  {http://adsabs.harvard.edu/abs/2017ApJ...834..208R} {834, 208}

\bibitem[\protect\citeauthoryear{{Salem} \& {Bryan}}{{Salem} \&
  {Bryan}}{2014}]{Sale14}
{Salem} M.,  {Bryan} G.~L.,  2014, \mn@doi [\mnras] {10.1093/mnras/stt2121},
  \href {http://adsabs.harvard.edu/abs/2014MNRAS.437.3312S} {437, 3312}

\bibitem[\protect\citeauthoryear{{Salem}, {Bryan}  \& {Corlies}}{{Salem}
  et~al.}{2016}]{2016MNRAS.456..582S}
{Salem} M.,  {Bryan} G.~L.,   {Corlies} L.,  2016, \mn@doi [\mnras]
  {10.1093/mnras/stv2641}, \href
  {https://ui.adsabs.harvard.edu/abs/2016MNRAS.456..582S} {456, 582}

\bibitem[\protect\citeauthoryear{{Schlickeiser}}{{Schlickeiser}}{1989}]{schlickeiser:89.cr.transport.scattering.eqns}
{Schlickeiser} R.,  1989, \mn@doi [\apj] {10.1086/167009}, \href
  {https://ui.adsabs.harvard.edu/abs/1989ApJ...336..243S} {336, 243}

\bibitem[\protect\citeauthoryear{Sharma, Colella  \& Martin}{Sharma
  et~al.}{2010}]{sharma.2010:cosmic.ray.streaming.timestepping}
Sharma P.,  Colella P.,   Martin D.~F.,  2010, \mn@doi [SIAM J. Sci. Comput.]
  {10.1137/100792135}, 32, 3564

\bibitem[\protect\citeauthoryear{{Simpson}, {Pakmor}, {Marinacci}, {Pfrommer},
  {Springel}, {Glover}, {Clark}  \& {Smith}}{{Simpson} et~al.}{2016}]{Simp16}
{Simpson} C.~M.,  {Pakmor} R.,  {Marinacci} F.,  {Pfrommer} C.,  {Springel} V.,
   {Glover} S.~C.~O.,  {Clark} P.~C.,   {Smith} R.~J.,  2016, \mn@doi [\apjl]
  {10.3847/2041-8205/827/2/L29}, \href
  {http://adsabs.harvard.edu/abs/2016ApJ...827L..29S} {827, L29}

\bibitem[\protect\citeauthoryear{{Skilling}}{{Skilling}}{1971}]{skilling:1971.cr.diffusion}
{Skilling} J.,  1971, \mn@doi [\apj] {10.1086/151210}, \href
  {http://adsabs.harvard.edu/abs/1971ApJ...170..265S} {170, 265}

\bibitem[\protect\citeauthoryear{{Skilling}}{{Skilling}}{1975}]{1975MNRAS.172..557S}
{Skilling} J.,  1975, \mn@doi [\mnras] {10.1093/mnras/172.3.557}, \href
  {http://adsabs.harvard.edu/abs/1975MNRAS.172..557S} {172, 557}

\bibitem[\protect\citeauthoryear{{Skinner} \& {Ostriker}}{{Skinner} \&
  {Ostriker}}{2013}]{2013ApJS..206...21S}
{Skinner} M.~A.,  {Ostriker} E.~C.,  2013, \mn@doi [\apjs]
  {10.1088/0067-0049/206/2/21}, \href
  {https://ui.adsabs.harvard.edu/abs/2013ApJS..206...21S} {206, 21}

\bibitem[\protect\citeauthoryear{{Strong} \& {Moskalenko}}{{Strong} \&
  {Moskalenko}}{2001}]{strong:2001.galprop}
{Strong} A.~W.,  {Moskalenko} I.~V.,  2001, \mn@doi [Advances in Space
  Research] {10.1016/S0273-1177(01)00112-0}, \href
  {https://ui.adsabs.harvard.edu/abs/2001AdSpR..27..717S} {27, 717}

\bibitem[\protect\citeauthoryear{{Su} et~al.,}{{Su}
  et~al.}{2019}]{su:2018.stellar.fb.fails.to.solve.cooling.flow}
{Su} K.-Y.,  et~al., 2019, \mn@doi [\mnras] {10.1093/mnras/stz1494}, \href
  {https://ui.adsabs.harvard.edu/abs/2019MNRAS.487.4393S} {487, 4393}

\bibitem[\protect\citeauthoryear{{Su} et~al.,}{{Su}
  et~al.}{2020}]{su:turb.crs.quench}
{Su} K.-Y.,  et~al., 2020, \mn@doi [\mnras] {10.1093/mnras/stz3011}, \href
  {https://ui.adsabs.harvard.edu/abs/2020MNRAS.491.1190S} {491, 1190}

\bibitem[\protect\citeauthoryear{{Thomas} \& {Pfrommer}}{{Thomas} \&
  {Pfrommer}}{2019}]{thomas.pfrommer.18:alfven.reg.cr.transport}
{Thomas} T.,  {Pfrommer} C.,  2019, \mn@doi [\mnras] {10.1093/mnras/stz263},
  \href {https://ui.adsabs.harvard.edu/abs/2019MNRAS.485.2977T} {485, 2977}

\bibitem[\protect\citeauthoryear{{Uhlig}, {Pfrommer}, {Sharma}, {Nath},
  {En{\ss}lin}  \& {Springel}}{{Uhlig}
  et~al.}{2012}]{uhlig:2012.cosmic.ray.streaming.winds}
{Uhlig} M.,  {Pfrommer} C.,  {Sharma} M.,  {Nath} B.~B.,  {En{\ss}lin} T.~A.,
  {Springel} V.,  2012, \mn@doi [\mnras] {10.1111/j.1365-2966.2012.21045.x},
  \href {http://adsabs.harvard.edu/abs/2012MNRAS.423.2374U} {423, 2374}

\bibitem[\protect\citeauthoryear{{Wiener}, {Oh}  \& {Guo}}{{Wiener}
  et~al.}{2013a}]{wiener:cr.supersonic.streaming.deriv}
{Wiener} J.,  {Oh} S.~P.,   {Guo} F.,  2013a, \mn@doi [\mnras]
  {10.1093/mnras/stt1163}, \href
  {http://adsabs.harvard.edu/abs/2013MNRAS.434.2209W} {434, 2209}

\bibitem[\protect\citeauthoryear{{Wiener}, {Zweibel}  \& {Oh}}{{Wiener}
  et~al.}{2013b}]{Wien13}
{Wiener} J.,  {Zweibel} E.~G.,   {Oh} S.~P.,  2013b, \mn@doi [\apj]
  {10.1088/0004-637X/767/1/87}, \href
  {http://adsabs.harvard.edu/abs/2013ApJ...767...87W} {767, 87}

\bibitem[\protect\citeauthoryear{{Wilson}, {Couch}, {Cochran}, {Le Blanc}  \&
  {Barkat}}{{Wilson} et~al.}{1975}]{wilson:1975.m1.closure}
{Wilson} J.~R.,  {Couch} R.,  {Cochran} S.,  {Le Blanc} J.,   {Barkat} Z.,
  1975, in {Bergman} P.~G.,  {Fenyves} E.~J.,   {Motz} L.,  eds,  Texas
  Symposium on Relativistic Astrophysics Vol. 262, Seventh Texas Symposium on
  Relativistic Astrophysics. New York Academy of Sciences, Annals, pp 54--64,
  \mn@doi{10.1111/j.1749-6632.1975.tb31420.x}

\bibitem[\protect\citeauthoryear{{Winner}, {Pfrommer}, {Girichidis}, {Werhahn}
  \& {Pais}}{{Winner} et~al.}{2020}]{winner:2020.cr.harps.transport}
{Winner} G.,  {Pfrommer} C.,  {Girichidis} P.,  {Werhahn} M.,   {Pais} M.,
  2020, \mn@doi [\mnras] {10.1093/mnras/staa2989}, \href
  {https://ui.adsabs.harvard.edu/abs/2020MNRAS.499.2785W} {499, 2785}

\bibitem[\protect\citeauthoryear{Yan \& Lazarian}{Yan \&
  Lazarian}{2002}]{yan.lazarian.02}
Yan H.,  Lazarian A.,  2002, \mn@doi [Phys. Rev. Lett.]
  {10.1103/PhysRevLett.89.281102}, 89, 281102

\bibitem[\protect\citeauthoryear{{Yan} \& {Lazarian}}{{Yan} \&
  {Lazarian}}{2004}]{yan.lazarian.04:cr.scattering.fast.modes}
{Yan} H.,  {Lazarian} A.,  2004, \mn@doi [\apj] {10.1086/423733}, \href
  {https://ui.adsabs.harvard.edu/abs/2004ApJ...614..757Y} {614, 757}

\bibitem[\protect\citeauthoryear{{Yan} \& {Lazarian}}{{Yan} \&
  {Lazarian}}{2008}]{yan.lazarian.2008:cr.propagation.with.streaming}
{Yan} H.,  {Lazarian} A.,  2008, \mn@doi [\apj] {10.1086/524771}, \href
  {http://adsabs.harvard.edu/abs/2008ApJ...673..942Y} {673, 942}

\bibitem[\protect\citeauthoryear{{Zank}}{{Zank}}{2014}]{zank:2014.book}
{Zank} G.~P.,  2014, {Transport Processes in Space Physics and Astrophysics}.
 Lecture Notes in Physics Vol. 877, Springer Science+Business Media New York,
  \mn@doi{10.1007/978-1-4614-8480-6}

\bibitem[\protect\citeauthoryear{{Zweibel}}{{Zweibel}}{2013}]{Zwei13}
{Zweibel} E.~G.,  2013, \mn@doi [Physics of Plasmas] {10.1063/1.4807033}, \href
  {http://adsabs.harvard.edu/abs/2013PhPl...20e5501Z} {20, 055501}

\bibitem[\protect\citeauthoryear{{Zweibel}}{{Zweibel}}{2017}]{zweibel:cr.feedback.review}
{Zweibel} E.~G.,  2017, \mn@doi [Physics of Plasmas] {10.1063/1.4984017}, \href
  {https://ui.adsabs.harvard.edu/abs/2017PhPl...24e5402Z} {24, 055402}

\bibitem[\protect\citeauthoryear{{de la Torre Luque}, {Mazziotta}, {Loparco},
  {Gargano}  \& {Serini}}{{de la Torre Luque}
  et~al.}{2021}]{delaTorre:2021.dragon2.methods.new.model.comparison}
{de la Torre Luque} P.,  {Mazziotta} M.~N.,  {Loparco} F.,  {Gargano} F.,
  {Serini} D.,  2021, arXiv e-prints, \href
  {https://ui.adsabs.harvard.edu/abs/2021arXiv210101547D} {p. arXiv:2101.01547}

\bibitem[\protect\citeauthoryear{{le Roux}, {Matthaeus}  \& {Zank}}{{le Roux}
  et~al.}{2001}]{2001GeoRL..28.3831L}
{le Roux} J.~A.,  {Matthaeus} W.~H.,   {Zank} G.~P.,  2001, \mn@doi [\grl]
  {10.1029/2001GL013400}, \href
  {https://ui.adsabs.harvard.edu/abs/2001GeoRL..28.3831L} {28, 3831}

\bibitem[\protect\citeauthoryear{{le Roux}, {Zank}, {Li}  \& {Webb}}{{le Roux}
  et~al.}{2005}]{2005ApJ...626.1116L}
{le Roux} J.~A.,  {Zank} G.~P.,  {Li} G.,   {Webb} G.~M.,  2005, \mn@doi [\apj]
  {10.1086/430088}, \href
  {https://ui.adsabs.harvard.edu/abs/2005ApJ...626.1116L} {626, 1116}

\bibitem[\protect\citeauthoryear{{le Roux}, {Zank}, {Webb}  \& {Khabarova}}{{le
  Roux} et~al.}{2015}]{2015ApJ...801..112L}
{le Roux} J.~A.,  {Zank} G.~P.,  {Webb} G.~M.,   {Khabarova} O.,  2015, \mn@doi
  [\apj] {10.1088/0004-637X/801/2/112}, \href
  {https://ui.adsabs.harvard.edu/abs/2015ApJ...801..112L} {801, 112}

\bibitem[\protect\citeauthoryear{{van Marle}, {Casse}  \& {Marcowith}}{{van
  Marle} et~al.}{2019}]{2019MNRAS.tmp.2249V}
{van Marle} A.~J.,  {Casse} F.,   {Marcowith} A.,  2019, \mn@doi [\mnras]
  {10.1093/mnras/stz2624}, \href
  {https://ui.adsabs.harvard.edu/abs/2019MNRAS.tmp.2249V} {p.~2249}

\makeatother
\end{thebibliography}

\begin{appendix}

\section{Summary of Key Equations}
\label{sec:equations}

We summarize some of the key equations derived herein, in compact form and with the consistent RSOL formulation (Eq.~\ref{eqn:rsol.1}) included. All variables are defined in the main text.

Eq.~\ref{eqn:df.general} is the general evolution equation valid for any gyrotropic CR DF $f=f({\bf x},\,p,\,x,\,s,\,t,\,...)$, including all QLT scattering terms, to leading $\mathcal{O}(u/c)$ in all terms, written in finite-volume form (suitable for methods which evolve the DF on a grid of $\mu$):
\begin{align}
\frac{1}{\tilde{c}}\,D_{t} f  &+  \nabla \cdot (\mu\,\beta\,f\,\bhat) =   \\
\nonumber & \frac{\partial}{\partial\mu}\left[ \chi\,\left\{ -f\, \beta\,\nabla\cdot\bhat + \frac{\nu}{c}\,\left( \frac{\partial f}{\partial \mu} + \frac{\bar{v}_{A}}{v} p\,\frac{\partial f}{\partial p} \right)  \right\} \right] + \\
\nonumber & \frac{1}{p^{2}} \frac{\partial }{\partial p}\left[ p^{3}\,\left\{ (\mathbb{D}:\nabla \boldsymbol{\beta}_{u})\, f + \frac{\nu\,\chi}{c}\,\left( \frac{\bar{v}_{A}}{v}\, \frac{\partial f}{\partial \mu} + \frac{v_{A}^{2}}{v^{2}}\,p\,\frac{\partial f}{\partial p} \right) \right\} \right] .
\end{align}
Eq.~\ref{eqn:f0.eqn.final}-\ref{eqn:f1.eqn.final} take the first-two pitch-angle moments $\bar{f}_{0}$, $\bar{f}_{1}$ to derive a two-moment set of equations for $f$ (akin to radiation moments methods that do not evolve the entire $\mu$ distribution explicitly):
\begin{align}
\frac{1}{\tilde{c}}\,D_{t} \bar{f}_{0} + & \nabla \cdot  (\beta\,\bhat\,\bar{f}_{1})  
-  \mathbb{D}:\nabla\boldsymbol{\beta}_{u} \left[ 3\,\bar{f}_{0} +  \,p\,\frac{\partial \bar{f}_{0}}{\partial p} \right] \\
\nonumber & \ \ \ =  
\frac{1}{c\,p^{2}}\frac{\partial }{\partial p}\left[ p^{2}\,\left( S\,\bar{f}_{0}  
+ \tilde{D}_{p \mu}\,\bar{f}_{1}
+ \tilde{D}_{p p}\, \frac{\partial \bar{f}_{0}}{\partial p} 
\right)\right]  + \frac{j_{0}}{c}, \\
\nonumber \frac{1}{\tilde{c}}\,D_{t} \bar{f}_{1} + & 
\beta\,\momentgrad(\bar{f}_{0}) = - \frac{1}{c}\left[ \tilde{D}_{\mu\mu}\,\bar{f}_{1} + \tilde{D}_{\mu p}\,\frac{\partial \bar{f}_{0}}{\partial p} \right]
+ \frac{j_{1}}{c}, \\
\nonumber
\tilde{D}_{p p}  = \chi\,&\frac{p^{2}\,v_{A}^{2}}{v^{2}}\,\bar{\nu} 
\  , \  \
\tilde{D}_{p \mu} = \frac{p\,\bar{v}_{A}}{v}\,\bar{\nu} 
\  , \ \
\tilde{D}_{\mu\mu} = \bar{\nu} 
\  , \  \
\tilde{D}_{\mu p} = \chi\,\frac{p\,\bar{v}_{A}}{v}\,\bar{\nu}.
\end{align}
The following relations complete the closure of the moments hierarchy:
\begin{align}
\nonumber \momentgrad(q) &\equiv \bhat \cdot \nabla  \left( [1-2\,\chi]\,q \right)  + (1-3\,\chi)\,q\,\nabla \cdot \bhat  \\
\nonumber &\ \ \ = \nabla \cdot \left( \mutwo\,q\,\bhat \right) - \chi\,q\,\nabla \cdot \bhat = \bhat\cdot\left[ \nabla \cdot \left( \mathbb{D}\,q \right) \right], \\ 
\nonumber \mathbb{D} &\equiv \chi\,\mathbb{I} + \left( 1-3\,\chi \right)\,\bhat\bhat,  \\
\nonumber \chi &\equiv \frac{1-\mutwo}{2} = \frac{1}{2}\,\left[ 1 - \frac{\bar{f}_{2}}{\bar{f}_{0}} \right], \\
\nonumber \muone &\equiv \frac{\bar{f}_{1}}{\bar{f}_{0}} = \frac{F_{q}}{q\,v}, \\
\nonumber \mutwo &\approx \closurefunction\left( \muone\right) = \frac{3+4\,\muone^{2}}{5 + 2\,[4-3\,\muone^{2}]^{1/2}}. 
\end{align}
Eqs.~\ref{eqn:n.specific}, \ref{eqn:e.specific}, \ref{eqn:ek.specific} integrate these moments equations over a finite range of $p$ to define corresponding moments equations for CR number $n^{\prime}=dn/dp$, energy $e^{\prime}=de/dp$, and kinetic energy $\epsilon^{\prime}=d\epsilon/dp$ density, for a narrow range of $p$:
\begin{align}
\frac{1}{\tilde{c}}\,D_{t} n^{\prime}  + \nabla \cdot \left( \frac{F_{n}^{\prime}}{c}\,\bhat \right) &= \frac{S_{n}^{\prime}}{c}, \\ 
\nonumber \frac{1}{\tilde{c}}\,D_{t}\left( \frac{F_{n}^{\prime}}{c} \right) + \momentgrad \left( \beta^{2}\,n^{\prime} \right) &= 
-\frac{\bar{\nu}}{c^{2}}\,\left[ F_{n}^{\prime} - 3\,\chi\,\bar{v}_{A}\,n^{\prime} \right] + \frac{S_{F_{n}}^{\prime}}{c^{2}}, \\
\nonumber \frac{1}{\tilde{c}}\,D_{t} e^{\prime}  + \nabla \cdot \left( \frac{F_{e}^{\prime}}{c}\,\bhat \right) &= \frac{1}{c}\,\left[ S_{e}^{\prime} + \tilde{S}^{\prime}_{\rm sc} - \mathbb{P}^{\prime}:\nabla{\bf u} \right], \\ 
\nonumber \frac{1}{\tilde{c}}\,D_{t}\left( \frac{F_{e}^{\prime}}{c} \right) + \momentgrad \left( \beta^{2}\,e^{\prime} \right) &= 
-\frac{\bar{\nu}}{c^{2}}\,\left[ F_{e}^{\prime} - 3\,\chi\,\bar{v}_{A}\,(e^{\prime}+P_{0}^{\prime}) \right] + \frac{S_{F_{e}}^{\prime}}{c^{2}}, \\
\nonumber \frac{1}{\tilde{c}}\,D_{t} \epsilon^{\prime}  + \nabla \cdot \left( \frac{F_{\epsilon}^{\prime}}{c}\,\bhat \right) &= \frac{1}{c}\,\left[ S_{\epsilon}^{\prime} + \tilde{S}_{\rm sc}^{\prime} - \mathbb{P}^{\prime}:\nabla{\bf u} \right], \\ 
\nonumber \frac{1}{\tilde{c}}\,D_{t}\left( \frac{F_{\epsilon}^{\prime}}{c} \right) + \momentgrad \left( \beta^{2}\,\epsilon^{\prime} \right) &= 
-\frac{\bar{\nu}}{c^{2}}\,\left[ F_{\epsilon}^{\prime} - 3\,\chi\,\bar{v}_{A}\,(\epsilon^{\prime}+P_{0}^{\prime}) \right] + \frac{S_{F_{\epsilon}}^{\prime}}{c^{2}} ,
\end{align}
with $\tilde{S}_{\rm sc}^{\prime} = -(\bar{\nu}/c^{2})\,\left[ \bar{v}_{A}\,F^{\prime}_{e} - 3\,\chi\,{v}^{2}_{A}\,\left( e^{\prime} + P_{0}^{\prime} \right)  \right]$, $\mathbb{P}^{\prime}\equiv3\,P_{0}^{\prime}\,\mathbb{D}$, $P_{0}^{\prime}\equiv \beta^{2}\,e^{\prime}/3$. The spectrally-integrated equations are then obtained by integrating the above over $\int dp$. Of particular relevance is Eq.~\ref{eqn:e.total}, the spectrally-integrated total energy equation assuming most of the CR energy is ultra-relativistic:
\begin{align}
\nonumber \frac{1}{\tilde{c}}\, D_{t}& e  + \nabla \cdot \left( \frac{F_{e}}{c}\,\bhat \right) \approx \frac{S_{e}}{c} - \mathbb{P}_{e}:\nabla\boldsymbol{\beta}_{u} -\frac{\bar{\nu}_{e}}{c}\left[ \frac{\bar{v}_{A}}{c}\,\frac{F_{e}}{c} - 4\,\chi_{e}\,\frac{{v}^{2}_{A}}{c^{2}}\,e  \right], \\
 \frac{1}{\tilde{c}}\, D_{t}&\left( \frac{F_{e}}{c} \right) + \bhat\cdot (\nabla \cdot \mathbb{P}_{e} ) \approx -\frac{\bar{\nu}_{e}}{c}\,\left[ \frac{F_{e}}{c} - 4\,\chi_{e}\,\frac{\bar{v}_{A}}{c}\,e \right] + \frac{S_{F_{e}}}{c^{2}},
\end{align}
where $\mathbb{P} = \int \mathbb{P}^{\prime}\,dp \approx e\,\mathbb{D}(\chi_{e})$ and $\chi_{e}$, $\bar{v}_{A}$, $\bar{\nu}_{e}$ and other terms are understood to be the appropriate spectrally-averaged values. 
Eqs.~\ref{eqn:gas.momentum.diff}-\ref{eqn:gas.momentum} give the DF-integrated CR force on gas: 
\begin{align}
\nonumber D_{t}(\rho\,{\bf u})+... &= \sum_{s} \int  4\pi\,p^{2}\,dp\, {\Bigl\{} -\left(\mathbb{I}-\bhat\bhat \right)\cdot \left[ \nabla \cdot  \left( \mathbb{D}\,p\,v\,\bar{f}_{0} \right) \right] \\ 
 & \ \ \ \ \ \ \ \ \ \ \ \ \ \ \ \ \ \   + \bhat\,\left[ \tilde{D}_{\mu\mu} \bar{f}_{1}\,p  + \tilde{D}_{\mu p} p^{2}\,\frac{\partial \bar{f}_{0} }{\partial p} \right]  {\Bigr\}}  \\
\nonumber & =-\nabla_{\bot} \cdot \mathbb{P} + \bhat\, \sum_{s} \int dp\, \frac{\bar{\nu}}{c^{2}}\,\left[ F^{\prime}_{e} - 3\,\chi\,\bar{v}_{A}\,(e^{\prime} + P^{\prime}_{0}) \right],
\end{align}
or alternatively from Eq.~\ref{eqn:gas.momentum.riemann},
\begin{align}
 D_{t}(\rho\,{\bf u}) &+ ... + \nabla \cdot  \mathbb{P}  =  \\
\nonumber &  \bhat\,\sum_{s} \int dp\,{\Bigl \{} \momentgrad(\beta^{2}\,e^{\prime}) +  \frac{\bar{\nu}}{c^{2}}\,\left[ F^{\prime}_{e} - 3\,\chi\,\bar{v}_{A}\,(e^{\prime} + P^{\prime}_{0}) \right] {\Bigr \}} .
\end{align}
Eq.~\ref{eqn:gas.energy} gives the corresponding gas energy equation $D_{t} e_{\rm gas} = {\bf u} \cdot \left[ D_{t}(\rho\,{\bf u})\,|_{\rm cr} \right] - \int dp\,[\tilde{S}_{\rm sc}^{\prime}+S_{e}^{\prime}]$.

Eqs.~\ref{eqn:individual.mu}-\ref{eqn:individual.momentum} use the results above to derive the evolution equations for the mean values $\langle {\bf U} \rangle$ of a group of CRs with identical state $p$, $\mu$, $s$, etc. The most relevant of these is the evolution equation for the mean momentum $p$ of a group with an initially-identical $p$, after gyro and pitch-angle averaging:
\begin{align}
\nonumber \left(\frac{c}{\tilde{c}}\right)\,\frac{\dot{\langle p \rangle}}{p} &= -\left( \mathbb{D}:\nabla{\bf u}  \right) 
- \frac{v_{A}}{v}\,\left[\muone\,\delta\bar{\nu} - \chi\,\frac{\partial \delta\nu}{\partial \mu}{\Bigr|_{\muone}} \right] \\
\nonumber & \ \ \ \ \ \ \ \ \ \ + \chi\,\frac{v_{A}^{2}}{v^{2}}\left[ 2\,\bar{\nu}\,\left( 1+\beta^{2} \right) +  p\,\frac{\partial \bar{\nu}}{\partial p}{\Bigr|_{p}} \right] .
\end{align}

\end{appendix}

\end{document}